\documentclass[twocolumn,amsmath,amssymb,aps, prd, 10pt, floatfix]{revtex4-2}
\usepackage{bm}
\usepackage{color}
\usepackage{natbib}
\usepackage{amsmath}
\usepackage{graphicx}
\def\CCL#1{{\color{blue}#1}}

\newcommand{\pasp}{Publ. Astron. Soc. Pac.}

\begin{document}
\title[Stacked Hilbert-Huang Transform]{Effectiveness of Stacks in the Stacked Hilbert-Huang Transform}
\author{Lupin Chun-Che Lin$^{1}$}
\author{Chin-Ping Hu$^{2}$}
	\thanks{E-mail: cphu0821@gm.ncue.edu.tw}
\author{Chien-Chang Yen$^{3}$}
	\thanks{E-mail: yen@math.fju.edu.tw}
\author{Kuo-Chuan Pan$^{4}$}
\author{C. Y. Hui$^{5}$}
\author{Kwan-Lok Li$^{1}$}
\author{Yu-Chiung 
Lin$^{4}$}
\author{Yi-Sheng 
Huang$^{1}$}
\author{Albert K. H. Kong$^{4}$}

\affiliation{$^{1}$Department of Physics, National Cheng Kung University,  70101 Tainan City, Taiwan}
\affiliation{$^{2}$Department of Physics, National Changhua University of Education, Changhua, 50007, Taiwan}
\affiliation{$^{3}$Department of Mathematics, Fu Jen Catholic University, 24205 New Taipei City, Taiwan}
\affiliation{$^{4}$Institute of Astronomy, National Tsing Hua University, Hsinchu 30013, Taiwan}
\affiliation{$^{5}$Department of Astronomy and Space Science, Chungnam National University, Daejeon, Korea (ROK)}

\date{\today}


\begin{abstract}
The Hilbert-Huang transform (HHT) consists of empirical mode decomposition (EMD), which is a template-free method that represents the combination of different intrinsic modes on a time-frequency map (i.e., the Hilbert spectrum). 
The application of HHT involves introducing trials by imposing white noise on the signal and then calculating the ensemble mean process of the corresponding EMD to demonstrate its significance on the Hilbert spectrum. 
In this study, we develop a stacked Hilbert-Huang Transform (sHHT) method that generates the Hilbert spectrum for each trial and compiles all results to enhance the strength of the real instantaneous frequency of the main signal on the time-frequency map. 
This new approach is more sensitive to detecting/tracing the nonlinear and transient features of a signal embedded in astronomical databases than the conventional HHT, particularly when the signal experiences dramatic frequency changes in a short time. 
We analytically investigate the consistency of HHT and sHHT and perform numerical simulations to examine the dispersion of the instantaneous frequency obtained through sHHT and compare its advantages and effectiveness with those of conventional HHT. 
To confirm the feasibility of the sHHT, we demonstrate its application in verifying the signal of superorbital modulation in X-ray and binary black hole mergers in gravitational waves.
\end{abstract}


\maketitle

\section{Introduction \label{sec:Introduction}}

The development of multi-messenger astronomy in the recent decade amplifies the scope of the application in dynamical timing analysis; for example, the strain data related to the gravitational wave (GW) or the light curve of observation only contain information on time variation, and only the dynamical timing analysis is expected to discover intriguing events or the background signal \cite{Chatterji2004,Lin2015,Yanagisawa2019,HuDK2023}.
Furthermore, in order to efficiently study the chirp pattern of a candidate GW event originated from the compact binary coalescence (CBC) on the time-frequency map, Q-transform \citep{Brown91} was treated as the most common algorithm applied for a fast scan to trace the behavior of the signal while the Omicron software \citep{Robinet2020} derived from the Omega scans pipeline \citep{Chatterji2004} based on the Q-transform were then developed to perform such a multi-resolution time-frequency analysis of data obtained from the L-V-K (i.e., aLIGO-aVirgo-KAGRA) observational network \citep{BLS2022}.
Compared with the aforementioned method, we also note that the Hilbert-Huang transform (HHT) is another novel dynamical analytical approach of time series widely applied to study the variation of a signal in many subjects, including medical applications \citep{PS2002,JP2009}, finance \citep{Huang2003}, earthquakes \citep{ZSH2003,Li2016}, ocean/water waves \citep{Schlurmann2001,Veltcheva2002}, and astronomy \citep{Hu2014,Lin2020}. 
Although HHT computation is more time-consuming, investigation of GW via HHT still has many benefits, and it also has great potential to investigate the CBC events \citep{Camp2007,Kaneyama2016,Sakai2017,Akhshi2021} or burst-type GW signal (e.g., core collapsed supernovae; \cite{Takeda2021,Hu2022}).
Based on HHT, \citet{Son2018} also developed an event trigger generator (i.e., EtaGen) with a performance comparable to the Omicron claimed by the developer.

Compared with the standard fast Fourier transform and the wavelet method, HHT is an adaptive algorithm that does not impose a prior basis set on the data.
HHT decomposes an original time series data set into different intrinsic mode functions (IMFs) and then identifies the instantaneous frequency (IF) along with the associated instantaneous amplitude (IA) to further obtain the Hilbert spectrum \citep{WH2009}.
The IF of HHT varies with time, distinguishing it from the constant value determined by frequency in the Fourier analysis. 
Therefore, HHT is suitable for analyzing non-linear and non-stationary signals.
In addition, the IF of HHT can indicate a modulation of a base frequency over a small portion of the wave cycle, characterizing it as a differentiation algorithm for extracting a signal's local waveform.
Compared to the Fourier transform, HHT was claimed to provide a relatively high-resolution spectrogram on the time-frequency map to dynamically trace a signal behavior (i.e., Fig.~5 in \citet{Camp2007}).
This viewpoint is firmly established by the findings from simulated CBC \citep{Camp2007} and core collapsed supernova explosion \citep{Takeda2021} signals, and a thorough comparison of the resolution between HHT and wavelet for a CBC event clearly shows the superior advantages of using HHT (Fig.~4 of \citet{Hu2022}\CCL{)}.
Through cross-correlation of the high-resolution IF resolved from the data, HHT is also expected to check the physical time delays of events between the GW detectors \citep{Akhshi2021} and the merger or post-merger phases of binary neutron star (NS) coalescences \citep{Kaneyama2016}. 
Due to its unique characteristics, HHT is anticipated to be a powerful tool for studying nonlinear and transient GWs.
 
The original decomposition method proposed for HHT is empirical mode decomposition (EMD; \cite{Huang98}), which extracts IMFs by removing the local mean values obtained from the average of the upper and lower envelopes of a time series.
A signal with a large/broad modulation time scale can be decomposed into multiple IMFs by such an iterative process to lead to the ``mode-mixing'' or ``mode-splitting'' problems.
These problems can be alleviated by introducing white noise with finite amplitude into the data and then taking the ensemble mean of the corresponding IMFs from all simulated datasets, and this improved approach is known as the ensemble EMD (EEMD; \cite{WH2009}) method. 
Nevertheless, the merger phase of the GW signal, which occurs during the coalescence of two compact objects, spans a wide frequency range.
This complicates the decomposition into a single IMF, even with EEMD, and spreads the information across multiple intrinsic modes.
We have developed an algorithm to stack the time-frequency maps by employing the HHT multiple times to determine the final Hilbert spectrum \citep{Hu2022}.  
The pattern of the GW signal can be traced by inspecting the evolutionary track of the Hilbert energy, forming a remarkable feature in a two-dimensional series of time and IMF indices.
In this paper, we aim to systematically prove the feasibility of this stacking algorithm and conduct more detailed tests using artificial and real signals.

\section{Method \label{sec:Method}}
The HHT is an empirical tool for time-frequency analysis and employs a nonlinear approach. 
It is well known that HHT is more effective for non-stationary signals than the traditional method, which assumes constant frequency and amplitude.

The Hilbert transform of a function $x(t)$ is 
\begin{eqnarray}
\label{eqn:basic_HHT}
y(t)=\frac{1}{\pi}\mbox{p.v.}\int^\infty_{-\infty}
\frac{x(t)}{t-\tau}\,d\tau
\end{eqnarray}
where $\mbox{p.v.}$ is the Cauchy principal value of the singular integral. 
While $y(t)$ is the transform from the function $x(t)$, the analytic function can be defined 
\begin{eqnarray}
\label{eqn:zt}
z(t)=x(t)+i\ y(t)=a(t)e^{i\theta(t)}
\end{eqnarray}
and
\begin{eqnarray}
\label{eqn:at}
a(t)=\sqrt{x^2+y^2},\quad
\theta=\arctan\frac{y}{x}.
\end{eqnarray}
Here, $a(t)$ is the IA, and $\theta$ is the instantaneous phase. 
So the IF is
\begin{eqnarray}
\label{eqn:omega}
f = \frac{\omega}{2\pi}=\frac{1}{2\pi}\frac{d\theta}{dt}
\end{eqnarray}
\citet{Huang98} noted that a purely oscillatory function (or a mono-component) with a zero reference level is necessary for the IF calculation method to function appropriately.

An arbitrary time series $x(t)$ consists of internal noise and a few numbers of purely oscillatory functions. 
Therefore, a method of EMD is proposed by \citet{Huang98} to probe the IFs through the Hilbert transform. 
Each component defined as an IMF should satisfy the following conditions: (1) In the entire data set, the number of extrema and zero crossings must either be equal or differ by at most one. (2) At any given data point, the mean value of the envelopes created from only the local maxima and the local minima is zero. Moreover, an EMD decomposes a time series $x(t)$
into a sum of IMFs ($c_k$) and the residual $r_{N_m}$ as 
\begin{eqnarray}
\label{eqn:x_t}
x(t)=\sum^{N_m}_{k=1}c_k(t) + r_{N_m}(t),
\end{eqnarray}
where $N_m$ is a positive integer that represents the number of IMFs.  
The EMD can separate data into different components by their scales. 
Given that some data often contains noise of significant amplitude, a pertinent question arises: does a component (an IMF for EMD) reveal a true signal, or is it merely a result of noise?
We notice that a single IMF can either consist of signals at different scales or contain signals of similar scales within various IMF components.
It is so-called mode-mixing and is also one of the major drawbacks of EMD.
To address the scale mixing problem, \citet{Huang2008} and \citet{Huang2014} proposed the EEMD algorithm, which defines the true IMF components as the average of an ensemble of trials, 
each consisting of the signal plus a finite amplitude white noise.
EEMD contains the following steps: 
(1) add a white noise series to the targeted data; 
(2) decompose the data with the added white noise into IMFs; 
(3) repeat steps 1 and 2 again and again, but with different white noise levels each time; and 
(4) obtain the (ensemble) means of corresponding IMFs from the decomposition as the final result.
In more detail, an arbitrary $x(t)$ function under the $j$th ``artificial'' observation will be 
\begin{eqnarray}
\label{eqn:x_jt}
x_j(t)=x(t)+n_j(t),
\end{eqnarray}
where $n_j(t)$ is the $j$th realization of the white noise series \citep{WH2009}. 
Furthermore, the intrinsic mode $c_k(t)$ is defined by 
\begin{eqnarray}
\label{eqn:cj}
c_k(t)=\lim_{N_t\to\infty}\frac{1}{N_t}\sum^{N_t}_{j=1}c_{k,j}(t)
\end{eqnarray}
with
\begin{eqnarray}
\label{eqn:cjk}
c_{k,j}(t)=c_k(t)+r_{k,j}(t).
\end{eqnarray}
$N_t$ in eq.~(\ref{eqn:cj}) refers to the total number of trials for adding the white noise series, and $r_{k,j}(t)$ in eq.~(\ref{eqn:cjk}) denotes the contribution to the $k$th IMF from the white noise added during the $j$th trial in the noise-added signal.
To detect a signal that experiences an abrupt frequency change within a short time (e.g., CBCs), the IMFs $c_k(t)$ and the effects of $r_{k,j}(t)$ must be revised. 
By applying the Hilbert transform to IMFs $c_k(t)$ along with the realization of frequency as described in equation (\ref{eqn:omega}), we can visualize a time-frequency plot, and this process is referred to as time-frequency realization. 
Meanwhile, the process defined in (\ref{eqn:cj}) is called the ensemble mean process. 
 

Here, we concentrate on modifying the conventional HHT with the application of only EMD.
Let's begin by introducing the stacked spectrum of the HHT for each trial $x_j(t)$.
$j=1,2,3,\ldots,N_t$ denotes the number of trials to stack the spectrum and $N_t$ is the total number of trials. $k=1,2,3,\ldots,N_m$ in sHHT corresponds to different IMFs, $c_{k,j}$, and $N_m$ is the total number of intrinsic modes for each $x_j$. 
Compared with HHT, the proposed method directly calculates the Hilbert spectrum on these intrinsic modes without an ensemble process,
\begin{eqnarray}
c^R_{k,j}+ic^I_{k,j}, \quad k=1,2,\ldots,N_m,
\end{eqnarray}
where $c^{R}_{k,j}$ and $c^{I}_{k,j}$ denote the real and imaginary parts of the IMF, respectively.
Regarding random variables for $c^{R}_{k,j}$ and $c^{I}_{k,j}$, the amplitude ($a_{k,j}(t)$) and the phase ($\theta_{k,j}(t)$) can be given by 
\begin{eqnarray}
\label{eqn:am}
a_{k,j}(t)=\sqrt{(c^R_{k,j})^2+(c^I_{k,j})^2}
\end{eqnarray}
and
\begin{eqnarray}
\label{eqn:thetam}
\theta_{k,j}(t)=\arctan(\frac{c^I_{k,j}(t)}{c^R_{k,j}(t)})
\end{eqnarray}
where $c^R_{k,j}$ and $c^I_{k,j}=g(c^R_{k,j})$ are random variables for some function $g$. 
In our simulations, both $c^R_{k,j}$ and $c^I_{k,j}$ have the Gaussian distribution with zero mean and variance in 1 (i.e., ${\cal N}(0,1)$).
Meanwhile, $c^R_{k,j}$ and $c^I_{k,j}$ are orthogonal. 
The motivation for this derivation is to understand the distribution of the stacked Hilbert spectrum. 
The instantaneous angular frequency can be calculated by the derivative of the angle function for each intrinsic mode
\begin{eqnarray}
\omega_{k.j}=\frac{d}{dt}\theta_{k,j}
=\frac{c^R_{k,j}\frac{d}{dt}c^I_{k,j}-c^I_{k,j}\frac{d}{dt}c^R_{k,j}}{(c^R_{k,j})^2+(c^I_{k,j})^2}
\end{eqnarray}
Based on it, we are going to explain the detailed calculations used in sHHT.
Compared with (\ref{eqn:omega}), it follows that
\begin{eqnarray}
\label{eqn:omgjk}
f_{k,j}(t) = \frac{\omega_{k,j}(t)}{2\pi}
=\frac{1}{2\pi}\frac{d}{dt}\theta_{k,j}(t)=\frac{1}{2\pi}\frac{{\cal B}_{k,j}}{{\cal A}_{k,j}},
\end{eqnarray}
where
\begin{eqnarray*}
{\cal A}_{k,j}=(c^R_{k,j})^2+(c^I_{k,j})^2,
\quad
{\cal B}_{k,j}=c^R_{k,j}\frac{d}{dt}c^I_{k,j}-c^I_{k,j}\frac{d}{dt}c^R_{k,j}
\end{eqnarray*}
Combining (\ref{eqn:am}) and (\ref{eqn:omgjk}), we obtain the IF and IA to construct the time-frequency presentation/realization, and the Hilbert spectrum for the $k$-th intrinsic mode of the $j$-th trial ($c^R_{k,j}$) can be represented as ${\cal I}_{k,j}$. 
The sHHT is to ensemble all of the time-frequency realization ${\cal I}_{k,j}$,
\begin{eqnarray}
\mbox{sHHT}=\frac{1}{N_tN_m}\sum^{N_m}_{k=1}\sum^{N_t}_{j=1}
{\cal I}_{k,j}
\end{eqnarray}
Therefore, there are $N_m$ time-frequency realization images for HHT and $N_m\times N_t$ time-frequency realization images for sHHT.

To estimate the deviation of IF measured between the HHT without the ensemble process and the sHHT, we assume that $c^R_{k,j}=G^R_k+n_{k,j}$ and $c^I_{k,j}=G^I_k+{\tilde n}_{k,j}$, where $G^R_k$ is a real signal, $n_{k,j}$
is white noise, and $\tilde n_{k,j}=g(n_{k,j})$. 
The IF measured from HHT without an ensemble process has the coefficients,
\begin{eqnarray}
\label{eqn:AB_k}
\nonumber
{\cal A}_k=(G^R_k)^2+(G^I_k)^2,
\\
{\cal B}_k=G^R_k\frac{d}{dt}G^I_k-G^I_k\frac{d}{dt}G^R_k
\end{eqnarray}
and the IF measured from sHHT has the coefficients,
\begin{eqnarray}
{\cal A}_{k,j}=
(G^R_k+n_{k,j})^2+(G^I_k+{\tilde n}_{k,j})^2,
\nonumber
\end{eqnarray}
and
\begin{eqnarray}
\label{eqn:AB_kj}
{\cal B}_{k,j}&=&(G^R_k+n_{k,j})\frac{d}{dt}(G^I_k+{\tilde n}_{k,j})
\nonumber\\  
&& -(G^I_k+{\tilde n}_{k,j})\frac{d}{dt}(G^R_k+n_{k,j})
\end{eqnarray}
respectively. 
If ${\cal A}_k$ and ${\cal B}_k$ are not vanished, then
\begin{eqnarray}
\frac{{\cal B}_{k,j}}{{\cal A}_{k,j}}
=\frac{{\cal B}_k}{{\cal A}_k}\times \frac{1+\beta_{k,j}/{\cal B}_k}{1+\alpha_{k,j}/{\cal A}_k}
\end{eqnarray}
where 
\begin{eqnarray}
\alpha_{k,j}
=2G^R_kn_{k,j}+n^2_{k,j}+2G^I_{k}\tilde{n}_{k,j}
+{\tilde n}^2_{k,j}.
\nonumber
\end{eqnarray}
and
\begin{eqnarray}
\label{eqn:beta_kj}
\beta_{k,j}
&=& G^R_k\frac{d}{dt}(\tilde{n}_{k,j})
+n_{k,j}\frac{d}{dt}G^I_k + n_{k,j}\frac{d}{dt}(\tilde{n}_{k,j})
\nonumber\\
&&- G^I_k\frac{d}{dt}(n_{k,j})
-{\tilde n}_{k,j}\frac{d}{dt}G^R_k - {\tilde n}_{k,j}\frac{d}{dt}(n_{k,j}).
\end{eqnarray}

Let us denote $E(X)$ as the expectation of a random variable $X$.
The deviation ratio of IF measured from the sHHT compared with the HHT without the ensemble process is 
\begin{eqnarray}
\label{eqn:calD}
{\cal D}_k=
E[\frac{1+\beta_{k,j}/{\cal B}_k}{1+\alpha_{k,j}/{\cal A}_k}]
\end{eqnarray}
This computation is to understand the dispersion of IF
(${\cal D}$) in (\ref{eqn:calD}) using sHHT. 
However, it is not linear, and there is no analytic solution. 
If 
\begin{eqnarray}
\label{eqn:E0E}
E(n_{k,j})=0=E(\tilde n_{k,j}),
\end{eqnarray}
and
\begin{eqnarray}
\label{eqn:E2E}
E(n^2_{k,j})=E(\tilde n^2_{k,j})=\sigma^2_k, 
\end{eqnarray}
According to $\alpha_{k,j}$ defined in eq.~(\ref{eqn:beta_kj}), then we have 
\begin{eqnarray}
\label{eqn:E1}
E[\alpha_{k,j}/{\cal A}_k]
=
2\frac{\sigma^2_k}{{\cal A}_k},
\end{eqnarray}
and
\begin{eqnarray}
\label{eqn:E2}
E[\beta_{k,j}/{\cal B}_k]
&=&\frac{1}{{\cal B}_k}
E[n_{k,j}\frac{d}{dt}({\tilde n}_{k,j})
-{\tilde n}_{k,j}\frac{d}{dt}(n_{k,j})]\nonumber\\
&=&\mu\frac{2\sigma^2_k}{{\cal B}_k}, 
\end{eqnarray}
where $\mu$ is an amplification factor independent with $\sigma^2_k$ and ${\cal B}_k$.
We note that the conditions in equations~(\ref{eqn:E0E}) and (\ref{eqn:E2E}) can be verified by numerical approaches.

From the eq.~(\ref{eqn:E2}) for the discretization $\Delta t$ in the difference, it can be yield
\begin{eqnarray}
\label{eqn:EE1}
E[\beta_{k,j}/{\cal B}_k]
&=&\frac{1}{{\cal B}_k}
E[n_{k,j}\frac{d}{dt}({\tilde n}_{k,j})
-{\tilde n}_{k,j}\frac{d}{dt}(n_{k,j})]\nonumber\\
&\approx&\lambda\frac{2\sigma^2_k}{\Delta t{\cal B}_k}.
\end{eqnarray}
Here, $\mu=\lambda/\Delta t$, and $\lambda$ is a factor to count for the amplification over the discretization. 
No analytical value can be provided to $\lambda$, and therefore, we further employ the numerical simulation to evaluate this value. 
We assume a Gaussian noise $n$ with ${\cal N}(0,1)$ and $\Delta t$=1. 
Then we calculate the Hilbert transform of $\tilde n$ for $n$ to evaluate the values
\[
n_{k,j}[\tilde n_{k,j+1}-\tilde n_{k,j}]-
\tilde n_{k,j}[ n_{k,j+1}-n_{k,j}]
\]
for each position. 
The expectations of the above formula are simulated based on the number of points $N$ and the number of trials $K$. 
The expectation $e$ and the standard deviation $d$ are shown in the top sub-table and bottom sub-table in Table~\ref{tbl:ed}. 
From the observation of Table~\ref{tbl:ed}, the value $\lambda$ is estimated to be $\approx 0.63\pm0.03$.
\begin{center}
\begin{table}[h!]
\begin{flushleft}
    \begin{tabular}{|c||c|c|c|c|c|c|}\hline
    $N\backslash K$ & $32$ & $64$ & $128$ & $256$ & $512$ & $1024$ \\ \hline\hline
    $32$  &0.6662  &0.6300  &0.6410  &0.6389  &0.6319   &0.6339\\ \hline
    $64$  &0.6446  &0.6482  &0.6299  &0.6342  &0.6408   &0.6362\\ \hline
    $128$ &0.6568  &0.6352  &0.6365  &0.6393  &0.6402   &0.6382\\ \hline
    $256$ &0.6471  &0.6423  &0.6363  &0.6379  &0.6390   &0.6369\\ \hline
    $512$ &0.6410  &0.6401  &0.6381  &0.6354  &0.6365   &0.6367\\ \hline
    $1024$&0.6476  &0.6415  &0.6396  &0.6385  &0.6367   &0.6364\\ \hline \hline
    $32$  &0.148  &0.117  &0.071  &0.045  &0.032   &0.028\\ \hline
    $64$  &0.155  &0.092  &0.072  &0.051  &0.037   &0.023\\ \hline
    $128$ &0.148  &0.096  &0.077  &0.054  &0.040   &0.029\\ \hline
    $256$ &0.155  &0.115  &0.074  &0.051  &0.032   &0.027\\ \hline
    $512$ &0.147  &0.108  &0.079  &0.048  &0.035   &0.026\\ \hline
    $1024$&0.144  &0.106  &0.074  &0.050  &0.038   &0.025\\ \hline \hline
    \end{tabular}
    \caption{{\footnotesize $e$ in the top sub-table and the standard deviation $d$ in the bottom sub-table. These values obtained by simulation are used to evaluate the value $\lambda\approx e\pm d$.}}
    \label{tbl:ed}
\end{flushleft}
\end{table}
\end{center}

Now, we can derive the deviation ratio ${\cal D}_k$ for each mode from Equation~(\ref{eqn:E1}), where
\begin{eqnarray}
{\cal D}_k&=&E[1+\beta_{k,j}/{\cal B}_k-\alpha_{k,j}\beta_{jk}/{\cal A}_k{\cal B}_k+\cdots]\nonumber\\
&\approx & 1+E[\beta_{k,j}/{\cal B}_k] \approx 1+1.26\frac{\sigma^2_k}{{\cal B}_k} 
\label{eqn:DK}
\end{eqnarray}
whenever $\alpha_{k,j}$ and $\beta_{k,j}$ are small enough, and $\Delta{t} =1$.
$\alpha_{k,j}$ and $\beta_{k,j}$ in equation~\ref{eqn:beta_kj} correspond to the imposed noise to perform sHHT, while the entire process reverts to the conventional HHT if these two terms vanish.
Moreover,  ${\cal D}_{k}-1$ can be treated as the average of all the trials at the position $(t,\omega_t)$ along the frequency axis. 
Once we select a great ${\cal B}_k$ or omit ${\sigma^2_k}$ -- indicating that the additional input noise has a minimal impact -- then ${\cal D}_k$ approaches unity.
This signifies that formula (\ref{eqn:AB_kj}) returns to formula (\ref{eqn:AB_k}), demonstrating the consistency between the conventional HHT and sHHT methods.
The additional noise introduced by sHHT leads to a frequency dispersion compared to the exact signal in the measurement; nevertheless, the appropriate choice in variance $\sigma^2$ and the strength of ${\cal B}_k$ can only limit deviations to a small range around the exact IF of the signal. 
The true behavior of the exact IF can be further clarified by accumulating a large number of trials, which enhances the results through a stacked effect.
To numerically constrain the additional input noise level associated with data exhibiting various signal-to-noise ratios (SNRs) to track the exact signal, we also illustrate this correlation in different signal examples in \S~\ref{sec:dispersion}.

\begin{figure}
    \centering
    \includegraphics[width=\linewidth]{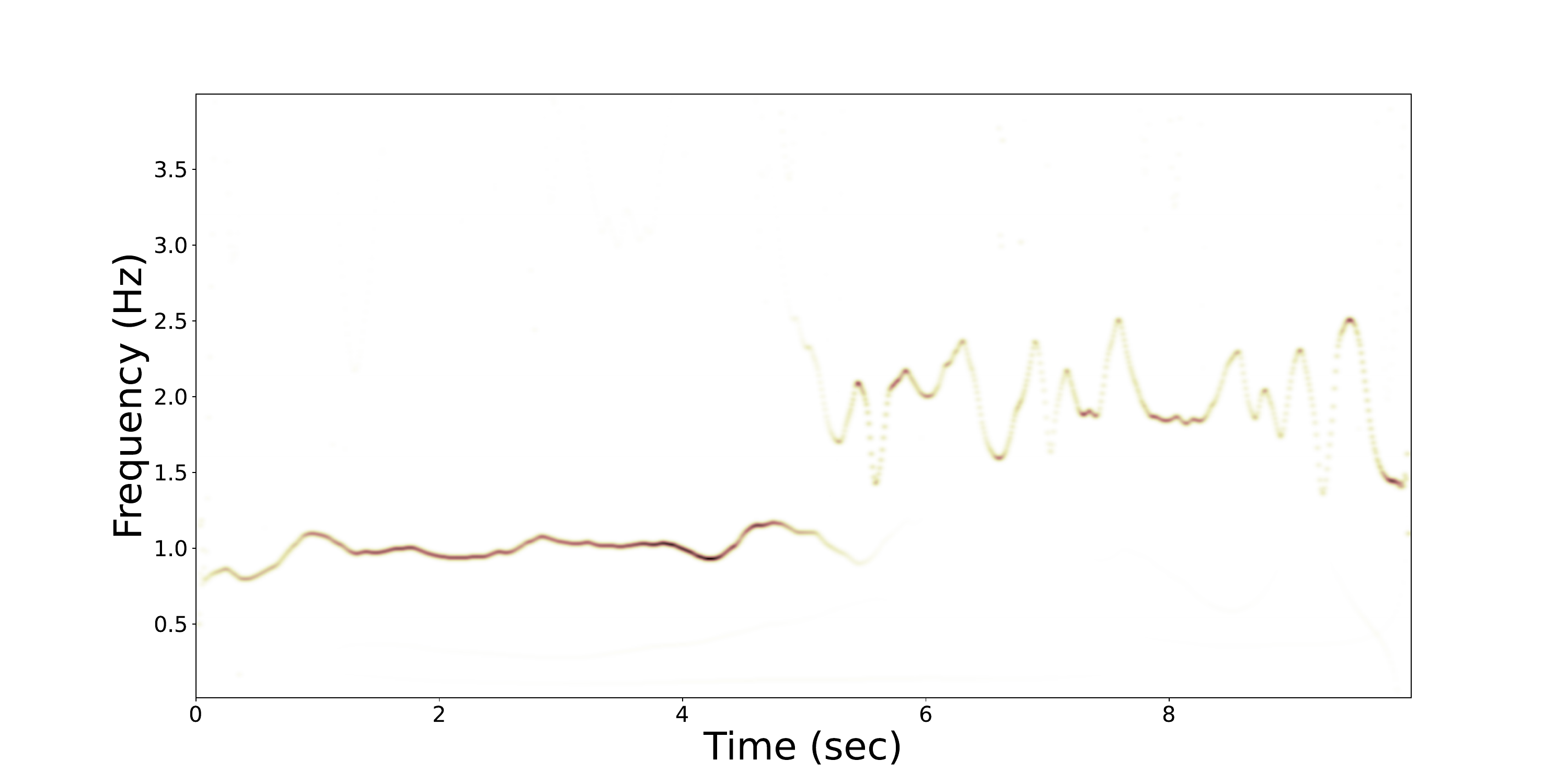}
    \includegraphics[width=\linewidth]{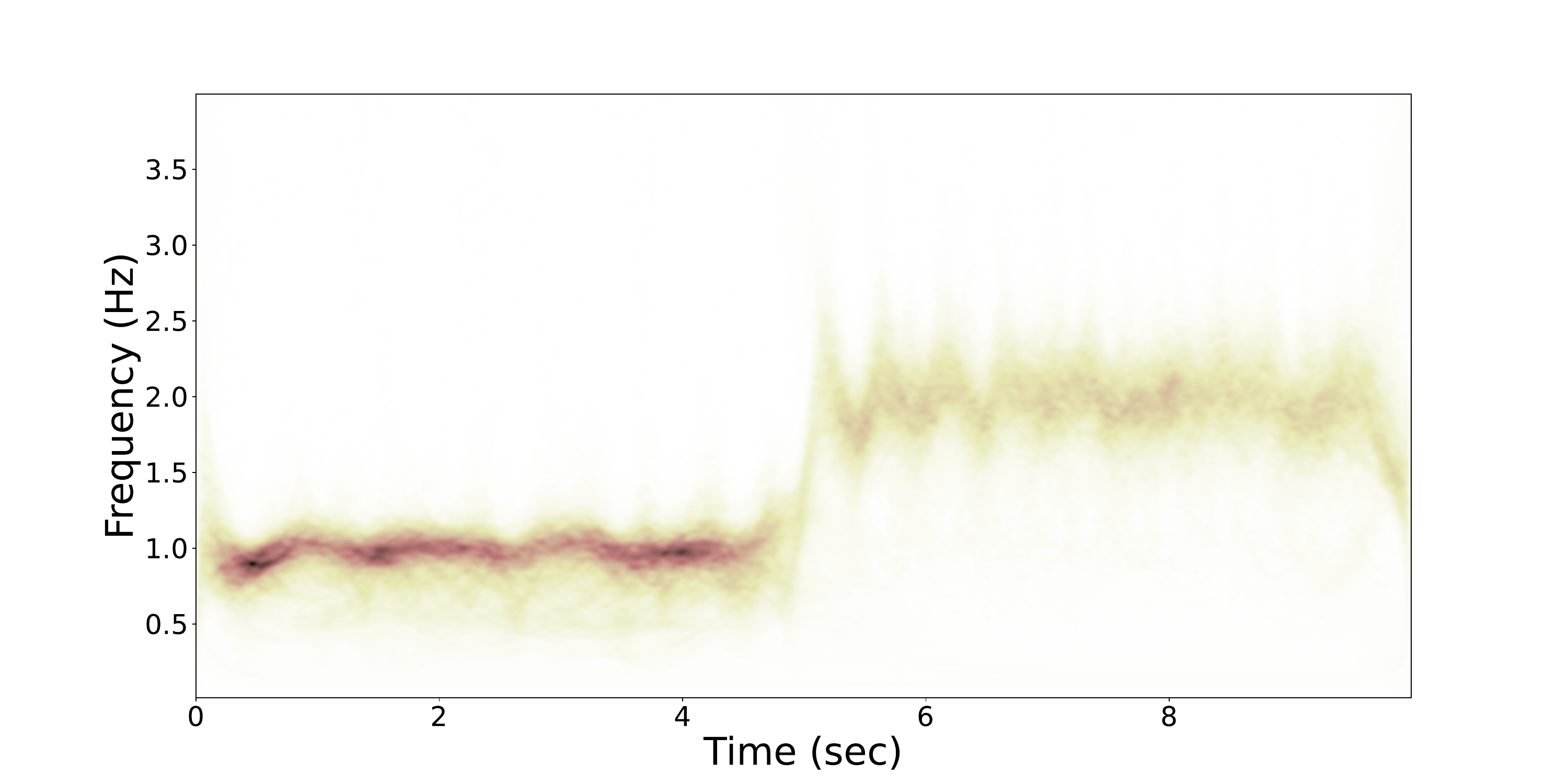}
    \caption{{\footnotesize A sinusoidal signal of the frequency change from 1\,Hz to 2\,Hz at 5\,s with $SNR=5$. The external input noise level of 0.5 was added to perform EEMD in HHT and sHHT. The Hilbert spectrum obtained via EEMD is shown in the top panel, and the stacked Hilbert spectrum is demonstrated in the bottom panel.}}
    \label{sHHT_discont}
\end{figure}

In the following section, we will outline the advantages of the stacked effect when utilizing sHHT. Additionally, we will highlight the limitations of the HHT in detecting abruptly changing waves within a short timeframe, specifically regarding the ensemble mean process prior to time-frequency realization. 
We have observed that the reverse process, employing time-frequency realization prior to the ensemble mean process, is more suitable than the original or conventional HHT. 
An example is provided below.
This simulation features a signal that experiences a sudden frequency jump from $f=$ 1\,Hz to 2\,Hz at $t=5$\,s. 
The signal is represented by the sinusoidal time series of $\sin (2\pi ft)$ with an SNR of 5, while the SNR was determined by the ratio of the signal's amplitude to the average level of internal noise. 
Please note that the simulated signal is given in an arbitrary unit with amplitude in $O(1)$. 
We added an external input noise level of 0.5 to perform sHHT to demonstrate the stacked effect compared to the EEMD applied in conventional HHT, and the noise level is universally used to describe the standard deviation ratio between the time series of the noise and signal throughout this manuscript.
The results for HHT and sHHT are shown in the top and bottom panels of Figure~\ref{sHHT_discont}, respectively. 
While we understand that the additional input noise influences the dispersion of the IF, it is evident that the 1\,Hz signal, which appears between 0 and 5 seconds, and the 2\,Hz signal, which appears after 5 seconds, exhibit a flatter profile when analyzed with the sHHT compared to the standard HHT. 
This flattening is attributed to the enhancement provided by the stack effect.
Additionally, the relationship between 1\,Hz and 2\,Hz is more evident for sHHT compared to HHT near the point $t=5$\,s.
Based on the observations above, we conclude that the visualization achieved through sHHT is clearer than that obtained through HHT.
In the next section, we will numerically analyze the effectiveness of stacks in sHHT using a signal with a stationary frequency, as well as linear and exponential chirp signals.

\section{Estimation of the Frequency Dispersion}
\label{sec:dispersion}
We have analytically derived the frequency deviation measured by conventional HHT and our improved method in the previous section.
The effects of this dispersion can be shown through various examples, emphasizing that sHHT is more effective than HHT for tracing signal evolution.
Here we present three examples: stable signals, linear chirp signals, and exponential chirp signals, to demonstrate the effectiveness of the sHHT method. 
Additionally, we investigate how frequency dispersion changes with variations in the SNRs of a dataset and different levels of external input noise while applying our algorithm.
The definition of the SNR and noise level for the mocked data set is the same as described in the previous section. The internal and input noise is assumed to be white Gaussian noise ${\cal N}(0,n\sigma)$, where $n$ denotes the significance of the input noise level.

\subsection{Stable Sinusoidal Signal}
\begin{figure*}[tp]
\centering
\includegraphics[width=8.0cm,height=5.2cm]{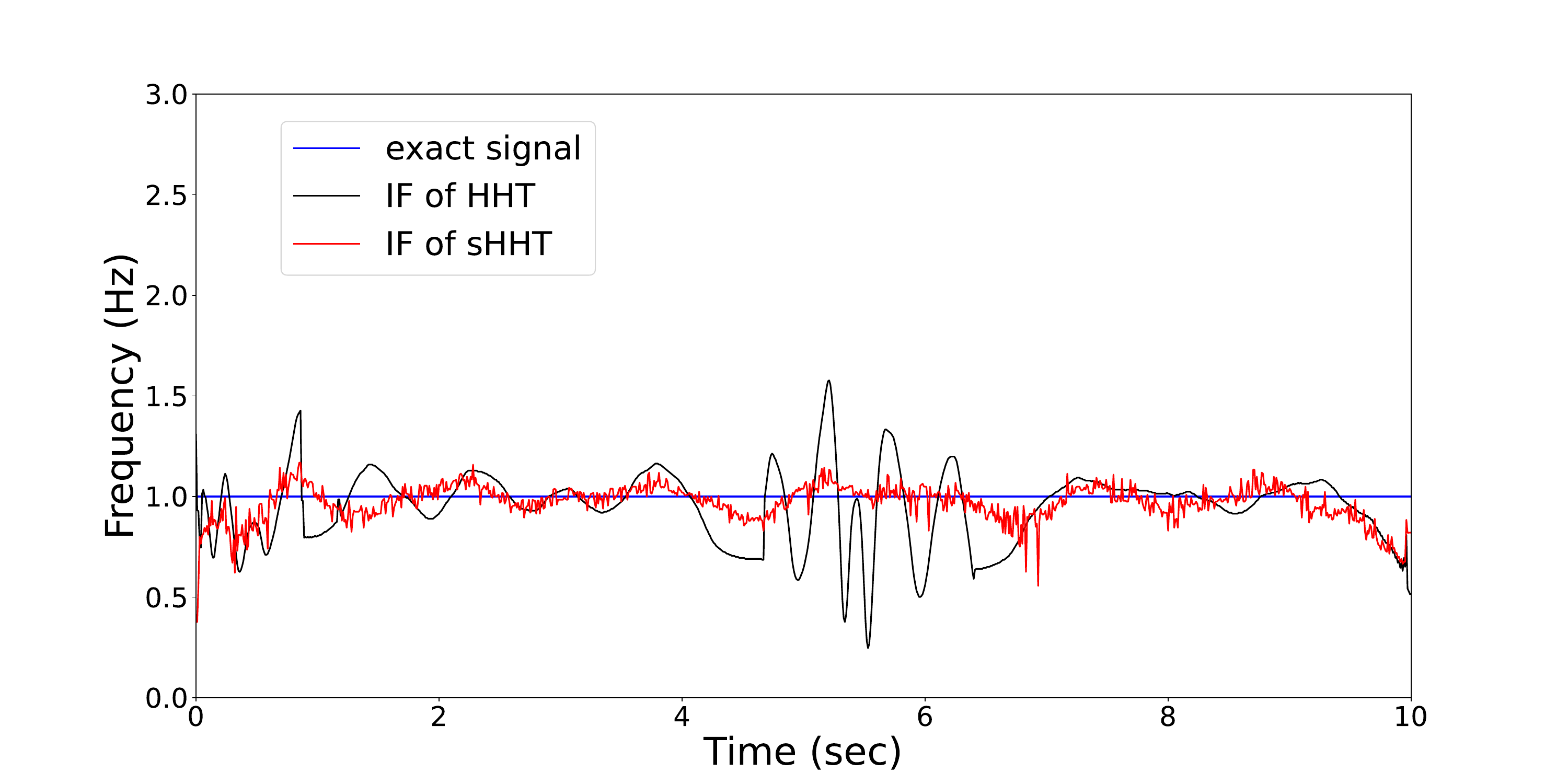}
\includegraphics[width=9.5cm,height=5.2cm]{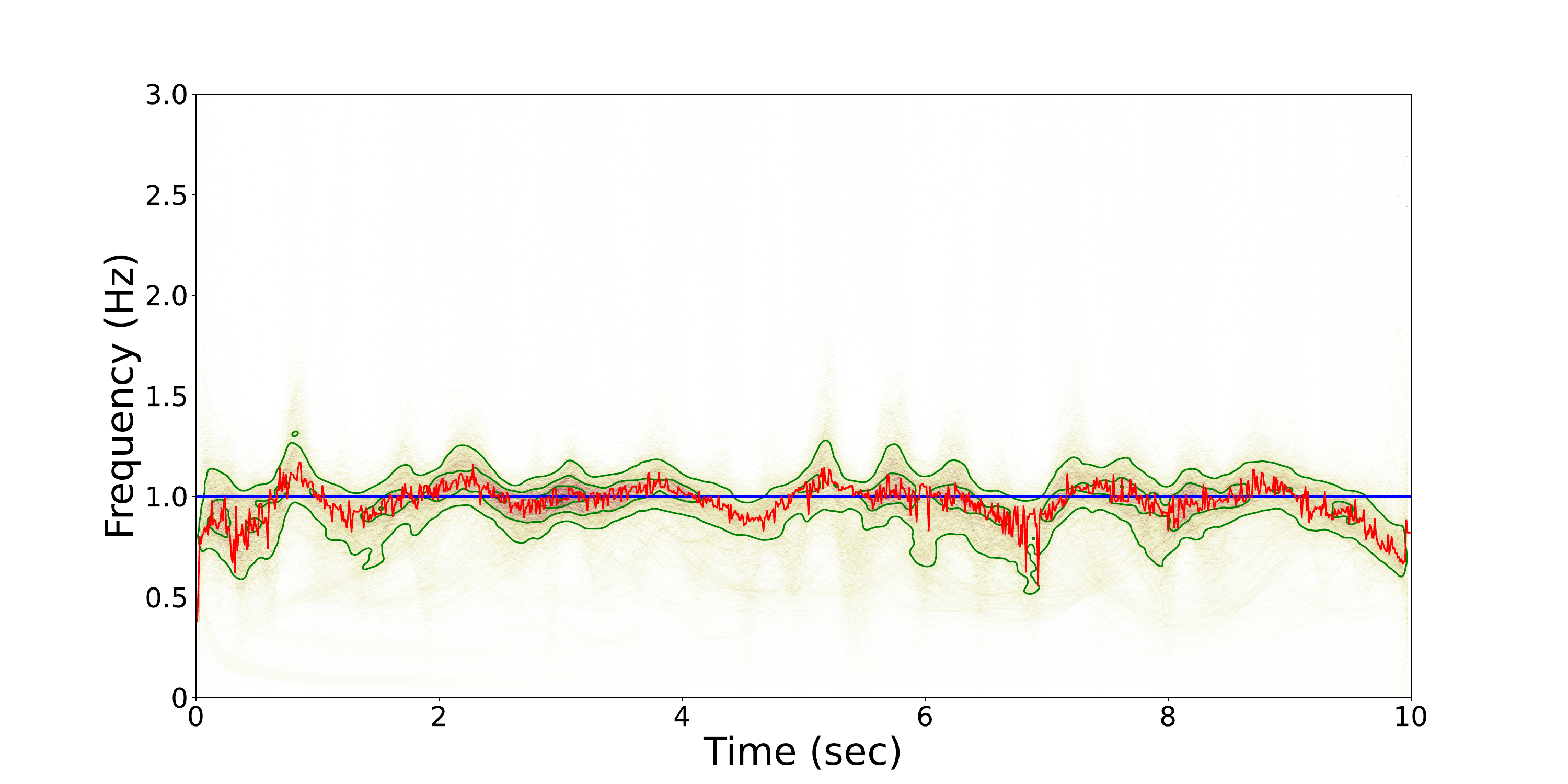}
\caption{{\footnotesize IFs of the stable sinusoidal signal obtained from the conventional HHT and sHHT. The stable signal was simulated with a frequency of 1\,Hz and SNR = 2. The external noise level imposed on this signal is 0.3. The left panel presents the real signal frequency labeled by the blue color and the IFs obtained from the conventional HHT and sHHT labeled by the black and red colors, respectively. The right panel presents the IF obtained from the sHHT imposed on the Hilbert spectrum. The green contour shows the Hilbert energy levels displayed on the sHHT time-frequency map.}} 
\label{sin_disp1}
\end{figure*}

The time series of the signal is generated by $\sin(\omega t)$, where $\omega = 2\pi f$ and $f$ is the frequency of the signal fixed at 1\,Hz. 
The evolving IF, derived from conventional HHT and sHHT, is illustrated in the left panel of Fig.~\ref{sin_disp1}.  
The IF determined by the sHHT method is closer to the actual frequency of our signal compared to the IF obtained from the conventional HHT. 
This comparison shows a clear deviation in the IFs between the traditional method and our new approach, thus verifying our derivation in the previous section.
The right panel of Fig.~\ref{sin_disp1} demonstrates the spectrum generated by sHHT, and the ``stacked'' method enhances the contour level to closely resemble the exact signal as depicted in the color map.

\begin{figure*}
\centering
\includegraphics[width=8.3cm,height=5.2cm]{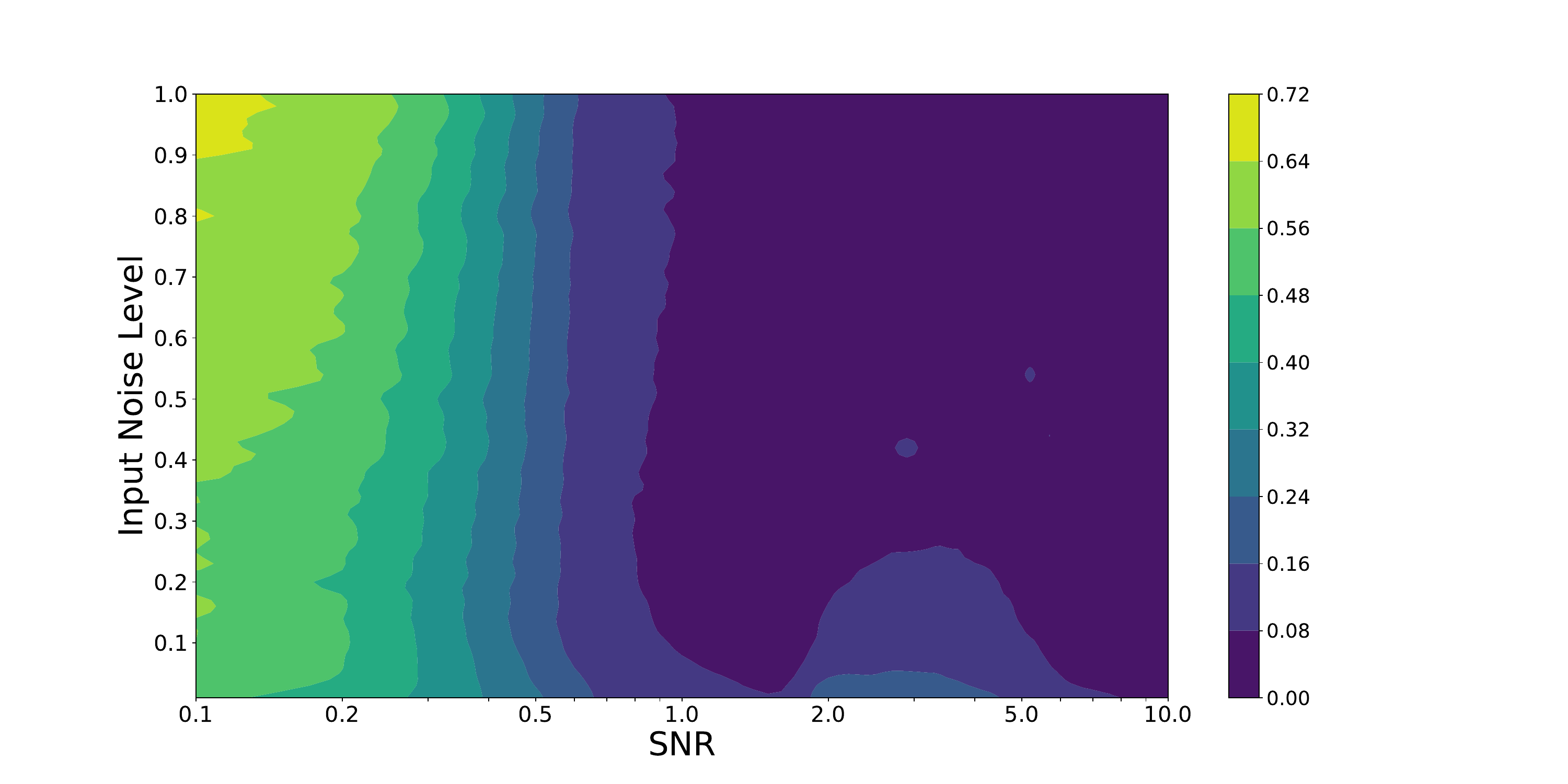}
\includegraphics[width=9.5cm,height=5.2cm]{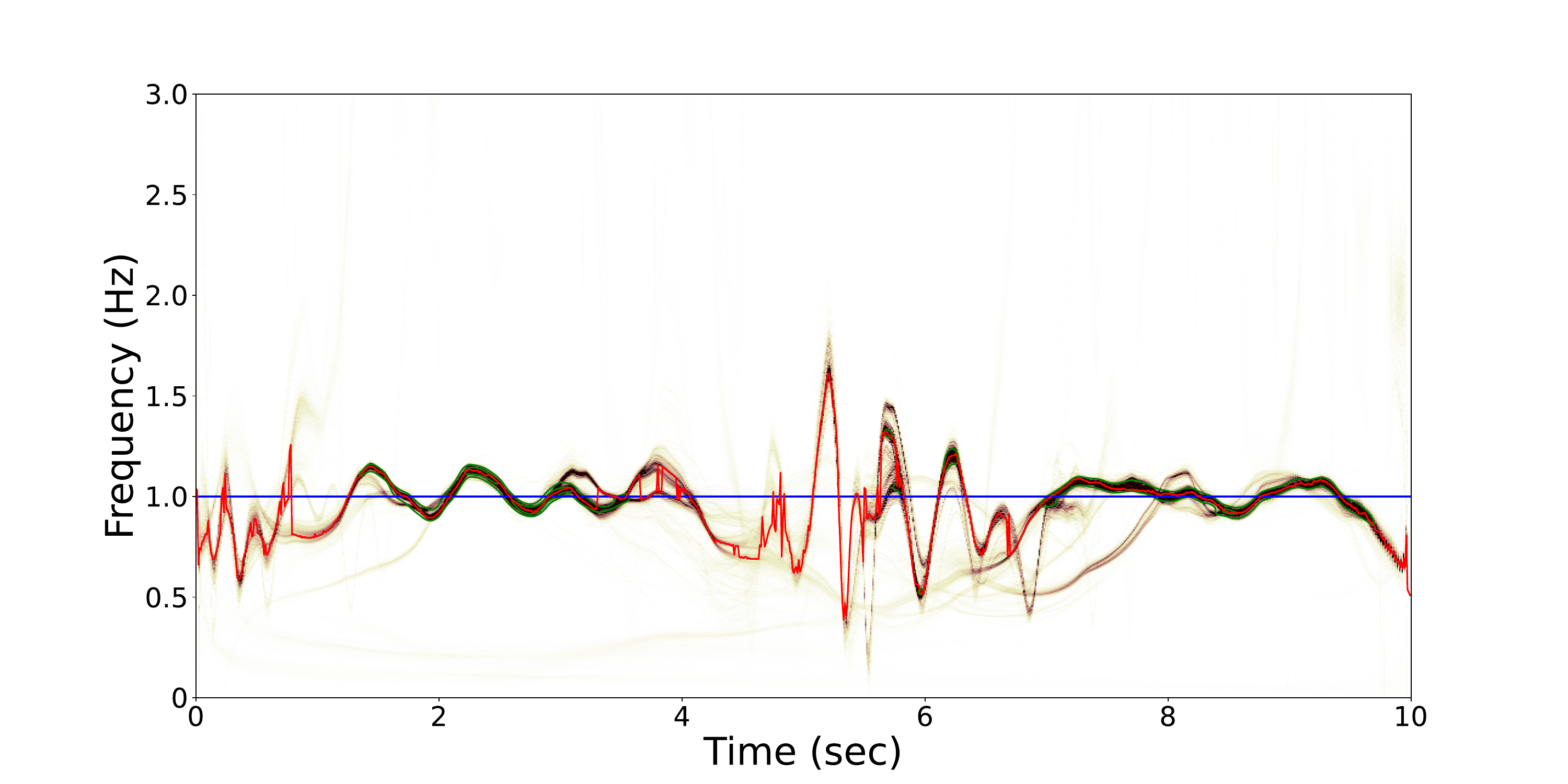}
\caption{{\footnotesize Frequency dispersion of a stationary sinusoidal signal measured by the sHHT and the IF of a specific stationary signal. The left panel demonstrates the frequency deviation ratio of the IF for a stable signal, as determined by the sHHT, compared to the exact frequency (i.e., 1\,Hz), under various SNRs and input noise levels. The right panel presents the IF obtained from sHHT imposed on the Hilbert spectrum. The signal was specifically simulated with the SNR = 2 and the imposed noise level of only 0.01.  The green contour shows the Hilbert energy levels on the sHHT time-frequency map.}} 
\label{sin_disp2}
\end{figure*}

The left panel of Fig.~\ref{sin_disp2} shows the frequency dispersion ratio (i.e., the frequency dispersion divided by the actual frequency) yielded from sHHT for a stable signal that varied with different SNRs and the imposed noise level.  
If we increase the noise level imposed on the signal, we expect that the frequency dispersion may become larger as the simulated sample maintains the same SNR. 
Nevertheless, intriguing variations in dispersion can be observed as the SNR of the signal increases. 
For instance, in the left panel of Fig.~\ref{sin_disp2}, the dispersion increases when the SNR of the signal is between 2 and 8 with a low input noise level (i.e., $\lesssim 0.3$). 
We therefore simulate a time series with an SNR of 2 and an imposed external noise level of only 0.01 to be analyzed using sHHT as shown in the right panel of Fig.~\ref{sin_disp2}.
Compared to Fig.~\ref{sin_disp1}, the mode mixing problem is more pronounced, even though the noise level is two orders of magnitude lower in this new example.

\begin{figure*}
\centering
\includegraphics[width=8.0cm,height=5.2cm]{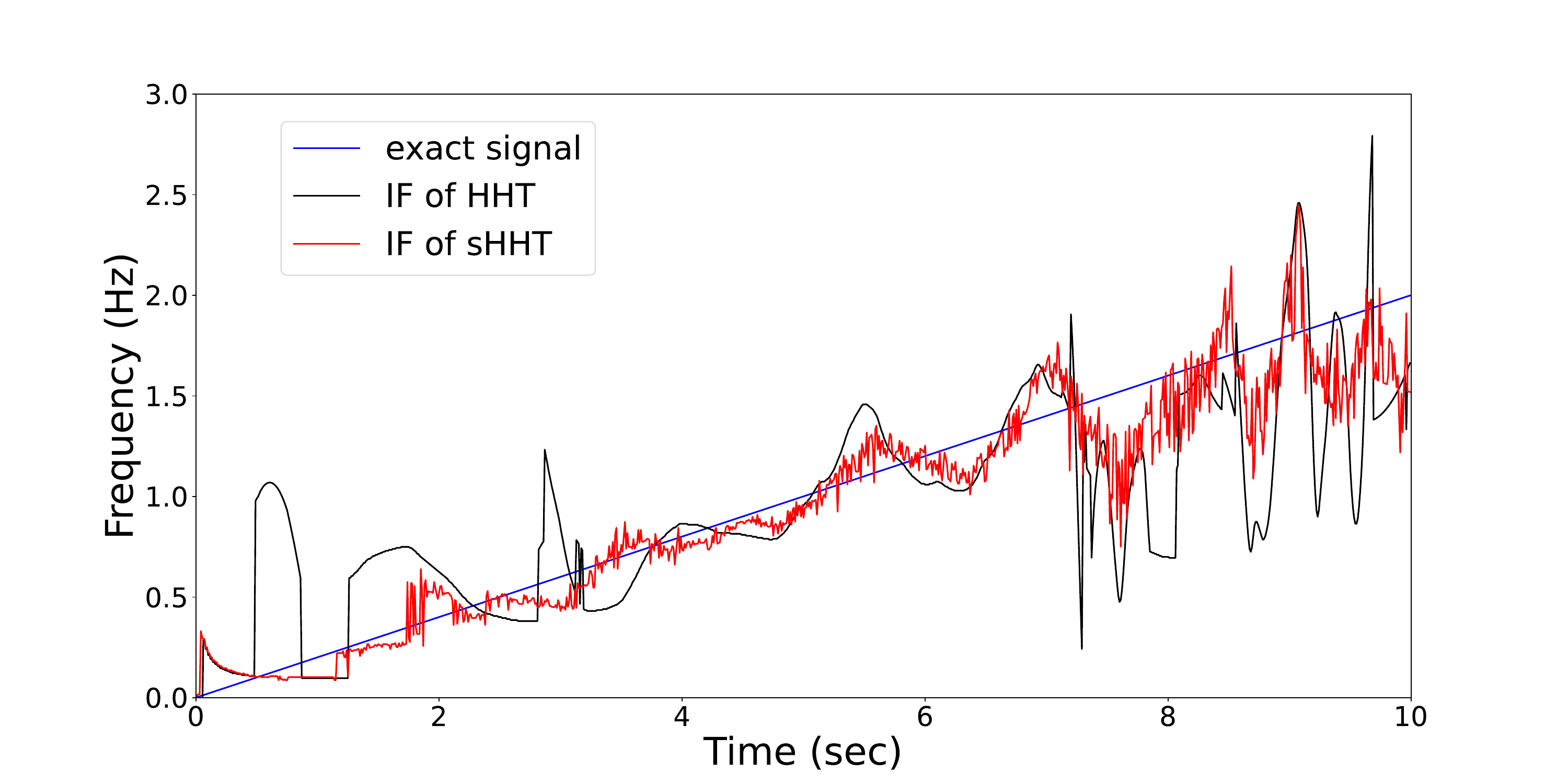}
\includegraphics[width=9.5cm,height=5.2cm]{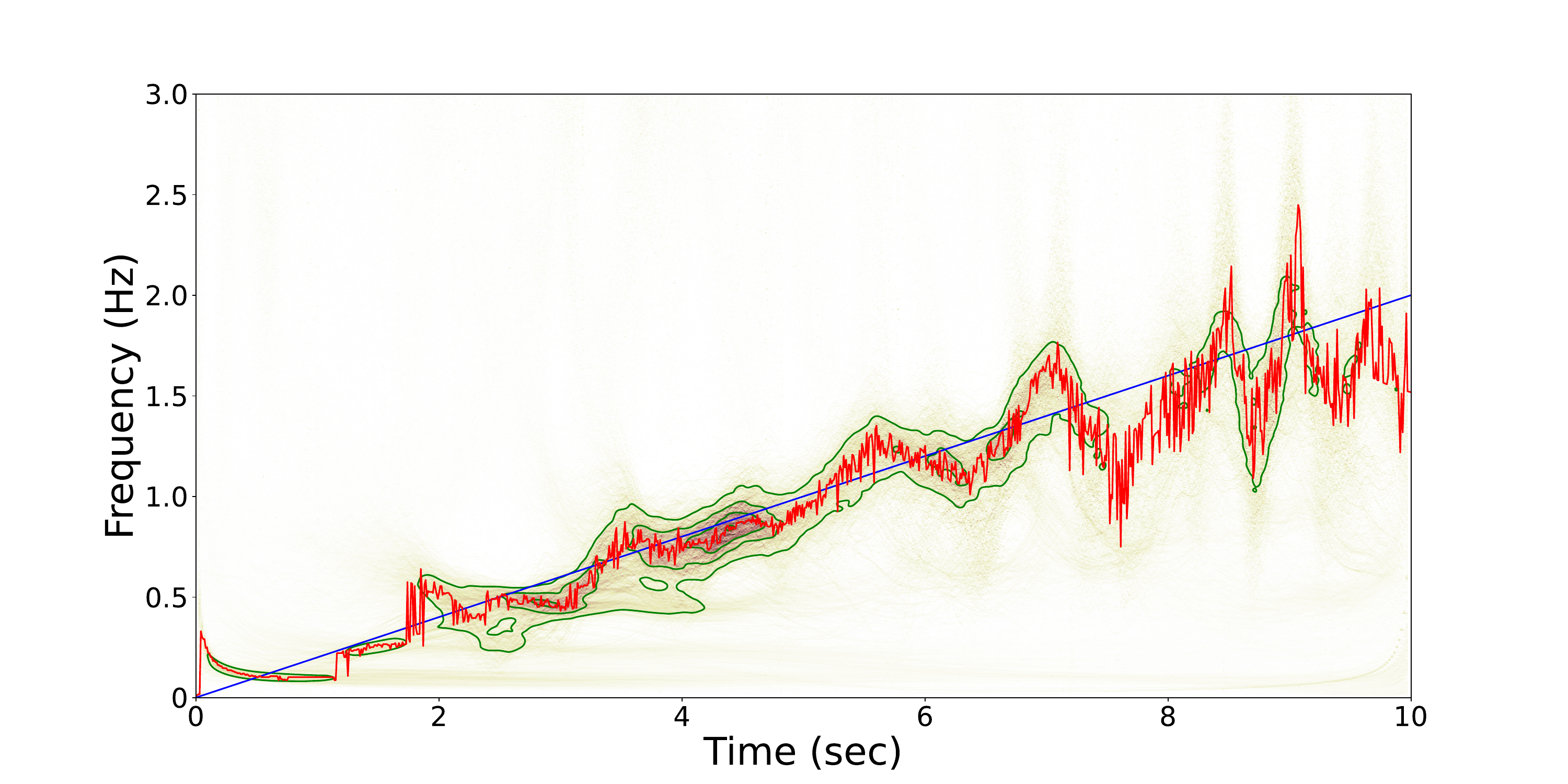}
\caption{{\footnotesize IFs of the linear chirp signal resolved by the conventional HHT and sHHT. The chirp signal was simulated with the SNR = 1 and the frequency of $0.2\cdot t+10^{-3}$, which evolves with time. The external noise level imposed on this signal is 0.6. The left panel shows evolving blue real signal frequency over time, along with IFs determined by the conventional HHT and sHHT, denoted by the black and red colors, respectively. The right panel presents the IF obtained from the sHHT imposed on the Hilbert spectrum. The green contour shows the Hilbert energy levels on the sHHT time-frequency map.}} 
\label{linearchirp_disp1}
\end{figure*}

\begin{figure*}[tp]
\centering
\includegraphics[width=8.3cm,height=5.2cm]{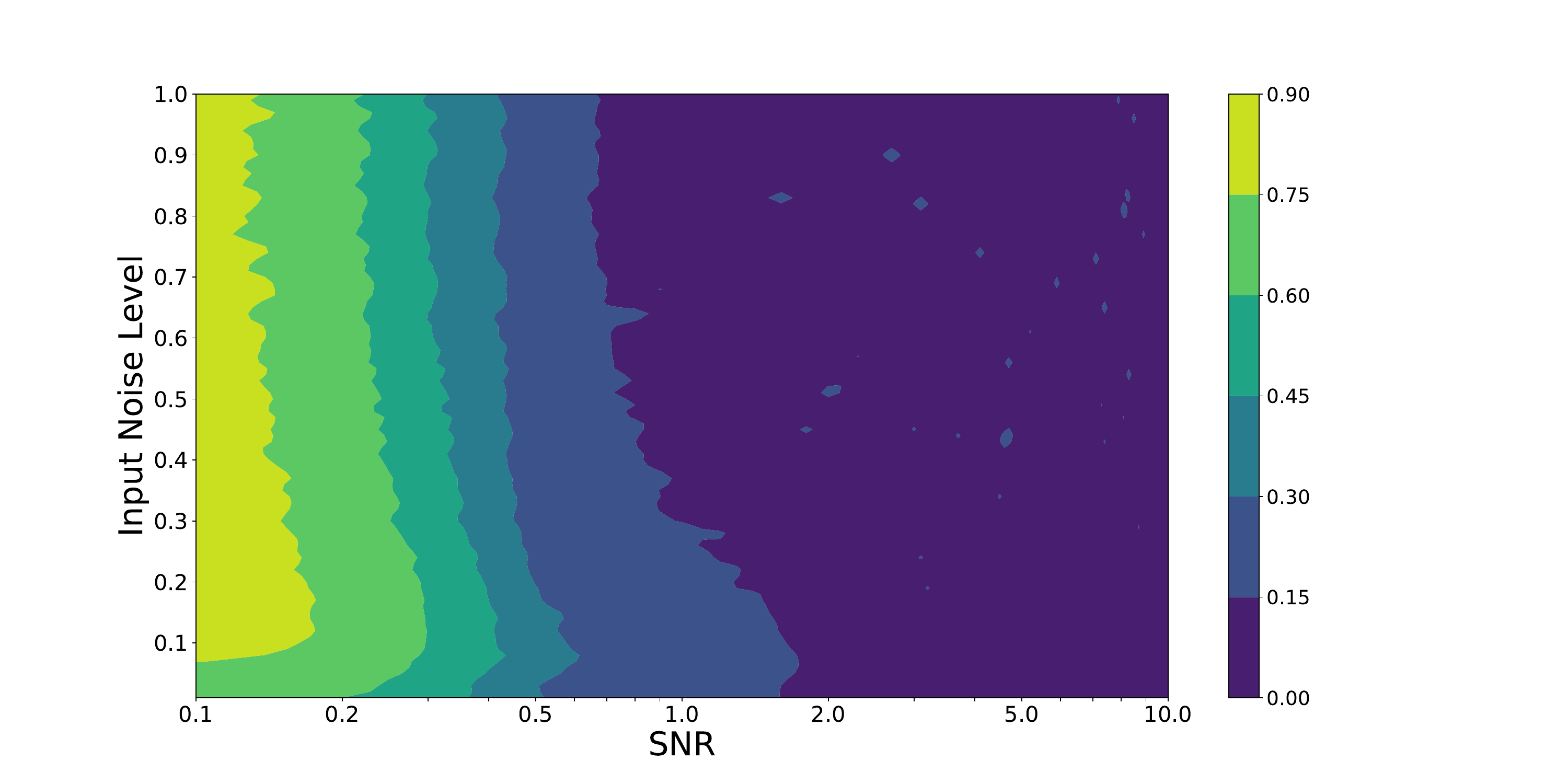}
\includegraphics[width=9.5cm,height=5.2cm]{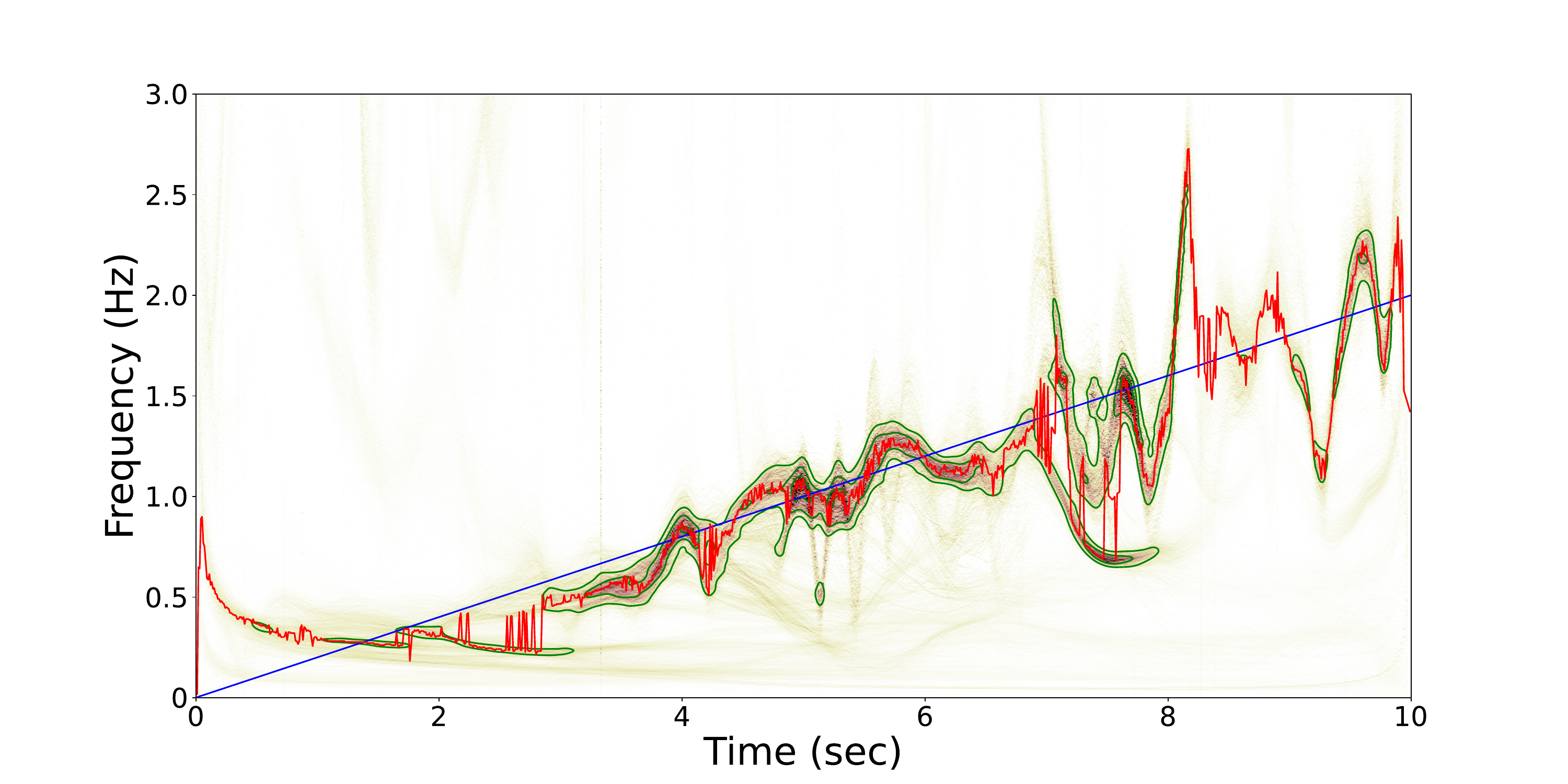}
\caption{{\footnotesize Frequency dispersion ratio of the linear chirp signal estimated by the sHHT and the IF of a specific linear chirp signal. The left panel illustrates the variation in frequency dispersion ratio for a linear chirp signal analyzed by the sHHT under different SNRs and input noise levels. The right panel shows the IF resolved by the sHHT imposed on the Hilbert spectrum. The chirp signal was specifically simulated with the SNR = 1 and the imposed noise level of only 0.1.  The green contour labels the Hilbert energy levels displayed on the sHHT time-frequency map.}} 
\label{linearchirp_disp2}
\end{figure*}
\begin{figure*}
\centering
\includegraphics[width=8.0cm,height=5.2cm]{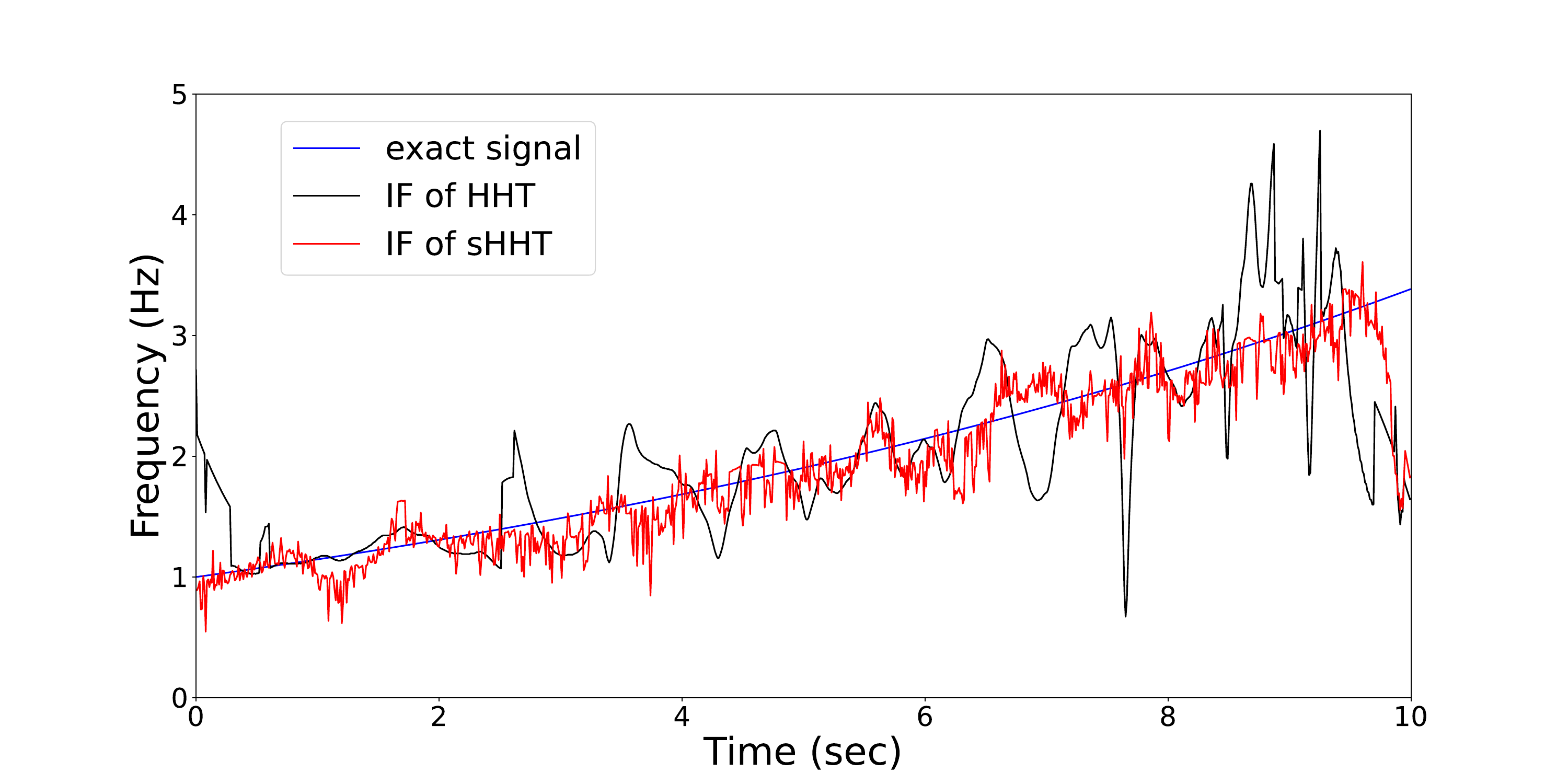}
\includegraphics[width=9.5cm,height=5.2cm]{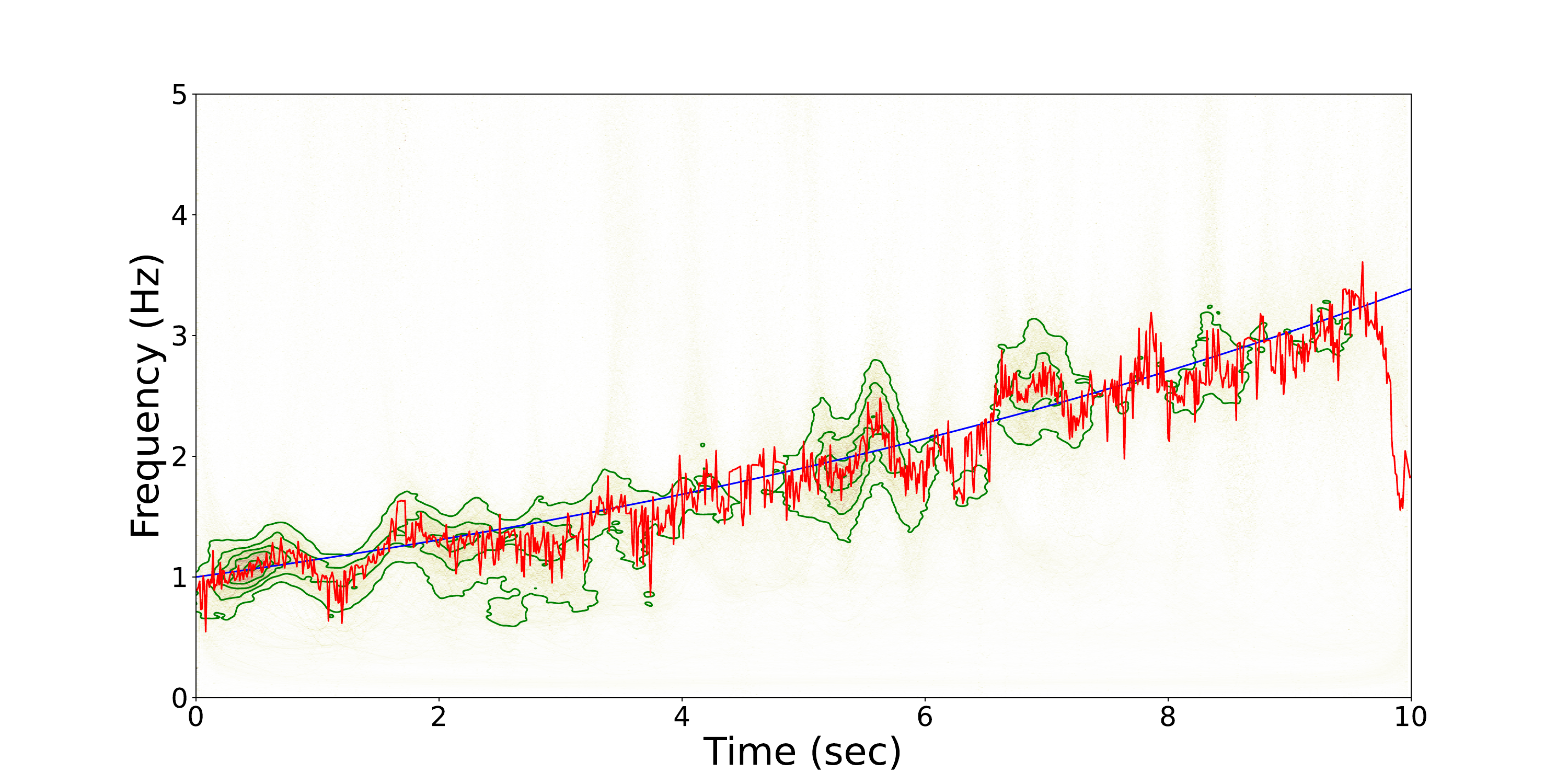}
\caption{{\footnotesize IFs of the exponential chirp signal resolved by the conventional HHT and sHHT. The chirp signal was generated with the SNR = 1 and the frequency of $2^{0.1 \cdot t}(1+0.1\cdot t\cdot \ln2)$. The external noise level imposed on this signal is 0.9. The left panel displays the actual signal frequency in blue, along with the IFs derived from conventional HHT and sHHT, indicated by black and red colors, respectively. The right panel presents the same IF resolved by the sHHT imposed on the Hilbert spectrum. The green contour exhibits the Hilbert energy levels displayed on the sHHT time-frequency map.}} 
\label{logchirp_disp1}
\end{figure*}
\begin{figure*}
\centering
\includegraphics[width=8.3cm,height=5.2cm]{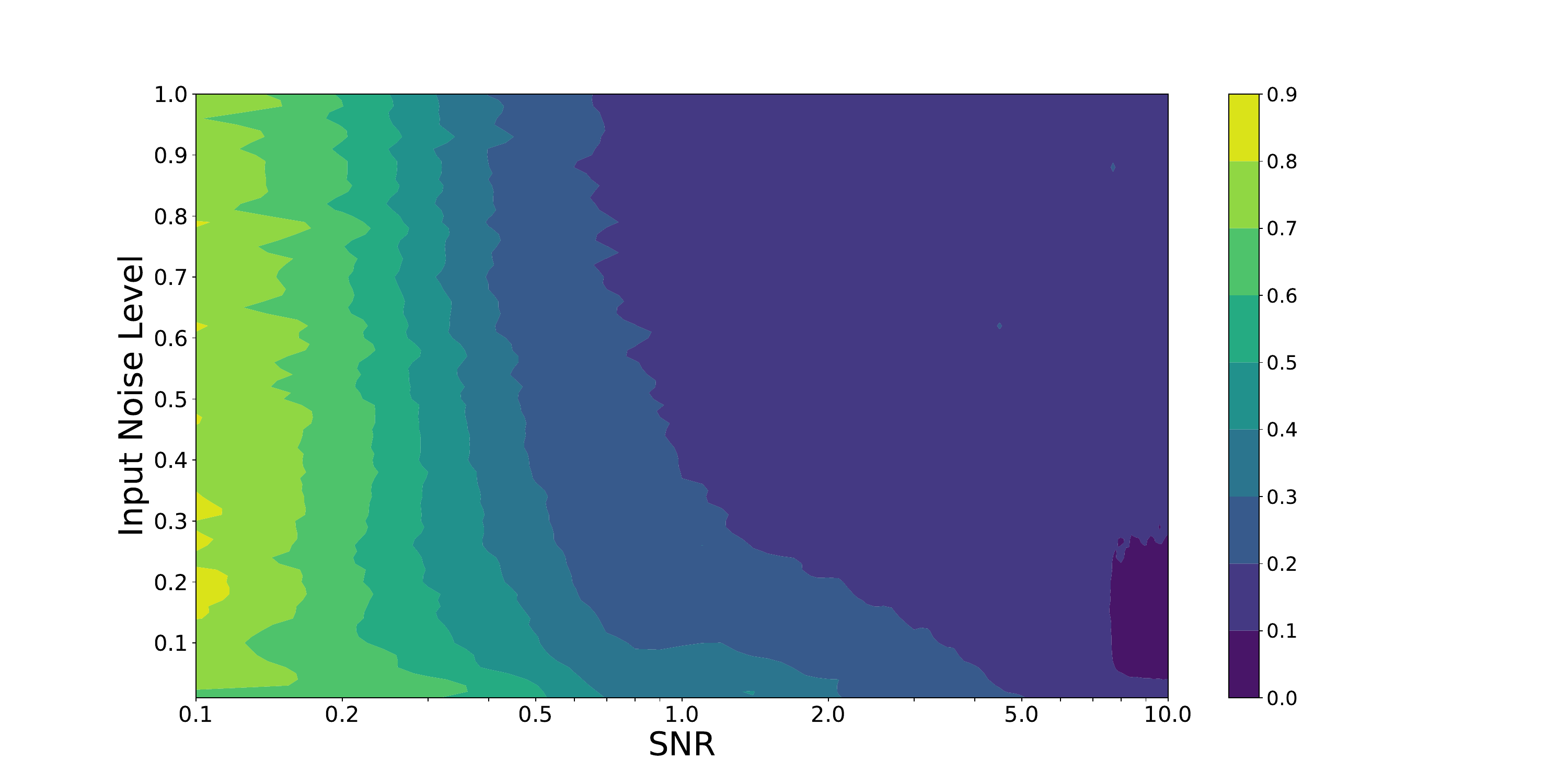}
\includegraphics[width=9.5cm,height=5.2cm]{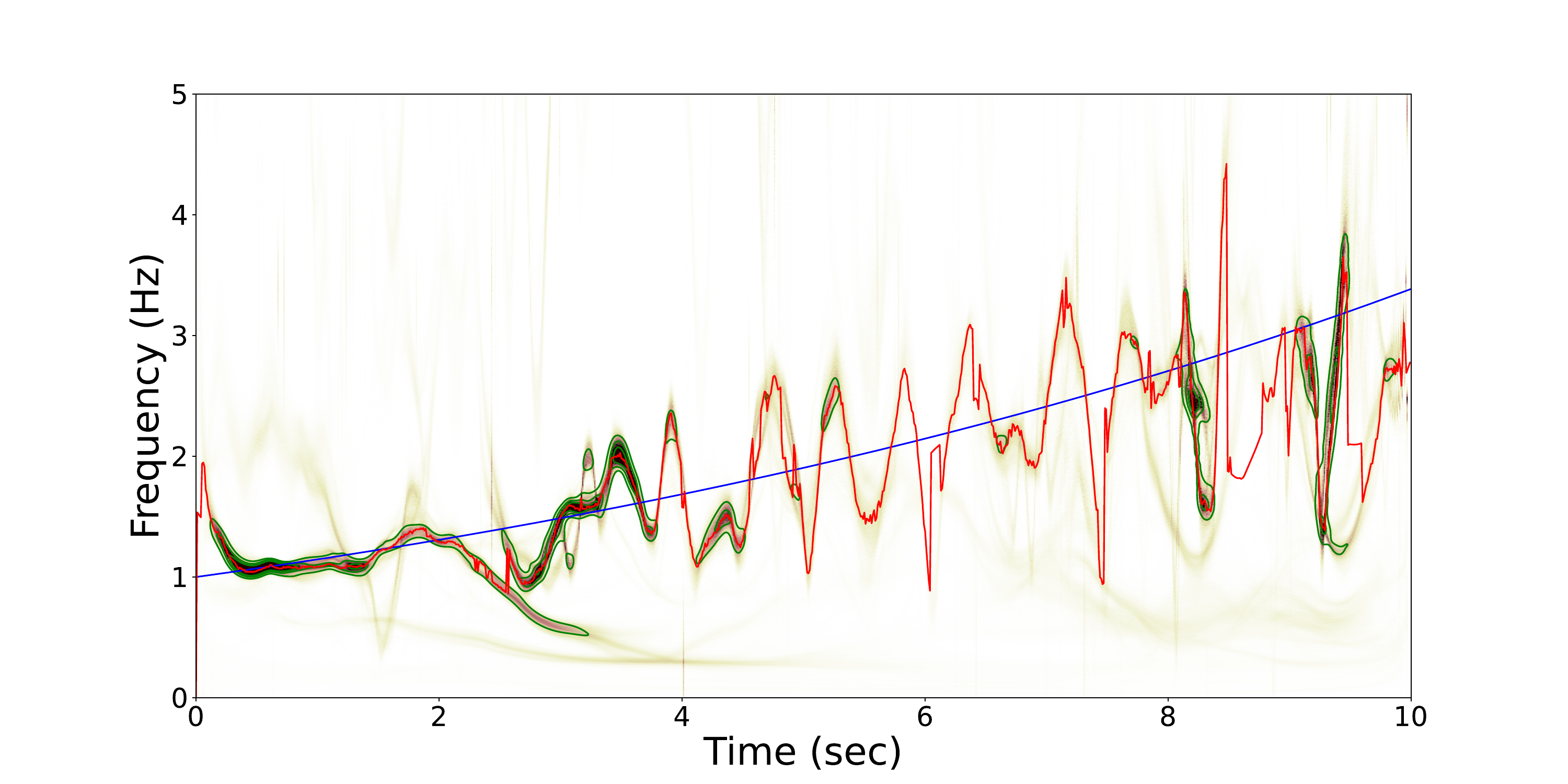}
\caption{{\footnotesize Frequency dispersion ratio of the exponential chirp signal estimated by the sHHT and the IF of a specific exponential chirp signal. The left panel exhibits the variation of the frequency dispersion ratio for an exponential chirp signal, determined by the sHHT, with different SNRs and input noise levels. The right panel shows the IF resolved by the sHHT imposed on the Hilbert spectrum. This chirp signal was specifically simulated with the SNR = 1 and the imposed noise level of only 0.05.  The green contour labels the Hilbert energy levels shown on the sHHT time-frequency map.}}

\label{logchirp_disp2}
\end{figure*}

\subsection{Linear Chirp Signal}
In the second example, we examine a signal where the frequency changes over time. 
In comparison to other timing methods, the HHT has an advantage in resolving signals of this type. 
This is because HHT utilizes an adaptive basis that adjusts to variations in the signal, enabling effective extraction of local information over time. 
Here we generate a variable sinusoidal signal $\sin(\omega t)$, where $\omega = 2\pi f'$ and $f'$ is defined as $0.1\cdot t+10^{-3}$, evolving with time ($t$). 
The exact frequency ($f$) of this signal is given by the equation $0.2\cdot t+10^{-3}$, which is represented by the blue straight line in the left panel of Fig~\ref{linearchirp_disp1}. 
The amplitude strength for both the signal and the internal noise was maintained at a consistent level to simulate mock data with an SNR of 1, and we arbitrarily introduced external white noise with an amplitude ratio of 0.6 to perform the sHHT.
According to the IFs determined by HHT and sHHT presented in the left panel of Fig.~\ref{linearchirp_disp1}, it is evident that the sHHT performs better in resolving the signal pattern, even though the boundary problem remains a challenge. 
The IF obtained from HHT may display some spiky peaks in the plot due to the singularity in the calculation; nevertheless, it can be smoothly moderated by our ``stacked'' algorithm.
The right panel of Fig.~\ref{linearchirp_disp1} presents the color spectrum of sHHT, and we can see that the deep color to denote the higher Hilbert energy accurately corresponds to the actual frequency of the simulated signal.

To examine the relationship between SNR and imposed noise level, which affects the frequency dispersion of the linear chirp signal, we analyze data sets with varying parameters. 
As shown in the left panel of Fig.~\ref{linearchirp_disp2}, the frequency dispersion ratio generally increases when we raise the imposed noise level or reduce the SNR of our data sample; however, we also notice some intriguing features when we focus on cases where the imposed noise level is less than 1.
If we fix the SNR of our dataset in 1, the ratio of frequency dispersion may decrease when we increase the imposed noise level. 
In the right panel of Fig.~\ref{linearchirp_disp2}, the imposed noise level to perform sHHT is fixed at 0.1, allowing for a comparison with the plot in Fig.~\ref{linearchirp_disp1}, where the noise level is set at 0.6.
From this comparison, it is evident that a higher noise level results in increased frequency dispersion and more pronounced mode-mixing issues.
We would like to emphasize that this phenomenon is likely to occur only when the imposed noise level is less than 1 and when the SNR of the data is between 0.2--1.6.
Otherwise, the frequency dispersion ratio will continue to increase as the imposed noise level increases. 

\subsection{Exponential Chirp Signal}
   
The signal pattern of a CBC GW event resembles an exponential chirp signal characterized by a rapid increase in frequency. 
For instance, during the coalescence of two black holes (BHs), the duration of the whitened waveform, when subjected to a bandpass filter, usually lasts less than one second.
During the merging phase, the frequency shifts from a few tens of hertz to several thousand hertz in just 0.1 seconds \citep{Mohapatra2012}.     
Although the signal pattern of a GW event is more complex, we can use an exponential chirp signal to illustrate it and examine its characteristics as recognized by HHT and sHHT. 
Here we generate an exponential chirp signal using the sinusoidal function $\sin(\omega t)$ with $\omega = 2\pi f'$, while $f'$ is a function of time to represent $2^{0.1\cdot t}$. 
The actual frequency of such a signal can be computed by $1/2\pi \cdot \rm{d}(\omega t)/\rm{d}t$, equivalent to $2^{0.1 \cdot t}(1+0.1\cdot t\cdot \ln2)$, which can be illustrated by the blue curve in Fig.~\ref{logchirp_disp1}.
In the generation of linear and exponential chirp signals, we multiply a small coefficient (i.e., 0.1) to retard the frequency increase, making it easier to compare with a stable sinusoidal signal; however, in order to match the GW signal, this coefficient can be much larger.
We simulate a time series with SNR = 1 to test HHT and arbitrarily impose an external noise level = 0.9 to identify the exponential chirp signal using sHHT. 
IFs determined by HHT and sHHT are represented by the black and red curves in the left panel of Fig.~\ref{logchirp_disp1}. 
As shown in the right panel, the IF obtained from sHHT more accurately approximates the actual frequency of the signal.

Similar to the linear chirp signal, the frequency dispersion ratio of the exponential chirp signal increases as the noise level rises and the SNR decreases, particularly when the input noise level exceeds one.
In the left panel of Fig.~\ref{logchirp_disp2}, we also find that the ratio of frequency dispersion within specific ranges of SNR decreases when we apply the sHHT while increasing the imposed noise level, as long as the noise level remains below one.
We demonstrate the situation by performing the sHHT with an external noise level of only 0.05 to resolve an exponential chirp signal with an SNR of 1, as shown in the right panel of Fig.~\ref{logchirp_disp2}.
In comparison to the data set displayed in the right panel of Fig.~\ref{logchirp_disp1}, the red IF in the right panel of Fig.~\ref{logchirp_disp2} exhibits a more significant deviation from the blue actual frequency, although it has an imposed noise level of one order of magnitude smaller.
In our simulation to perform the sHHT, we found that the frequency dispersion derived from sHHT increases directly with a higher imposed noise level on a lower SNR dataset. 
However, this is only true when the external noise level is sufficiently large (i.e., $\gtrsim$ 1). 
Additionally, more severe mode-mixing issues can still arise in datasets with extremely low imposed noise levels, provided certain special conditions are met.  

\subsection{Comparison with Q-transform}

\begin{figure}[bp]
    \centering   \includegraphics[width=0.49\textwidth]{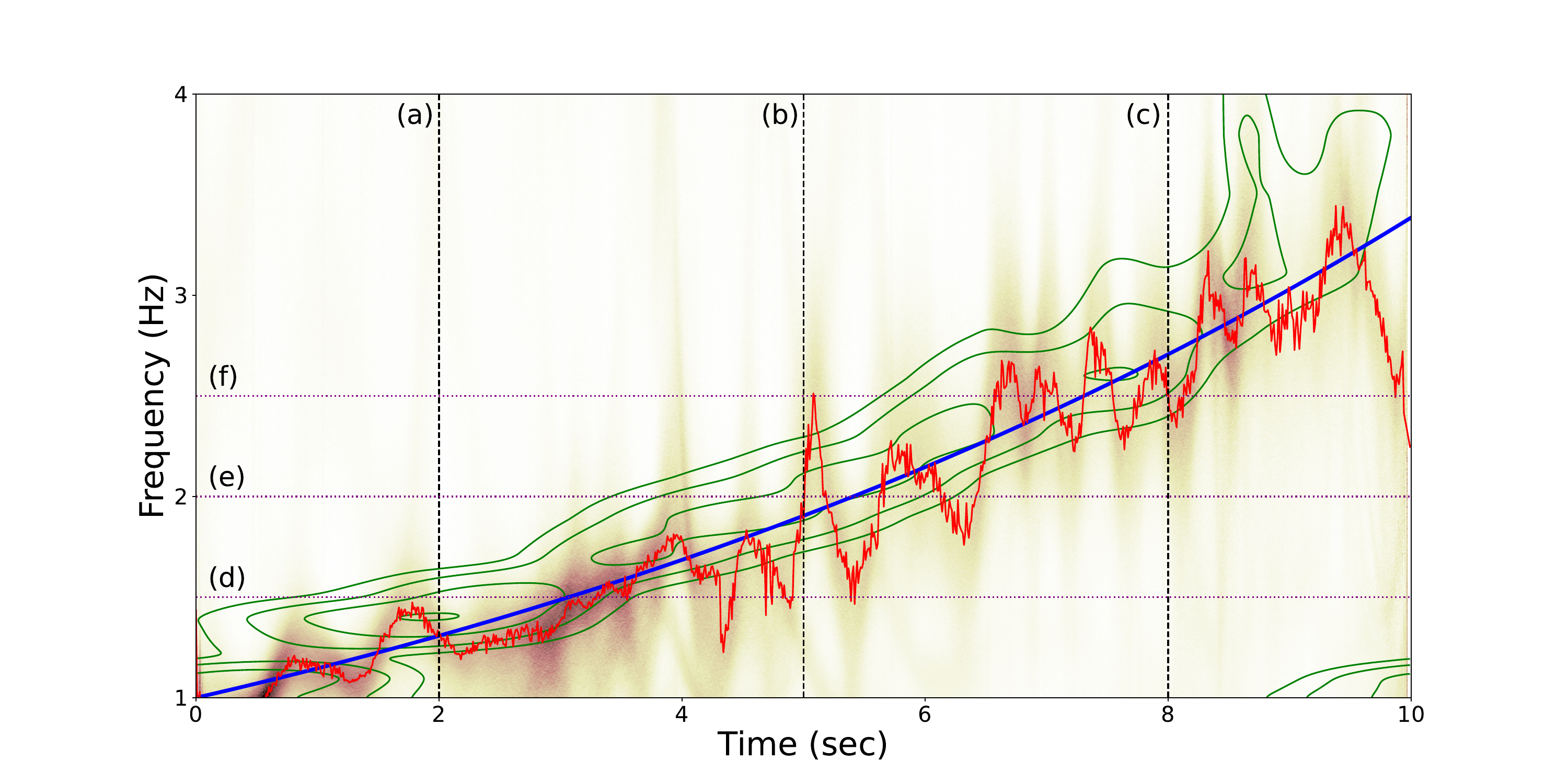}
    \centering   \includegraphics[width=0.49\textwidth]{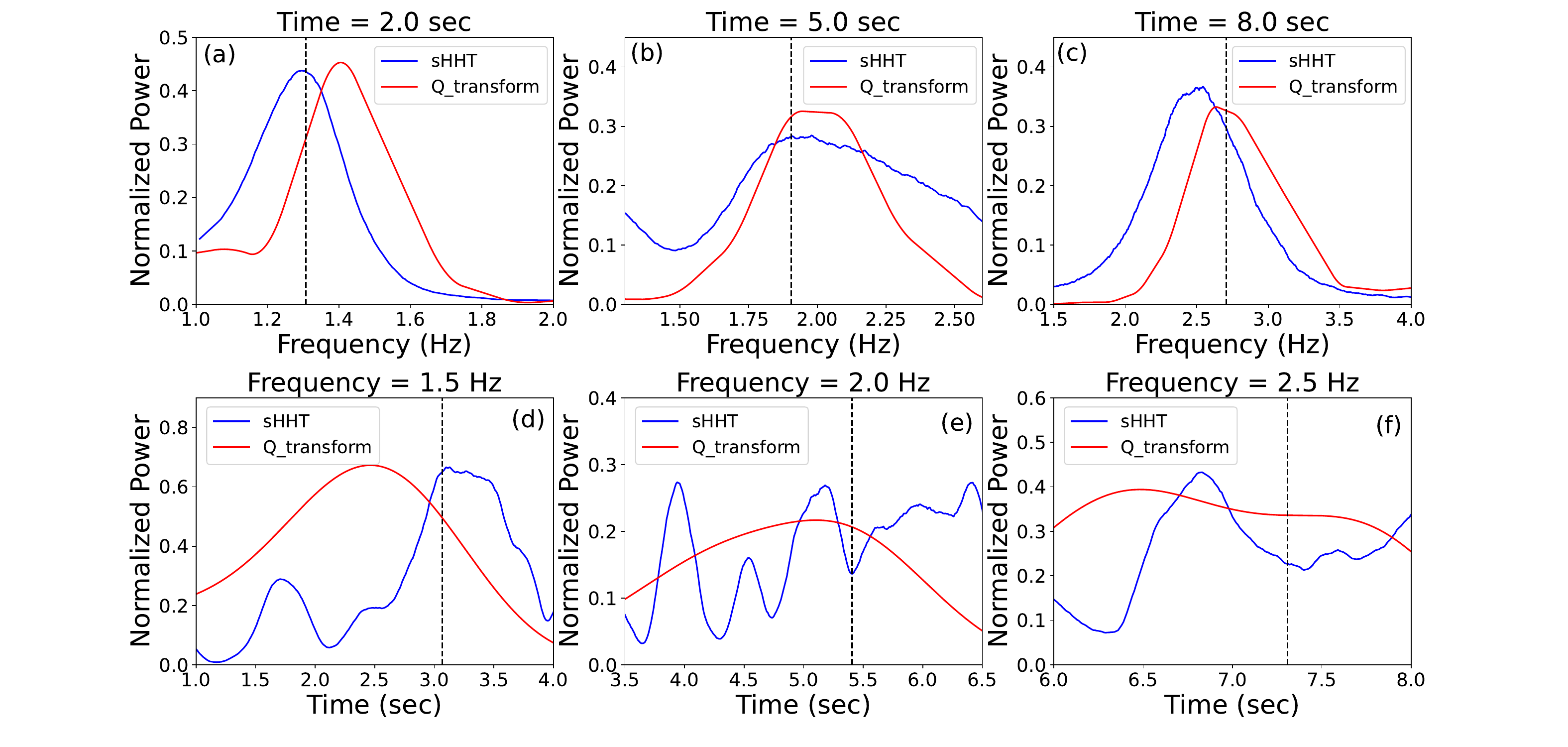}
    \caption{Comparison between the stacked HHT spectrum (color map) and the spectrum of Q-transform (green contour). The red curve in the top panel denotes the IF derived from the sHHT, while the blue one presents the actual signal. Six slices are labeled by the black-dashed/purple-dotted lines for (a) time = 2.0\,s, (b) time = 5.0\,s, (c) time = 8.0\,s, (d) frequency = 1.5\,Hz, (e) frequency = 2.0\,Hz, and (f) frequency = 2.5\,Hz. The amplitude vs. frequency/time plot (i.e., the power spectral density) is presented in the lower panels. The blue curve is smoothed to demonstrate the averaged profile of the instantaneous frequency/time. The green curve denotes a similar result generated from the Q-transform, where the black dashed line in the bottom panel indicates the actual frequency/time of the simulated signal. The HHT and Q-transform profiles in panels (a)–(f) are scaled for display purposes.}
    \label{fig:shht-Q}
\end{figure}

Comparing the investigation of exponential chirp signals between sHHT and the Q-transform is interesting because such signals behave like GW signals, and the Q-transform is commonly used for fast scanning.
The Q-transform effectively tracks signals with abrupt frequency changes by allowing the quality factor Q to be adjusted during computation, enabling the detection of instant changes in the frequency domain.
In the previous section, we added a small exponential factor to reduce the rate of signal variation in the frequency domain. 
This modification allows us to compare the results obtained with the conventional HHT without significant mode splitting.
For such a signal with a frequency change between 1 and 4\,Hz during the first 10\,s, the constant Q-transform \citep{Brown91}, which closely resembles the Morlet (or Gabor) wavelet \citep{bernardino2005}, can effectively transform our data into a train of sinusoid Gaussians.
Please note that a small Q factor results in better time resolution but poorer frequency resolution, while a high Q value has the opposite effect. 
Therefore, it is important to choose an appropriate Q value when performing the transformation to balance the visual time and frequency resolutions on the power spectral map.
Our investigation shows that the method with a fixed Q factor of 10 provides the best visual clarity on the power spectral density map (as shown by the green contour in the top panel of Fig.~\ref{fig:shht-Q}) compared to other Q factors in the constant Q-transform or the variable Q-transform.

In contrast to investigating the best relationship between the SNR of the data and the imposed noise level for performing sHHT, we stacked more spectra (i.e., 50000 iterations) here to smooth the distribution of the derived IF for a further check of frequency and time resolutions. 
Based on our studies in the previous section, we set the imposed noise amplitude between 0.5 and 1 to ensure a level comparable to the SNR of the data to minimize the derived frequency deviation, and we overlapped the results obtained from the Q-transform on the stacked Hilbert spectrum, as presented in Fig.~\ref{fig:shht-Q}.
Both Q-transform generated with a Python-based parameter estimation toolkit for CBC signals (PyCBC v2.4.0\footnote{https://pycbc.org/}; \cite{Biwer2019}) and sHHT provide the capability to track signal behavior. 
For further comparison, we selected six slices at different times (i.e., 2\,s, 5\,s, and 8\,s) and frequencies (i.e., 1.5\,Hz, 2\,Hz, and 2.5\,Hz), and this analysis focuses on the illustration of the related time and frequency resolutions as shown in the bottom panel of Fig.~\ref{fig:shht-Q}.
The full widths at half maximum for the red and blue profiles to label the Q-transform and sHHT in the (a) to (c) panels of Fig.~\ref{fig:shht-Q} are similar, while it shows that the two approaches give a comparable frequency resolution.  
We can also find that the frequency corresponding to the signal resolved by the Q-transform in the earlier stage is slightly overestimated (i.e., panel (a)), although it remains within a tolerable range.

Panels (d) to (f) show that the time resolution of the sHHT is superior to that of the Q-transform.
This is evident from the narrower full widths at half maximum, which better constrain the time intervals in the blue profile, even though multiple blue peaks may appear in the plot.
Please note that the top panel of Fig.~\ref{fig:shht-Q} showcases the optimal visual performance of the Q-transform. 
While opting for a smaller Q factor can improve time resolution, it sacrifices frequency resolution, making it more challenging to track the signal accurately.
The multiple peaks observed in panels (d) to (f), which are also reflected as oscillations in all the resolved signals shown on the Hilbert spectra, result from intra- or inter-wave modulations. 
These modulations are caused by significant variations in the cycle lengths of the intrinsic noise compared to the exact signal.
This effect is particularly pronounced when we impose a noise level comparable to the SNR of the datasets to perform sHHT, and it is expected to moderate this effect once we implement the zero-crossing algorithm in the sHHT method.
We note that the frequency change in this simulation occurs at a much slower rate compared to a real CBC signal during the merging phase, suggesting that the Q-transform (or wavelet method) still has advantages in accurately tracking the evolution of such a signal, but these oscillations can be ignored once the ratio of their amplitude to the entire rapid frequency changes is small.

\section{Applications\label{sec:Applications}}

HHT is widely applied in many academic and engineering fields, and here we would like to demonstrate its feasibility in studying the nonlinear signal embedded in astronomical observations. 
Because no priori basis is required for such an adaptive method to trace the signal pattern, HHT was confirmed as a practical algorithm to investigate the unstable astronomical event (e.g., QPO; \cite{Hu2014}, or superorbital modulation; \cite{Hu2011, Lin2015, Lin2020}), especially for the specific GW signal, which can have an immediate change in the frequency domain with a short time scale (i.e., CBC; \cite{Son2018, Akhshi2021}).
We will present two examples to demonstrate the practicability of using the sHHT, including investigations into superorbital modulation and the GW signal.
To trace the GW signal, we only include LIGO data in the following analysis.
This is because the currently known GW events detected by LIGO possess relatively high SNRs, making them significant enough to be recognized in the dynamic power spectra.

In general, we had no information on the SNR for one dataset, even though we quantitatively constrained the relationship between it and the required external noise level to perform sHHT, leading to the least dispersion in IF in \S~\ref{sec:dispersion}. 
Please note that the conventional method to determine the network SNR of a GW event is through the matched filtering algorithm \citep{2012Cannon}.
This approach differs from the simpler definition of SNR, which is based on the amplitude ratio between the signal and the noise, as universally applied throughout this manuscript.
Without information about the strength of the internal noise embedded in a dataset, we can only blindly check the external noise level required to generate an acceptable stacked Hilbert spectrum.
Typically, this external noise level is set between 0.01 and 10 to multiply the standard deviation of the time series.
If the noise level incorporated is significantly smaller than the intrinsic noise of the data, the mode-mixing problem persists, resulting in several unphysical features on the stacked Hilbert spectrum.
Conversely, the true signal in the Hilbert spectra can be obscured by the significant external noise involved in executing the sHHT.
The best external noise level incorporated to perform sHHT can be determined after a series of investigations on the obtained stacked Hilbert spectrum, and this algorithm is definitely more time-consuming than other dynamical timing methods.   
Since the current known GW events with low SNRs ($\lesssim 8$, determined by matched filtering) cannot provide a clear image of the time-frequency representation and the signal is only sensitive to the correlation with a theoretical template, we therefore only demonstrate the GW signals recorded by two LIGO detectors in  \S~\ref{sec:Applications}.B.

\begin{figure}[bp]
\centering   
\includegraphics[width=0.49\textwidth]{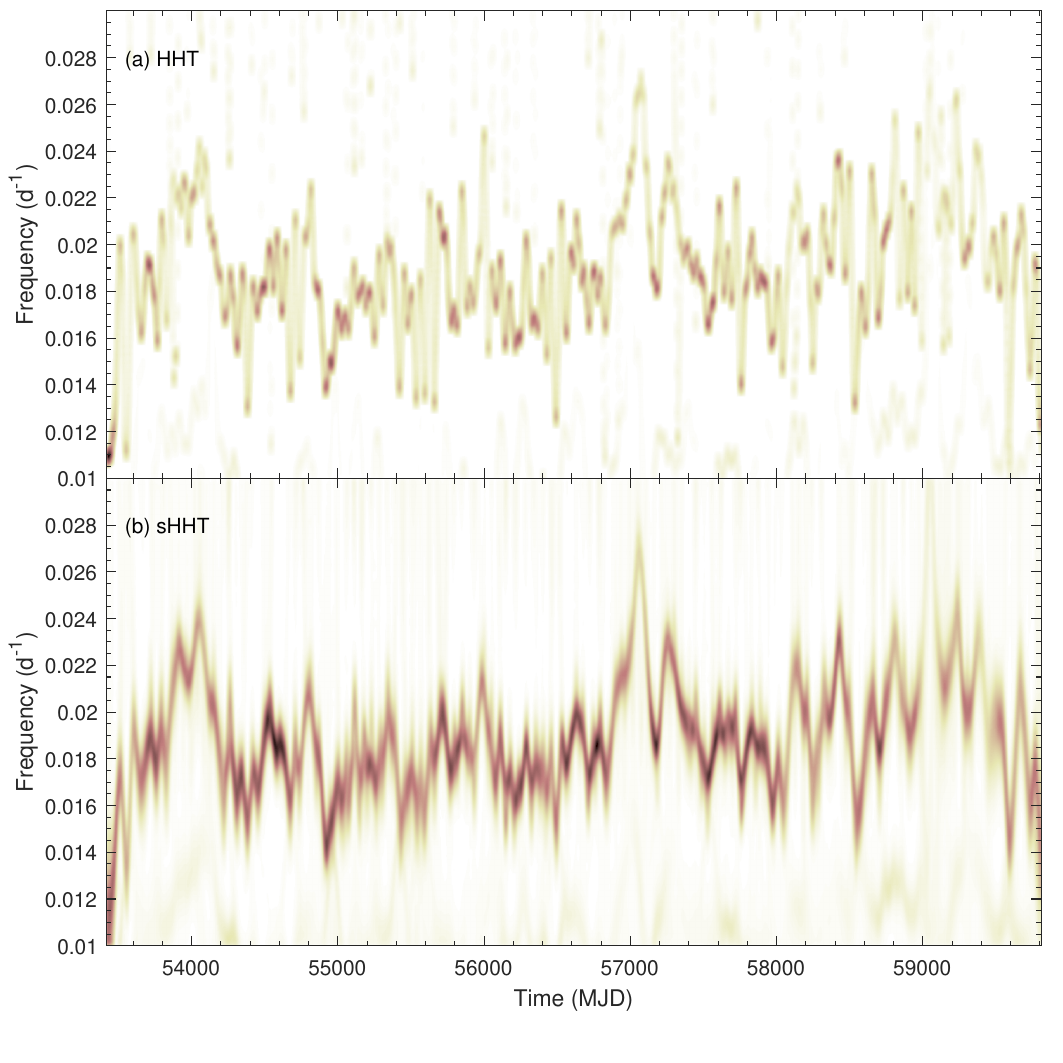}
\caption{{\footnotesize Time-frequency maps of the superorbital modulation of SMC X-1 obtained using (a) the HHT and (b) the sHHT.}}
\label{fig:smcx1}
\end{figure}
\begin{figure*}
\centering
\includegraphics[width=8.9cm,height=5.2cm]{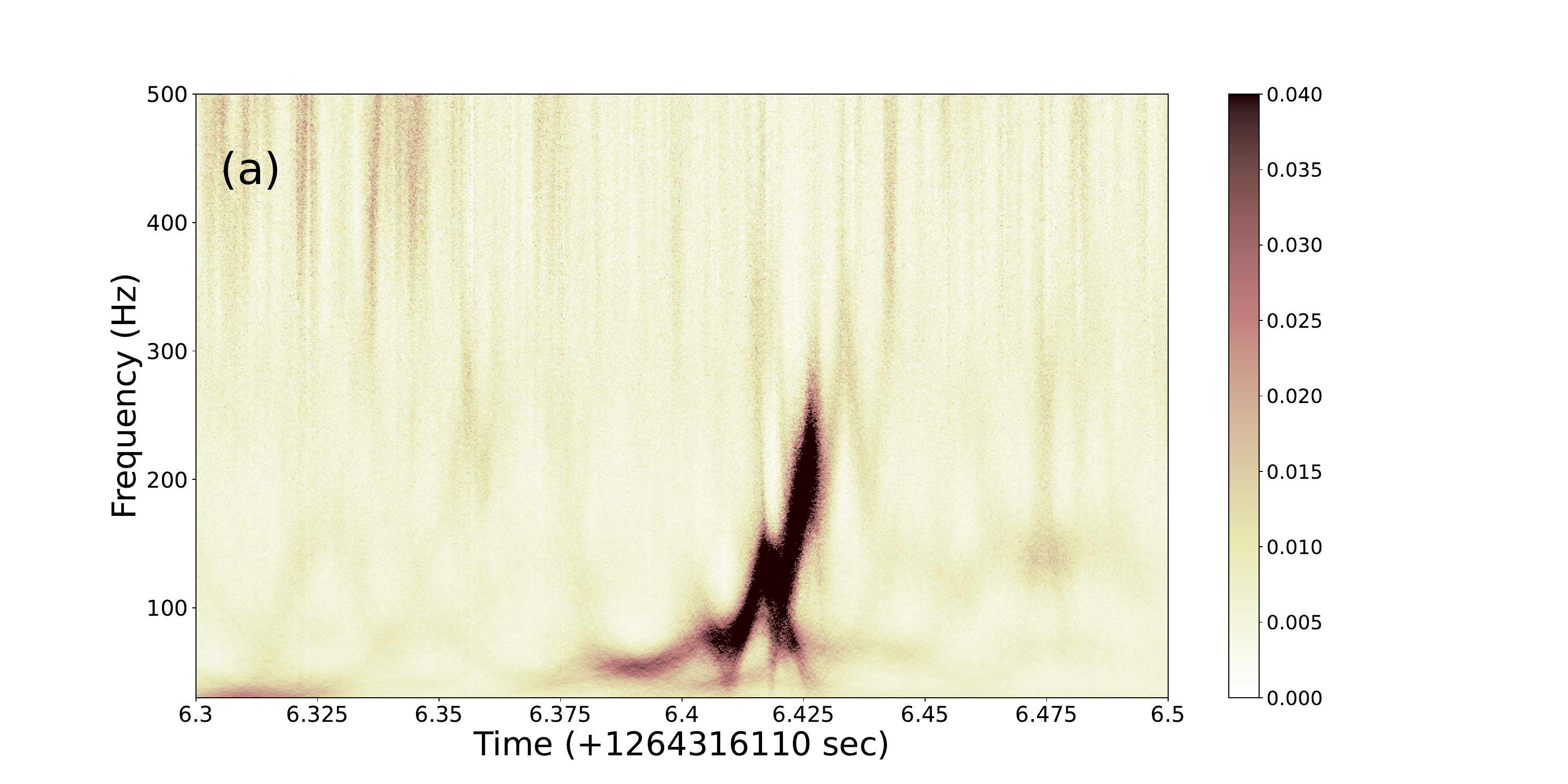}
\includegraphics[width=8.9cm,height=5.2cm]{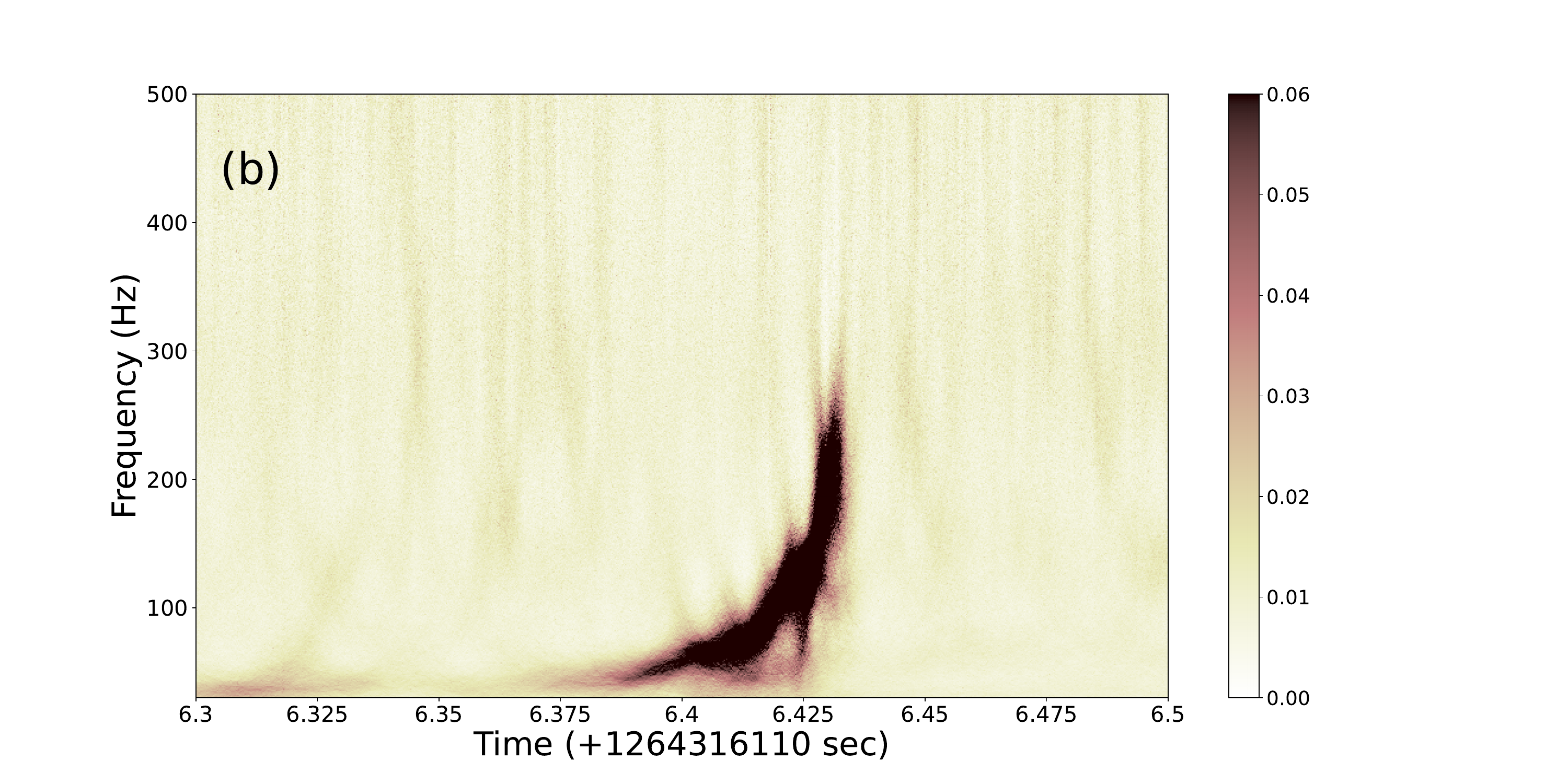}
\includegraphics[width=8.9cm,height=5.2cm]{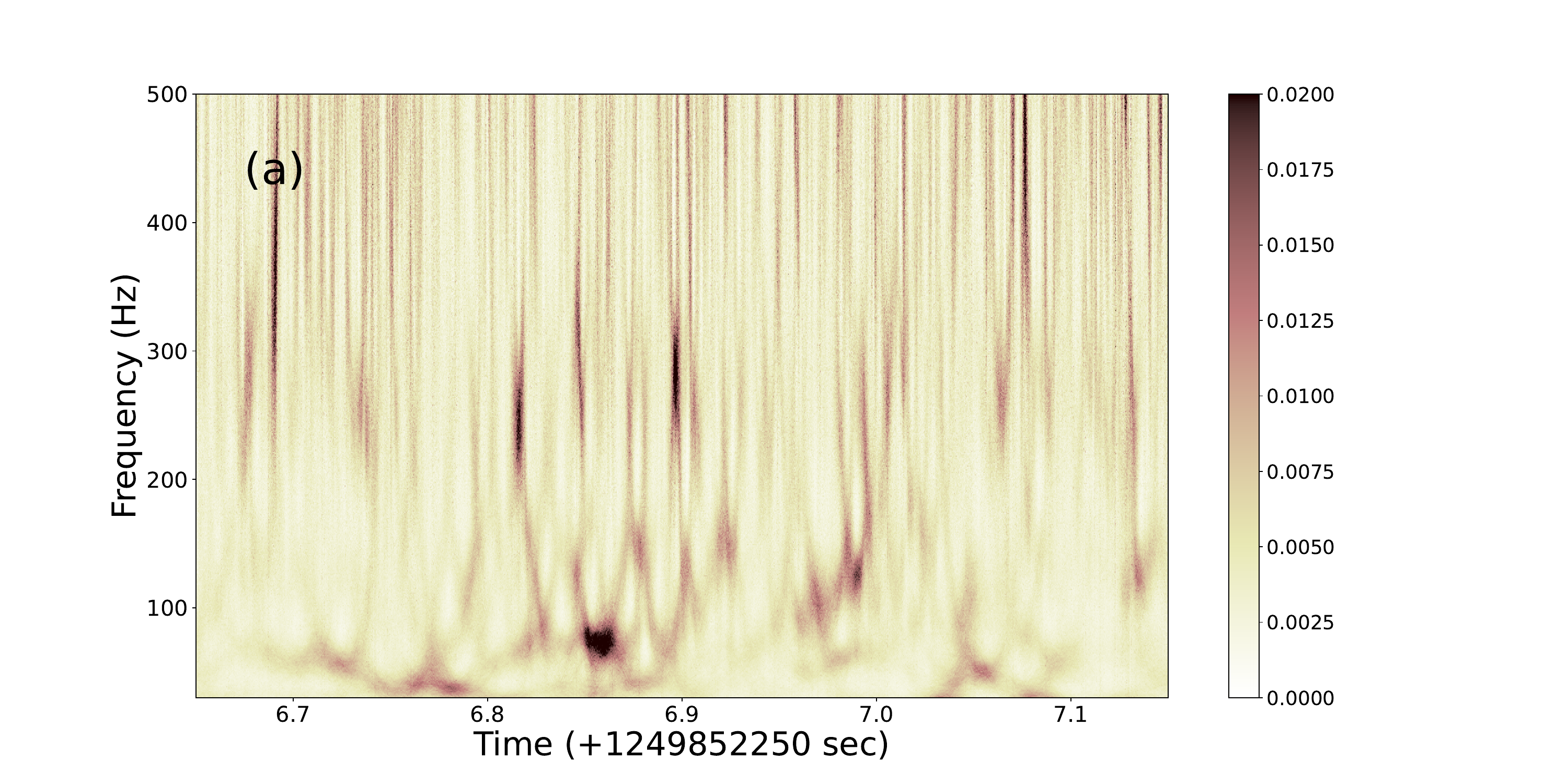}
\includegraphics[width=8.9cm,height=5.2cm]{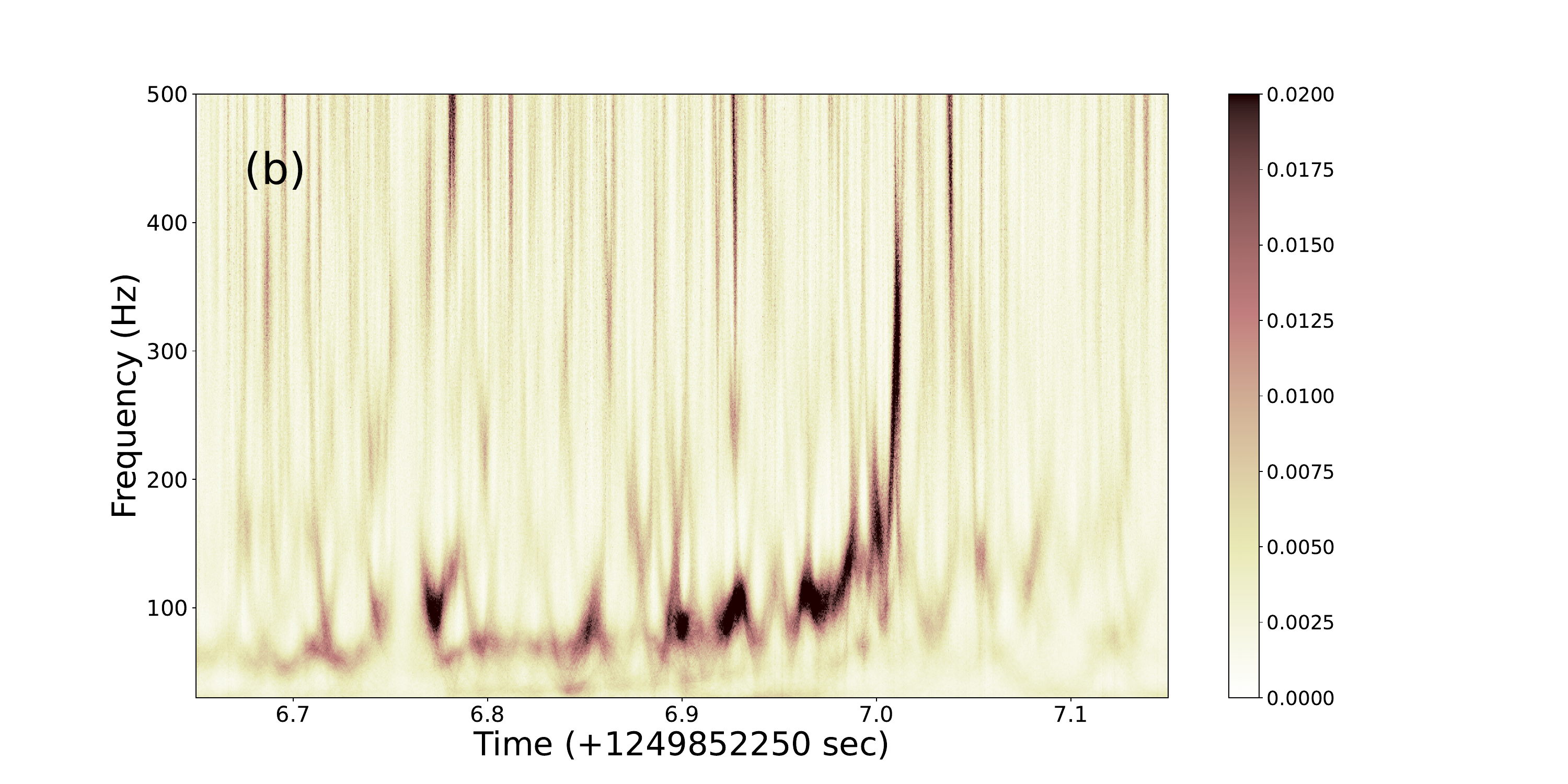}
\caption{{\footnotesize Time-frequency representation of GW200129\_065458 (upper panels) and GW190814 (bottom panels) obtained by sHHT. We present the stacked Hilbert spectra from LIGO Hanford (a) and Livingston (b) observations, generated from 10000 iterations to decompose the IMF affected by randomly distributed noise levels using EMD.}} 
\label{GW_example}
\end{figure*}

\subsection{SMC X-1} 
The HHT has been successfully used to monitor the superorbital modulation of SMC X-1, as reported in \citet{Hu2011}. 
The modulation period varies between approximately 35 to 70 days, which is just below a factor of 2. 
The EEMD, which is a dyadic filter, can decompose the superorbital modulation signal into a single IMF that facilitates the calculation of the instantaneous frequency. 
However, the modulation period has evolved to a shorter regime, which challenges the tracking of the superorbital frequency during the latest ``excursion'' event, identified by a reduction in the superorbital modulation period to 45 days or less (around MJD 54000, 57200, and 59200 in Figure \ref{fig:smcx1}). 
The light curve data applied to generate the figure were obtained from the hard X-ray monitoring program using the Neil Gehrels Swift Observatory \citep{KrimmHC2013}. 
To maintain the modulation within a single IMF, it becomes necessary to incorporate a significant amount of noise when performing the EEMD. 
This results in further fluctuations in the instantaneous frequency, as shown in Fig.~\ref{fig:smcx1}a, contrasting with Fig. 3 in \citet{Hu2011}. 
On the other hand, the sHHT provides a more distinct time-frequency map of the superorbital modulation, minimizing the potential spurious frequency variability, as demonstrated in Fig.~\ref{fig:smcx1}b and \citet{HuDK2023}.

\subsection{Investigation of the Gravitational wave}

\citet{Akhshi2021} used the conventional HHT as the basis to develop a template-free approach to extract the GW signal, and they derive the physical time delays between the arrivals of GWs at different detector locations by analyzing the waveforms given by the HHT.
However, the GW signal pattern obtained from the conventional HHT has a serious mode-mixing problem, especially when the EEMD was not employed in the related investigations.
In addition to the conventional HHT, sHHT has demonstrated its feasibility to investigate the GW from the CBC and a burst-typed event with a better performance \cite{Hu2022}.

We would like to provide additional examples to demonstrate the feasibility of sHHT in investigating confident GW events using the open data obtained from the O3 observing run of L-V-K-GEO \citep{O3data}, which were not included in \citet{Hu2022}.
We obtain the 32\,s LIGO data of 4k\,Hz released from the GW Open Science Center\footnote{https://gwosc.org/eventapi/html/}. 
Both data sets recorded at the Hanford and Livingston sites are included in our analysis. 
The GW data are the time series to record a long-term strain variation with a peak intensity of $\sim 10^{-21}$, and therefore, the serious contamination of the colored noise or line features from instruments and background can be seen from the simple Fourier transform.
The effective frequency range of the current L-V-K-GEO data is restricted from 30\,Hz to a few hundred Hz, depending on different interferometers \cite{2016LV,2017LV}, and we can only search for GW signals of specific origins in this frequency band.
To mitigate the effects of colored noise and spectral lines caused by instruments and background, we followed the standard procedure to whiten the data using a Python-based GW analysis library(GWPY v3.0.8\footnote{https://gwpy.github.io/docs/stable/}; \cite{gwpy}).
This process is intended to normalize power across all frequencies and enhance higher-frequency content.
The whitening process was performed using a fast Fourier transform integration lasting 4\,s, with a 2\,s window overlap for Hann smoothing. 
We concentrate on the frequency range of 30--500\,Hz in our analysis of LIGO data since the noise is high in the low- and high-frequency ends.

To decide the optimal external noise level incorporated to perform the sHHT, we blindly consider different input noise levels within 0.01--10 as one example provided in Appendix~\ref{app:EN_investigation}, and pick up the value to achieve the best visual performance.
We finally impose the external noise levels of 0.6 and 0.8 to generate the Hilbert spectra of GW200129\_065458 measured by Handford and Livingston detectors, respectively. 
For GW190814, we incorporate the external noise level of 0.9 to perform the sHHT on the data obtained from both detectors.
As shown in Fig.~\ref{GW_example}, we present the stacked Hilbert spectra of GW200129\_065458 and GW190814 \citep{GW190814}.
GW200129\_065458 is a significant event with a high network SNR ($26.8\pm 0.2$) inferred from the IMRPhenomXPHM model; \cite{Pratten2021}), and it was recorded in the GW Transient Catalog-3 (GWTC-3; \cite{GWTC-3}) with detection in all three detectors including Virgo.
This event is supposed to be a binary BH merger with a mass of approximately $\sim 35$\,$M_{\odot}$ and $\sim 29$\,$M_{\odot}$, but the SEOBNRv4PHM template \citep{Ossokine2020} indicates a preference for a merger with nearly equal masses, with a mass ratio of $q \sim 0.9$. 
In addition, there is weak evidence of precession, characterized by $\chi_{\rm p} \sim 0.36$.
Due to the data quality issues identified in the Livingston dataset as mentioned in \citet{GWTC-3}, only the GstLAL pipeline \citep{Messick2017} is considered for its trigger. 
This analysis yields a higher signal-to-noise ratio (SNR) for the Livingston data (i.e., 21.2) compared to the Hanford data (i.e., 14.6) and the Virgo data (i.e., 6.3).
By the sHHT, we can clearly observe signal patterns for GW200129\_065458 in the upper panel of Fig.~\ref{GW_example} while the power (i.e., Hilbert energy) accumulated in the Hilbert spectra for the Livingston data is also more substantial, aligning with the results provided by the GstLAL pipeline.

\begin{figure}[bp]
\centering   
\includegraphics[width=9.5cm,height=5.0cm]{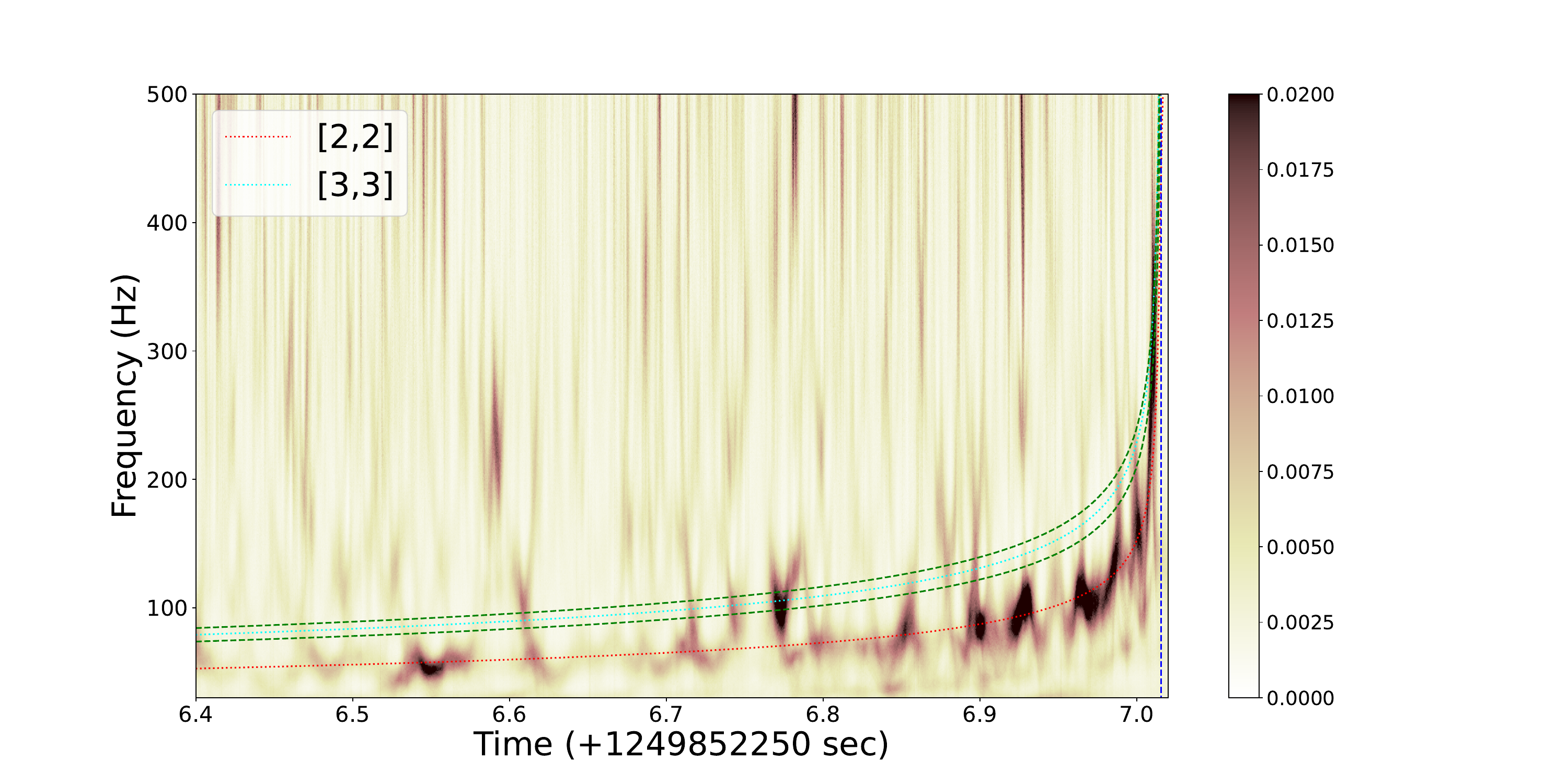}
\caption{{\footnotesize stacked Hilbert spectrum of the GW190814 event obtained from LIGO Livingston detector. The dashed blue vertical line shows the merger time (GPS time: 1249852257.015\,s) determined by Bayesian inference. The red and cyan curves correspond to the ($l$,$\lvert m \rvert$) = (2,2) and (3,3) multipoles of the waveform, respectively. Two green dashed curves label the time-frequency slice with $\alpha$=1.5 (related to the (3,3) HM) used to evaluate the residual power in the map.)}}
\label{fig:HMs_demo}
\end{figure}

Compared to generating a GW template for merging binary BHs, the amplitude of NS mergers is considerably lower, and the coherent signal can persist for a much longer duration during the inspiral phase.
According to \citet{Hu2022}, the time-frequency characteristics of a coalescing binary NSs event near the merging phase are smeared due to the wide Gaussian window, and the SNR is insufficient within this large window to accurately identify the intrinsic time-frequency properties of the signal.
The wavelet method can present the signal more clearly because we can tune the Gaussian envelope's parameter to one order smaller than the value used to examine the event of binary BH mergers.
Except for the coalescence of two NSs, we notice that the GW template corresponding to the coalescence of a BH and an NS or a lower mass gap (i.e.,  2.5 -- 5\,$M_{\odot}$; \cite{Bailyn98,Ozel2012}) can also have a longer coherence time. 
We therefore select the GW190814, categorized as a BH-NS (with the secondary mass $< 3$\,$M_{\odot}$) in GWTC-2.1 \citep{GWTC-2.1}, as our second example.
This GW event is supposed to be caused by the merger of  $\sim 23.3$\,$M_{\odot}$ black hole and $\sim 2.6$\,$M_{\odot}$ lower mass gap inferred from a set of posterior samples drawn with equal weight from the IMRPhenomXPHM \citep{Pratten2021} and SEOBNRv4PHM \citep{Ossokine2020} analyses.
We want to clarify that the Hanford interferometer was out of observation mode during the trigger time of GW190814. 
However, the strain dataset was still collected and carefully calibrated by the detector characterization team to identify useful time intervals for public release.
In our analysis, we also examine the sHHT using the Hanford data as shown in the bottom panel of Fig.~\ref{GW_example}.
The significance of the signal pattern is weaker than the performance shown for the Livingston data; however, its signal identification is still stronger than the GW170817 (NS-NS event; \cite{GW170817}) presented in \citet{Hu2022}.
The interwave modulation can be seen in the inspiral phase in the low-frequency range on the time-frequency map, and it is expected to improve if we use the zero-crossing algorithm in sHHT, which is not supported in the current EMD package in Python (emd v0.6.2\footnote{https://emd.readthedocs.io/en/stable/index.html}; \cite{Quinn2021}). 

We note that the waveform models including higher-order multipoles (HMs, e.g., IMRPhenomXPHM \cite{Pratten2021} and SEOBNRv4PHM \cite{Ossokine2020}) were used to provide more precise measurements of the parameters for a CBC event, while the importance of a subdominant multipole moment increases with the mass ratio.    
GW190814 was detected with strong evidence of HMs \citep{GW190814} affecting the network SNR, and the coherent Wave-Burst pipeline, which was constructed based on minimally modeled burst algorithms \citep{2008Klimenko,2016Klimenko}, can identify HMs through the time-frequency representation by detecting coherent excess power in the chirp-like regions that correspond to various HMs \citep{2022high-order}.
In Fig.~\ref{fig:HMs_demo}, we use the SEOBNRv4PHM\_ROM template to approximate the waveform of GW190814 with the dominant ($l$,$\lvert m \rvert$) = (2,2) quadrupole and the ($l$,$\lvert m \rvert$) = (3,3) subdomaint multipole and overlapped their IFs on the stacked Hilbert spectrum of the same event detected from the Livingston detector.
Because the IF of the generic multipole ($l$,$\lvert m \rvert$) can be approximated to a scaled version of the dominant (2,2) quadrupole, we can examine the generic HM gravitational radiation with the following IF dependence \citep{2018London,2021Roy}.
\begin{eqnarray}
\label{eqn:HM_IF}
{f}_{l,\lvert m \rvert}(t)\approx \frac {\lvert m \rvert}{2} {f}_{22}(t)\approx \alpha' {f}_{22}(t)
\end{eqnarray}
$\alpha'$ is a dimensionless parameter, which allows us to constrain a range on the time-frequency map to investigate the residual power/energy of a specific multipole's IF along the time-frequency track.
The green dashed curves in Fig.~\ref{fig:HMs_demo} label the time-frequency slice corresponding to [$\alpha'-0.1$, $\alpha'+0.1$]$\times f_{22}(t)$ with $\alpha' =1.5$ of (3,3) multipole. 
We can find significant energy/power excesses close to 0.2 and 0.4~s before the merger time constrained in the above region shown in the stacked Hilbert spectrum, and it is consistent with the feature presented by the discrete Wilson–Daubechies–Meyer wavelet transform \citep{2012NKM}.
Similar features obtained from sHHT on the time-frequency representation are expected to further yield a similar result in the waveform-agnostic approach \citep{2022high-order} to provide evidence for the presence of HM.
Please note that the time resolution for generating the stacked Hilbert spectrum aligns with the sampling rate (4k) of the data, enabling the capability for more precise measurements between different phases. 
Further quantitative assessment to examine the HM GW radiation using sHHT could be an interesting work in future development.

\section{Conclusion\label{sec:Conclusion}}
In this study, we have introduced the analytical properties of the sHHT, a novel improvement over the traditional HHT for time-frequency analysis, and evaluated its applicability using simulated signals and real data. 
Through a series of numerical simulations, we have demonstrated that sHHT can effectively capture signals experiencing rapid frequency changes (the relevant codes can be referred to github\footnote{https://github.com/linlupin/sHHT}). 
This is commonly seen in GW signals associated with CBCs and other burst-type events \citep{Hu2022}. 
Our exploration has shown that the sHHT method greatly enhances the accuracy of tracing these signals compared to the conventional HHT, especially in instances of mode-mixing issues.
We notice that the intra- or inter-wave modulations caused by the intrinsic noise within the data may produce the oscillation features on the time-frequency representation, and this can be confirmed by analyzing a pure signal using the sHHT method (details in Appendix~\ref{app:PS_investigation}).
While these oscillation features can become somewhat smeared when we consume more time to accumulate the Hilbert spectra, they cannot be completely eliminated.
Nevertheless, the related effect is minor when we track a signal crossing over a wide frequency range within a short time interval.

Similar to the HHT, the sHHT, a template-free method, does not require the imposition of a priori basis sets on the data. 
Therefore, the sHHT offers an adaptive approach to time-frequency analysis, making sHHT suitable for analyzing non-stationary signals. 
We have explored the frequency dispersion through examples of stationary sinusoidal, linear chirp, and exponential chirp signals. 
Our analysis reveals how the frequency dispersion evolves with changes in SNRs, highlighting the sensitivity of sHHT to these factors. 
By the investigation of simulations, we recommend incorporating an additional noise with a level between 0.5 and 1 times the standard deviation of the signal strength to perform this approach. 
Moreover, our application of sHHT to real astronomical data, including the superorbital modulation of SMC X-1 and the signals from confirmed GW events, further highlights its capability to extract meaningful physical insights. 
The aforementioned external noise level between 0.5 -- 1 is also suitable to generate the stacked Hilbert spectrum of most known GW events.
These results point out the potential of sHHT to serve as a powerful tool in the field of astrophysics and beyond, given the growing importance of the complexity of the signals observed in the multi-messenger era.
For example, it will be remarkable to further quantitatively assess the GW events including HMs to compare with Wilson–Daubechies–Meyer wavelet transform \citep{2012NKM} in the future.

\begin{acknowledgments}
This research has used data or software obtained from the Gravitational Wave Open Science Center (gwosc.org), a service of the LIGO Scientific Collaboration, the Virgo Collaboration, and KAGRA. 
This material is based upon work supported by NSF's LIGO Laboratory which is a major facility fully funded by the National Science Foundation, as well as the Science and Technology Facilities Council (STFC) of the United Kingdom, the Max-Planck-Society (MPS), and the State of Niedersachsen/Germany for support of the construction of Advanced LIGO and construction and operation of the GEO600 detector. 
Additional support for Advanced LIGO was provided by the Australian Research Council. 
Virgo is funded, through the European Gravitational Observatory (EGO), by the French Centre National de Recherche Scientifique (CNRS), the Italian Istituto Nazionale di Fisica Nucleare (INFN), and the Dutch Nikhef, with contributions by institutions from Belgium, Germany, Greece, Hungary, Ireland, Japan, Monaco, Poland, Portugal, and Spain. 
KAGRA is supported by the Ministry of Education, Culture, Sports, Science and Technology (MEXT), Japan Society for the Promotion of Science (JSPS) in Japan; National Research Foundation (NRF) and the Ministry of Science and ICT (MSIT) in Korea; Academia Sinica (AS) and National Science and Technology Council (NSTC) in Taiwan.
C.-P.H.~acknowledges support from the NSTC in Taiwan through grants 109-2112-M-018-009-MY3 and 112-2112-M-018-004-MY3. 
K.C.P. is supported by the Ministry of Science and Technology of Taiwan through grant NSTC 113-2811-M-007-026.
L.C.C.L., Y.S.H., and K.L.L. both acknowledge the support of the Yushan Fellow Program from the Ministry of Education (MOE) of Taiwan through grant MOE-109-YSFMS-0005-001-P1.
L.C.C.L., and K.L.L. are also supported by the NSTC of Taiwan through grant NSTC 113-2636-M-006-003.
C.-C.Y. is supported by the NSTC under 112-2115-M-030-002- and 113-2115-M-030-002-.
A.K.H.K. and Y.C.L. are supported by the NSTC of Taiwan under grant 113-2112-M-007-001. 
C.Y.H. is supported by the research fund of Chungnam National University and by the National Research Foundation of Korea grant 2022R1F1A1073952.
\end{acknowledgments}

{\sl Software:}  GWPY (Python-based GW analysis library, v3.0.82; \cite{gwpy}); PyCBC (A Python-based parameter estimation toolkit for CBC signals, v2.4.0; \cite{Biwer2019}); emd v0.6.23 \cite{Quinn2021}

\appendix
\section{Different input noise levels imposed on the real GW data}\label{app:EN_investigation}

\begin{figure*}[tp]
\centering
\includegraphics[width=4.4cm,height=4.2cm]{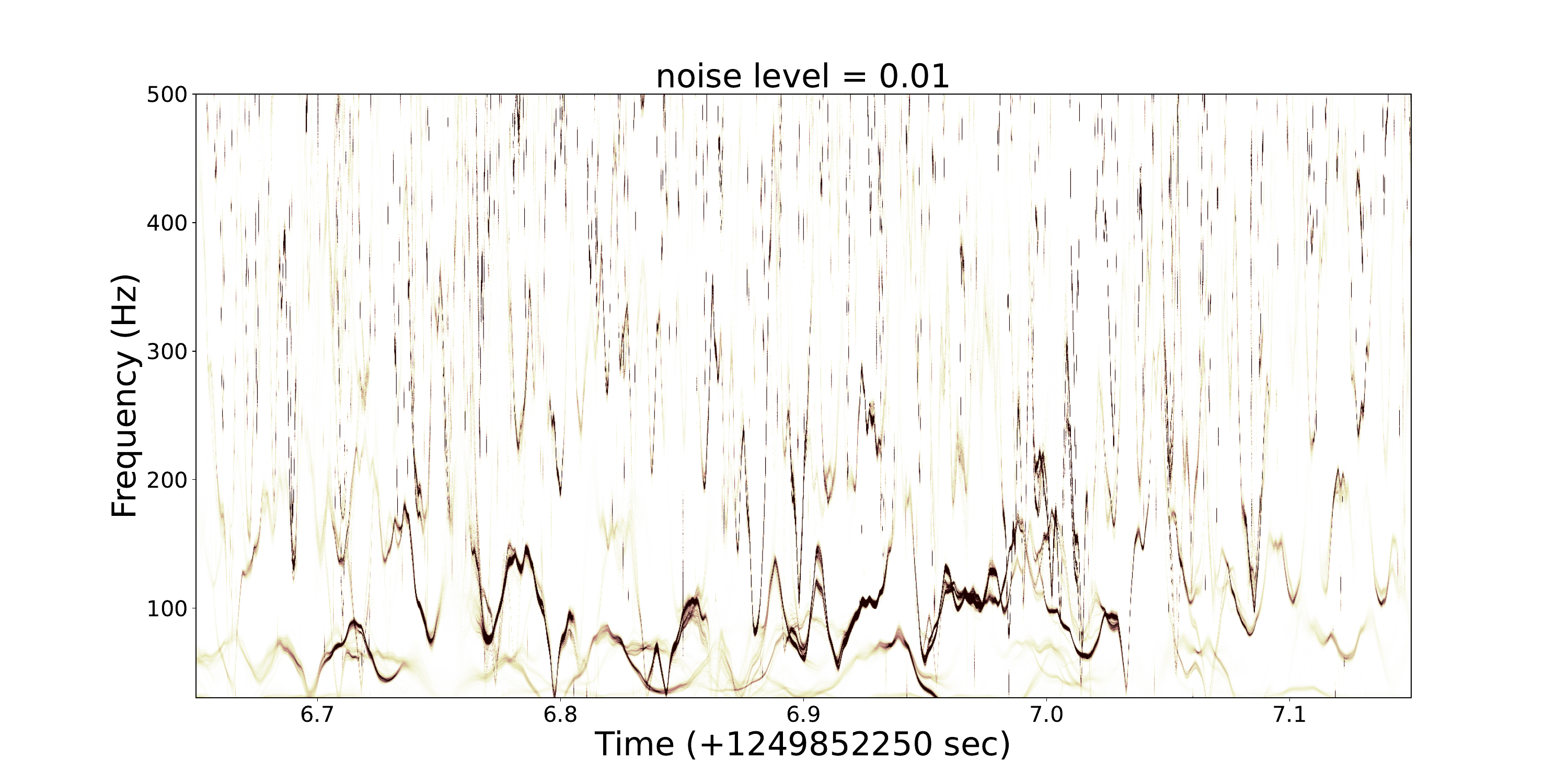}
\includegraphics[width=4.4cm,height=4.2cm]{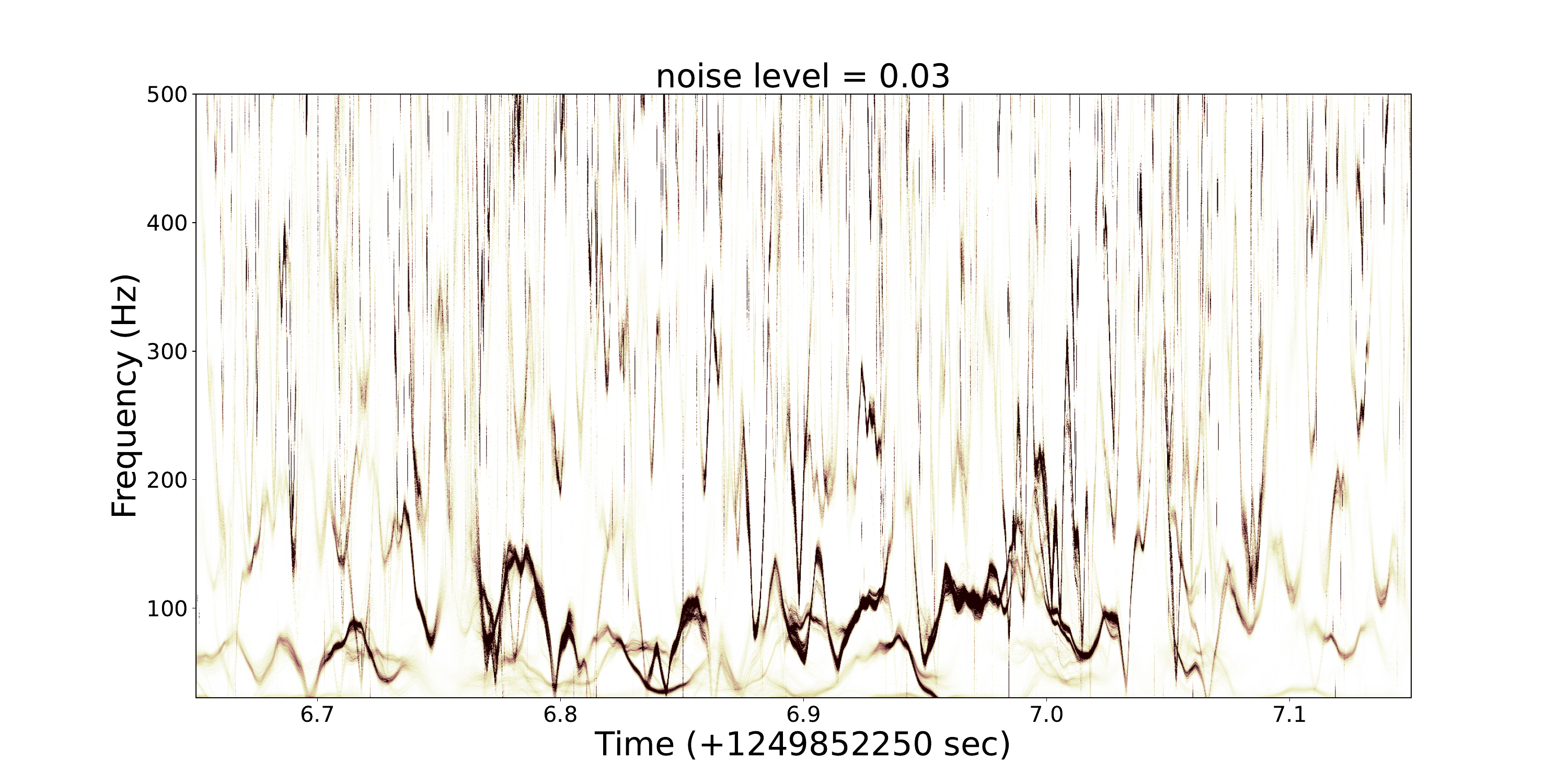}
\includegraphics[width=4.4cm,height=4.2cm]{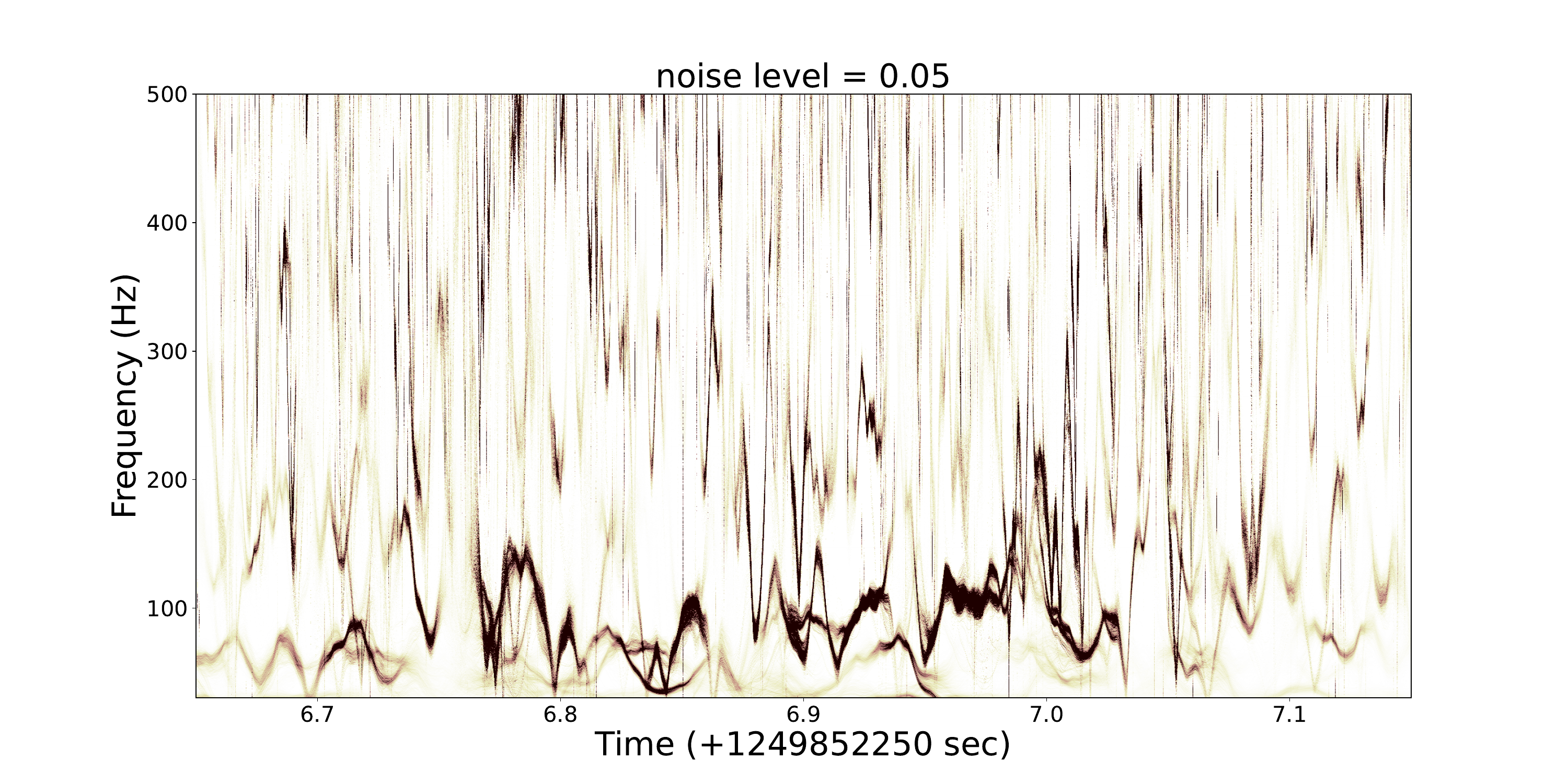}
\includegraphics[width=4.4cm,height=4.2cm]{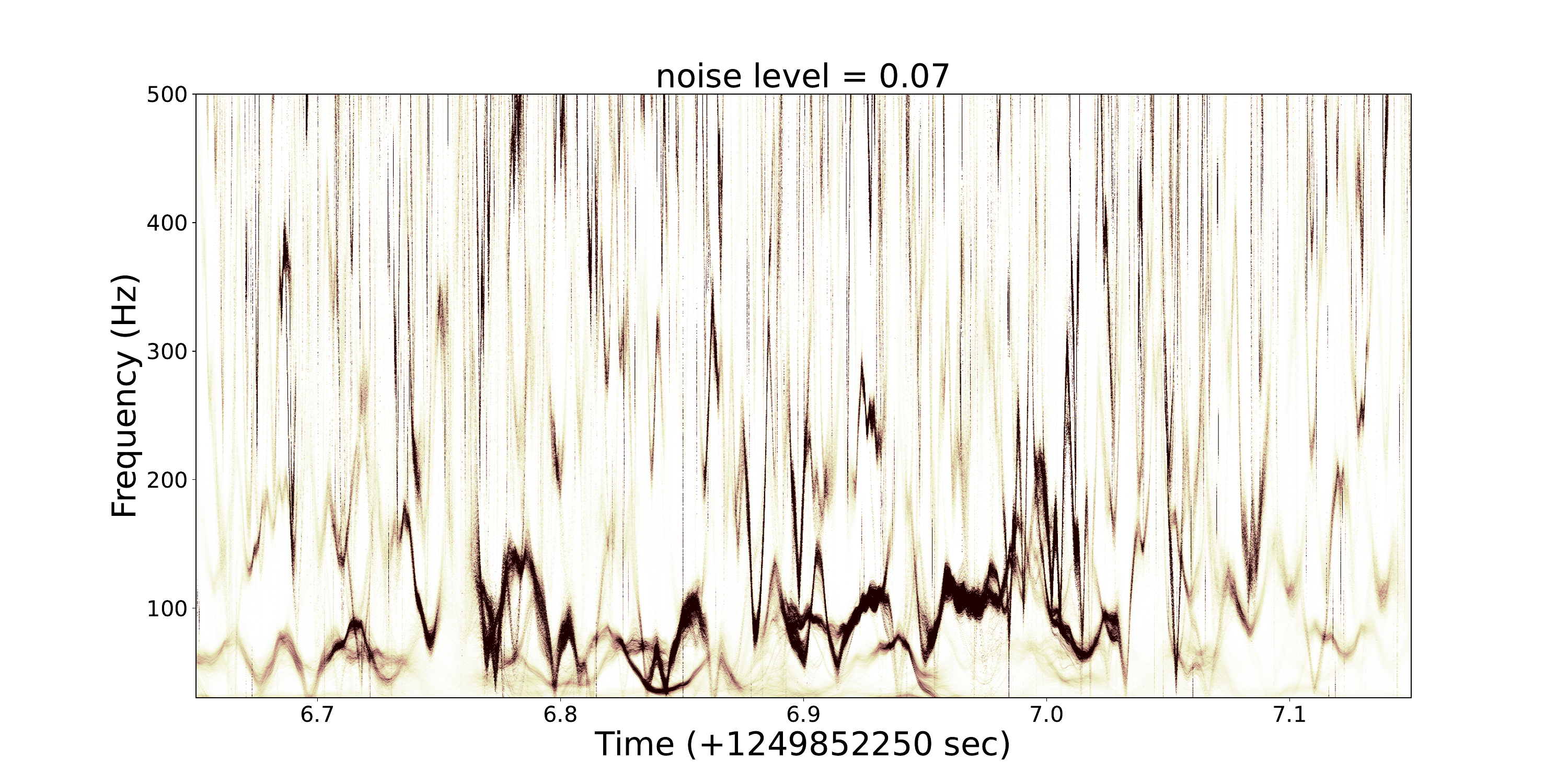}
\includegraphics[width=4.4cm,height=4.2cm]{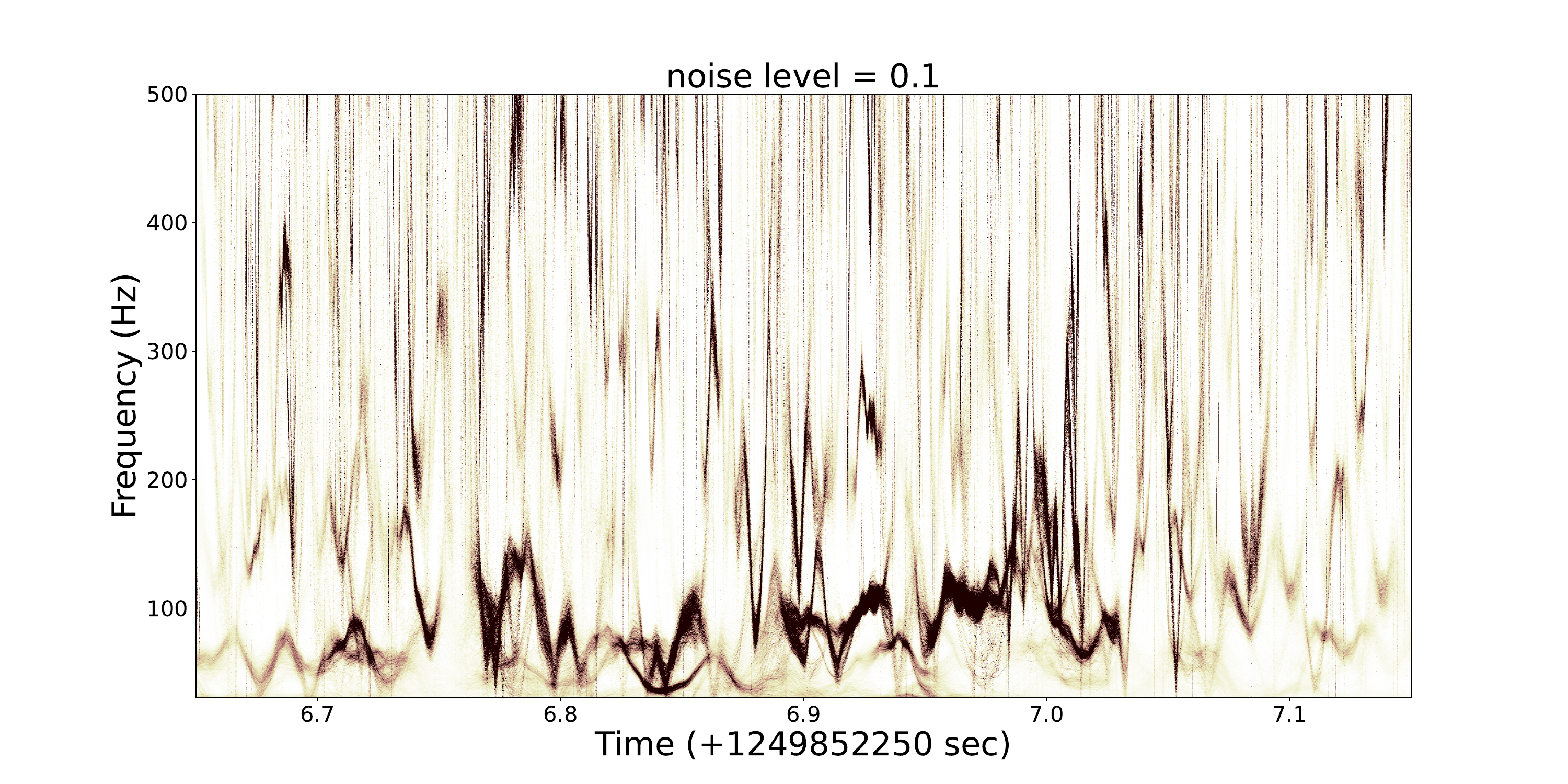}
\includegraphics[width=4.4cm,height=4.2cm]{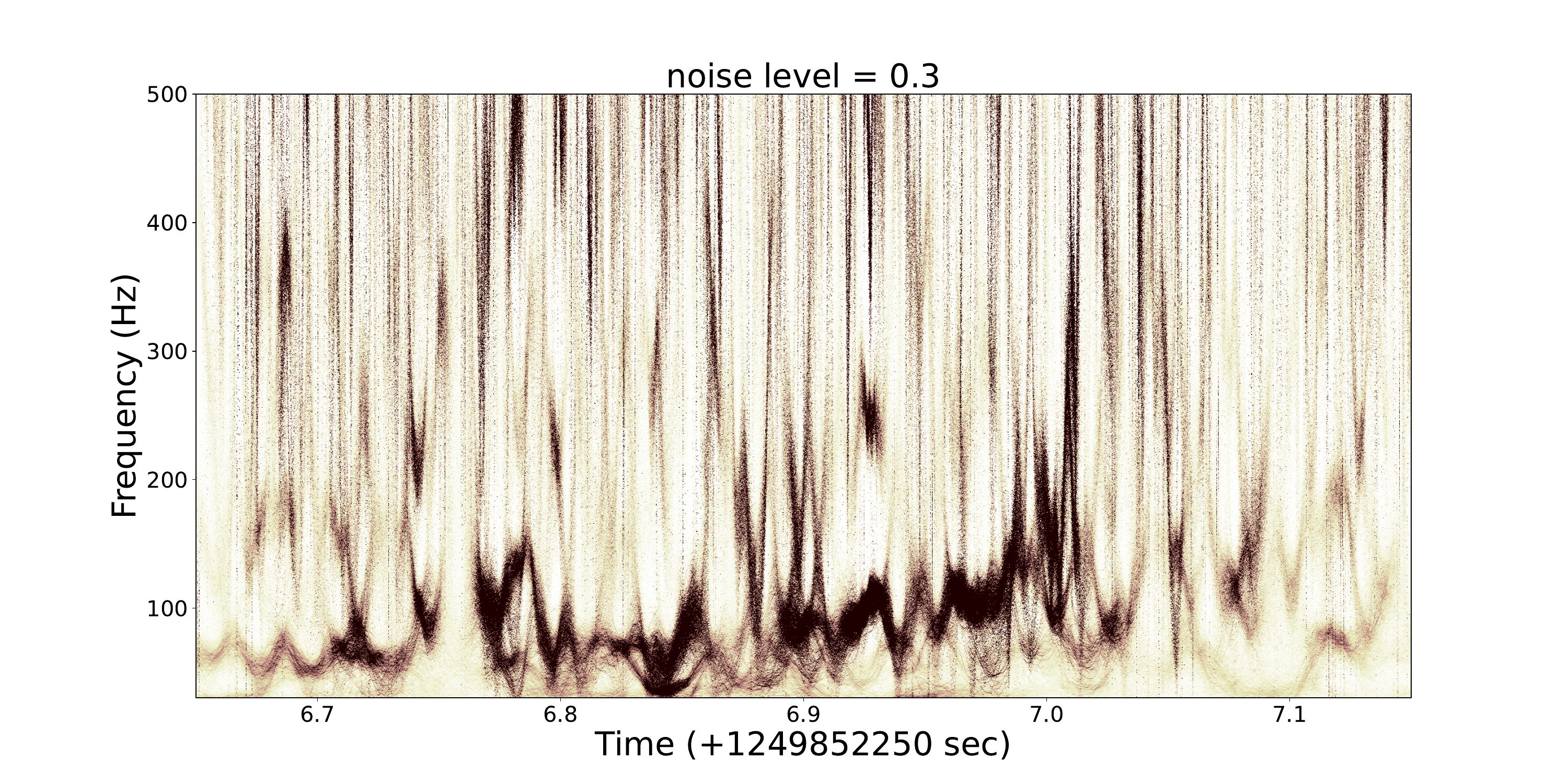}
\includegraphics[width=4.4cm,height=4.2cm]{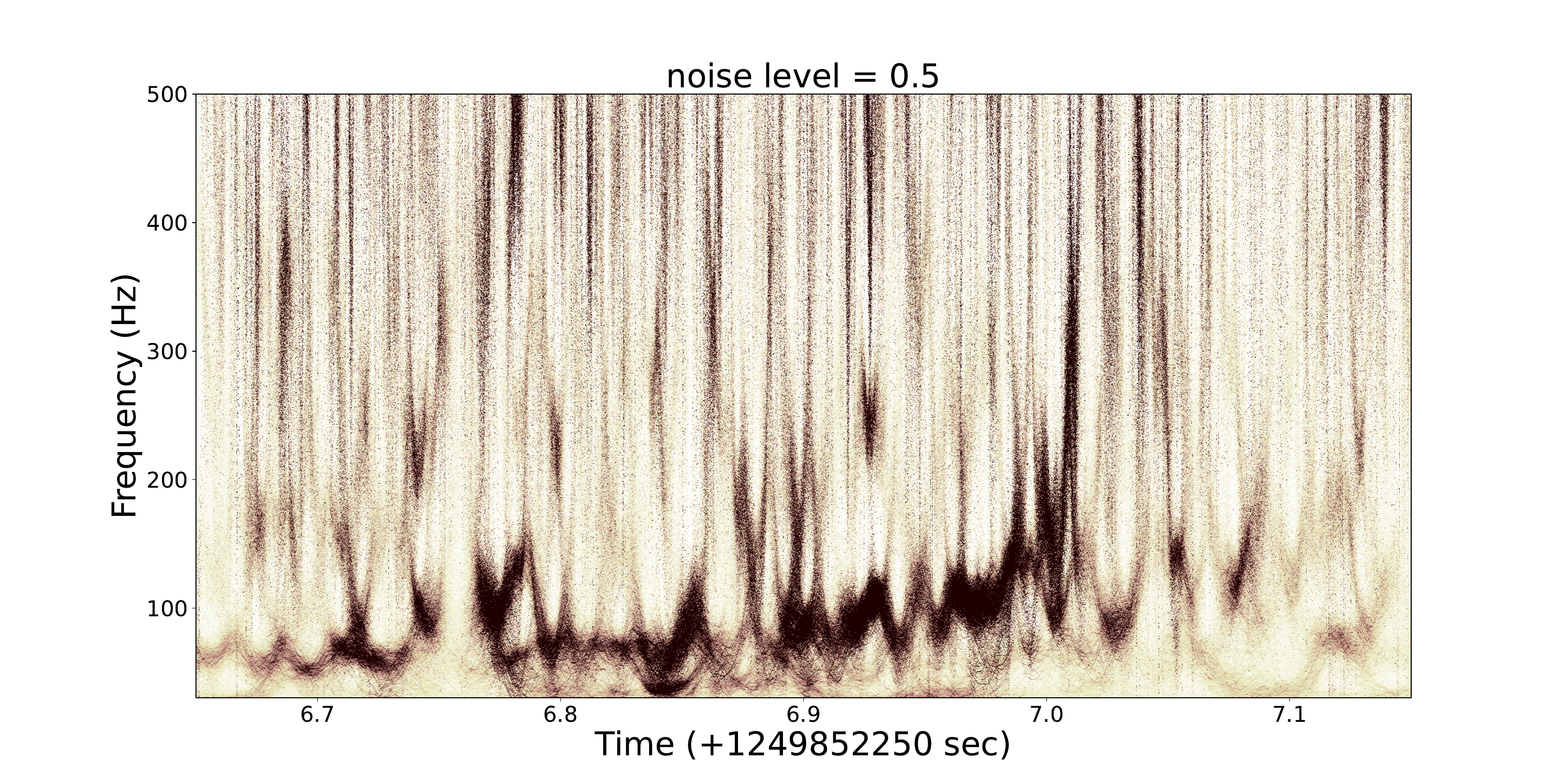}
\includegraphics[width=4.4cm,height=4.2cm]{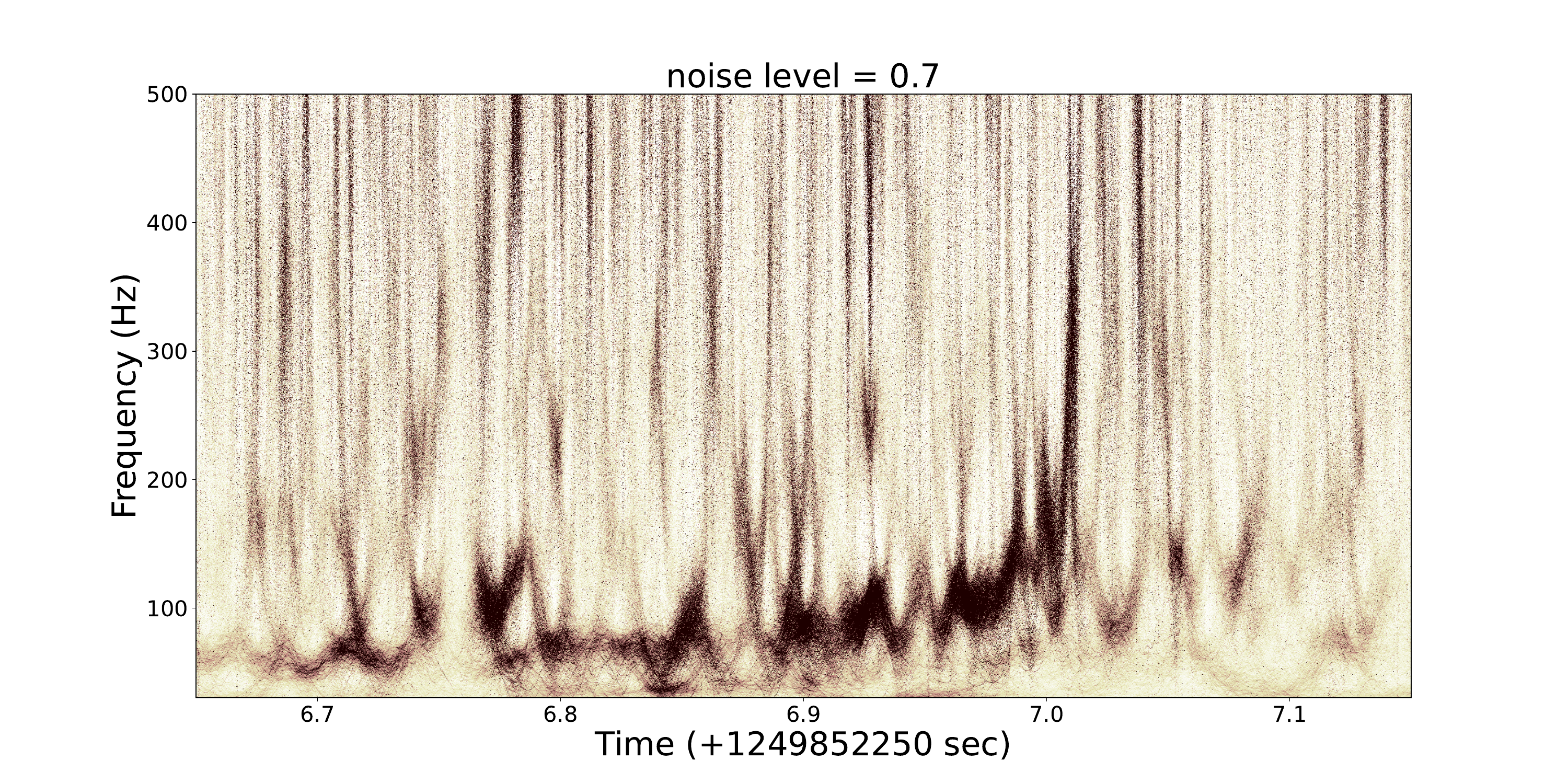}
\includegraphics[width=4.4cm,height=4.2cm]{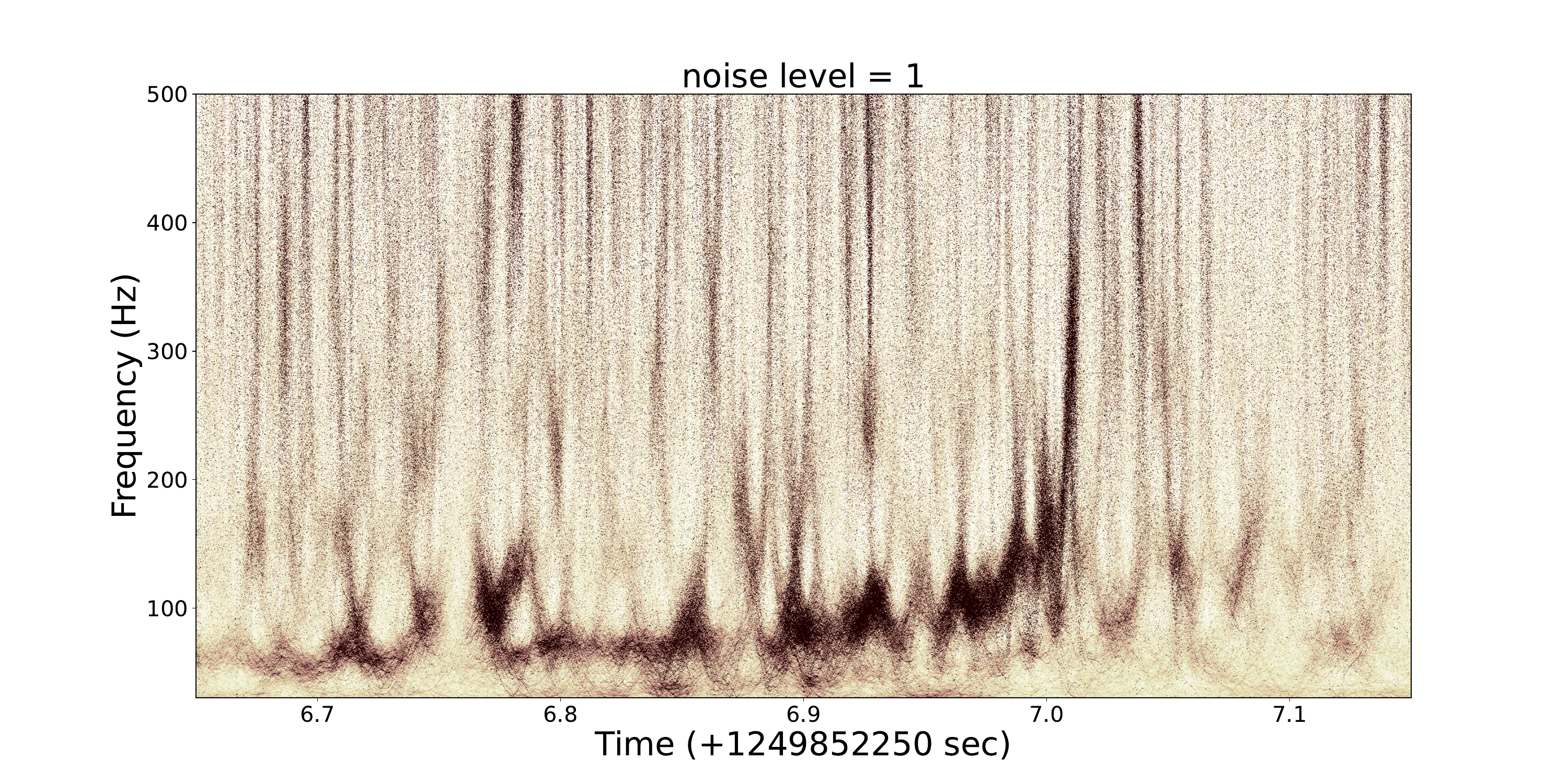}
\includegraphics[width=4.4cm,height=4.2cm]{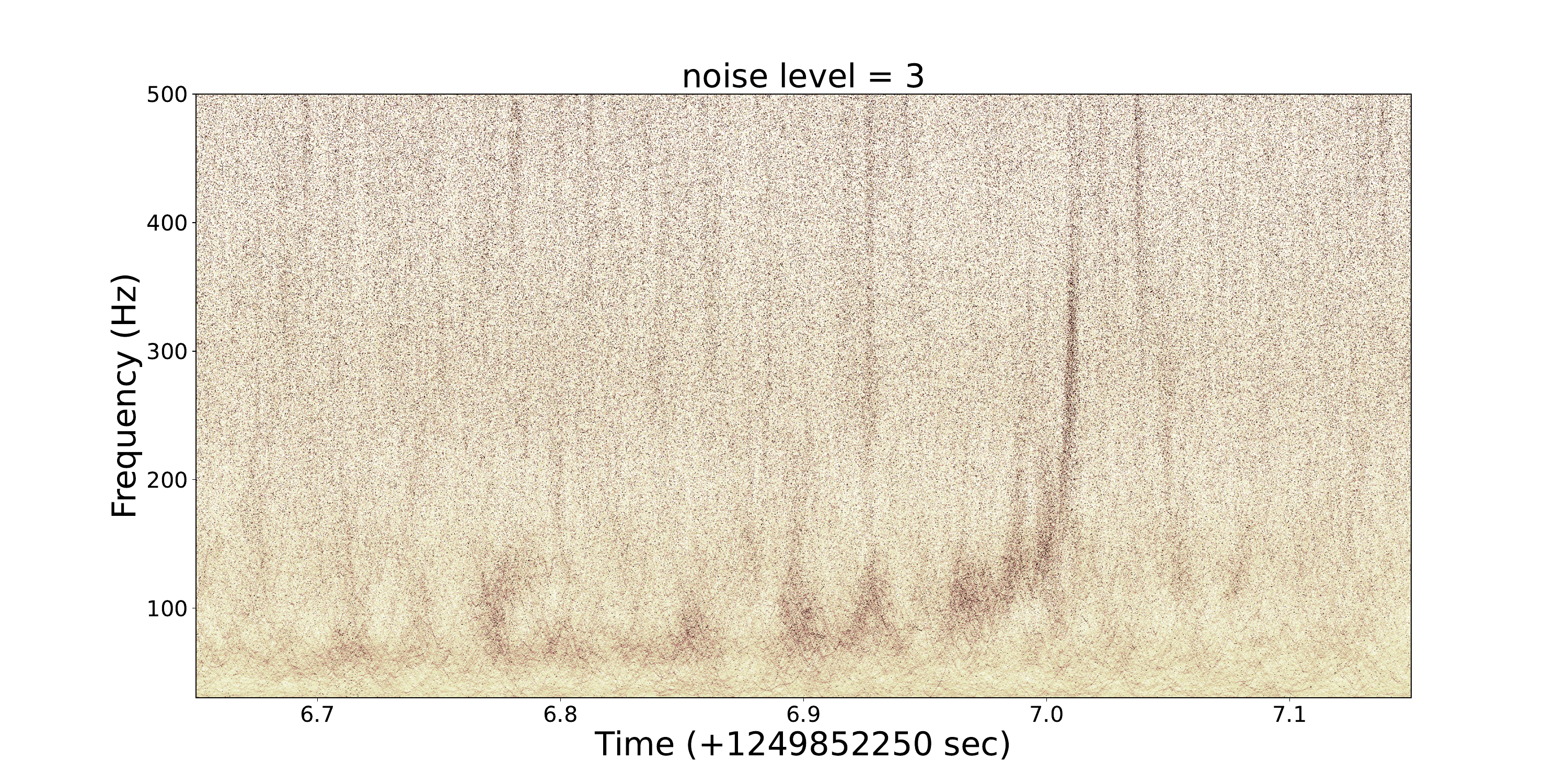}
\includegraphics[width=4.4cm,height=4.2cm]{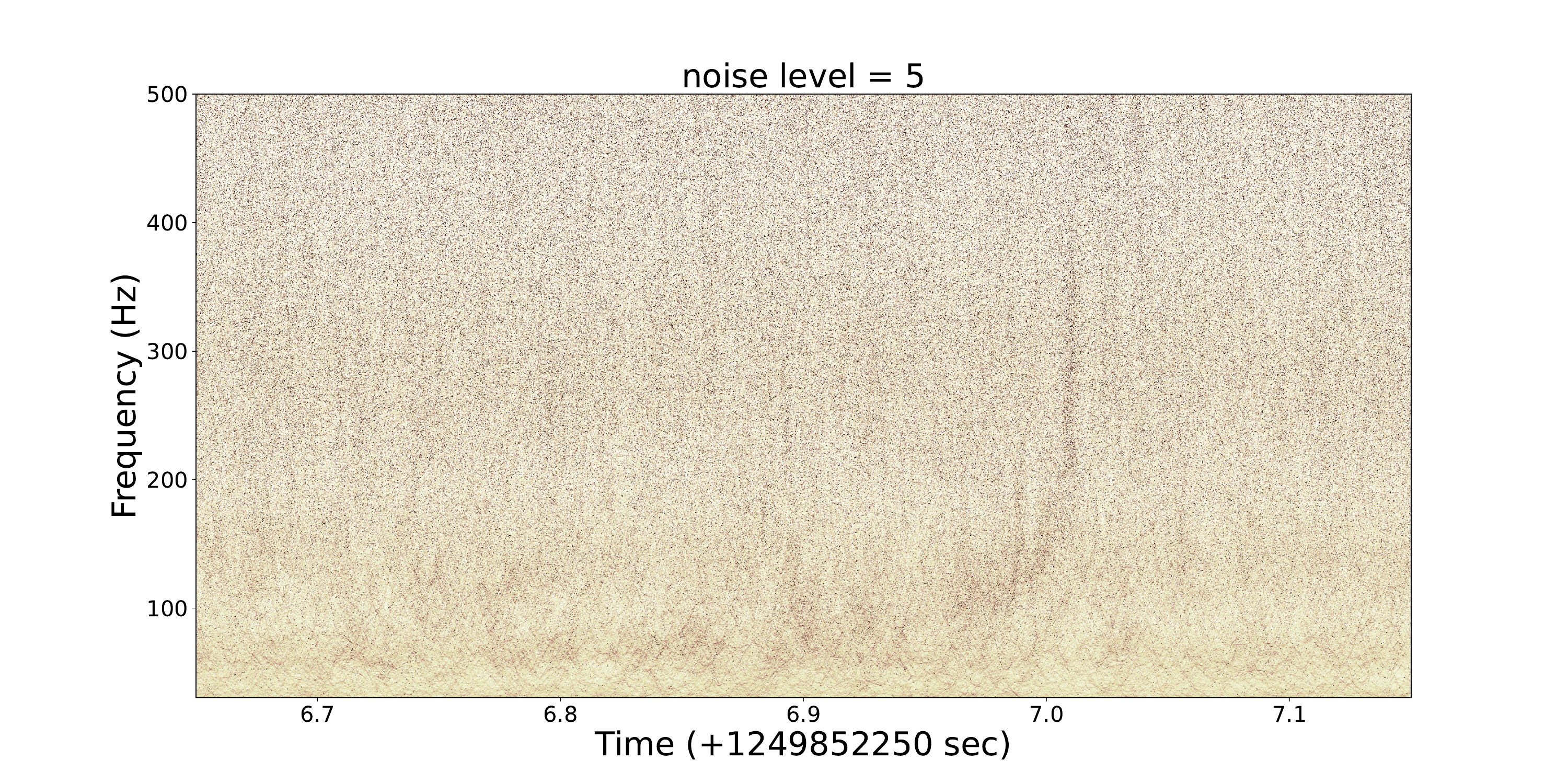}
\includegraphics[width=4.4cm,height=4.2cm]{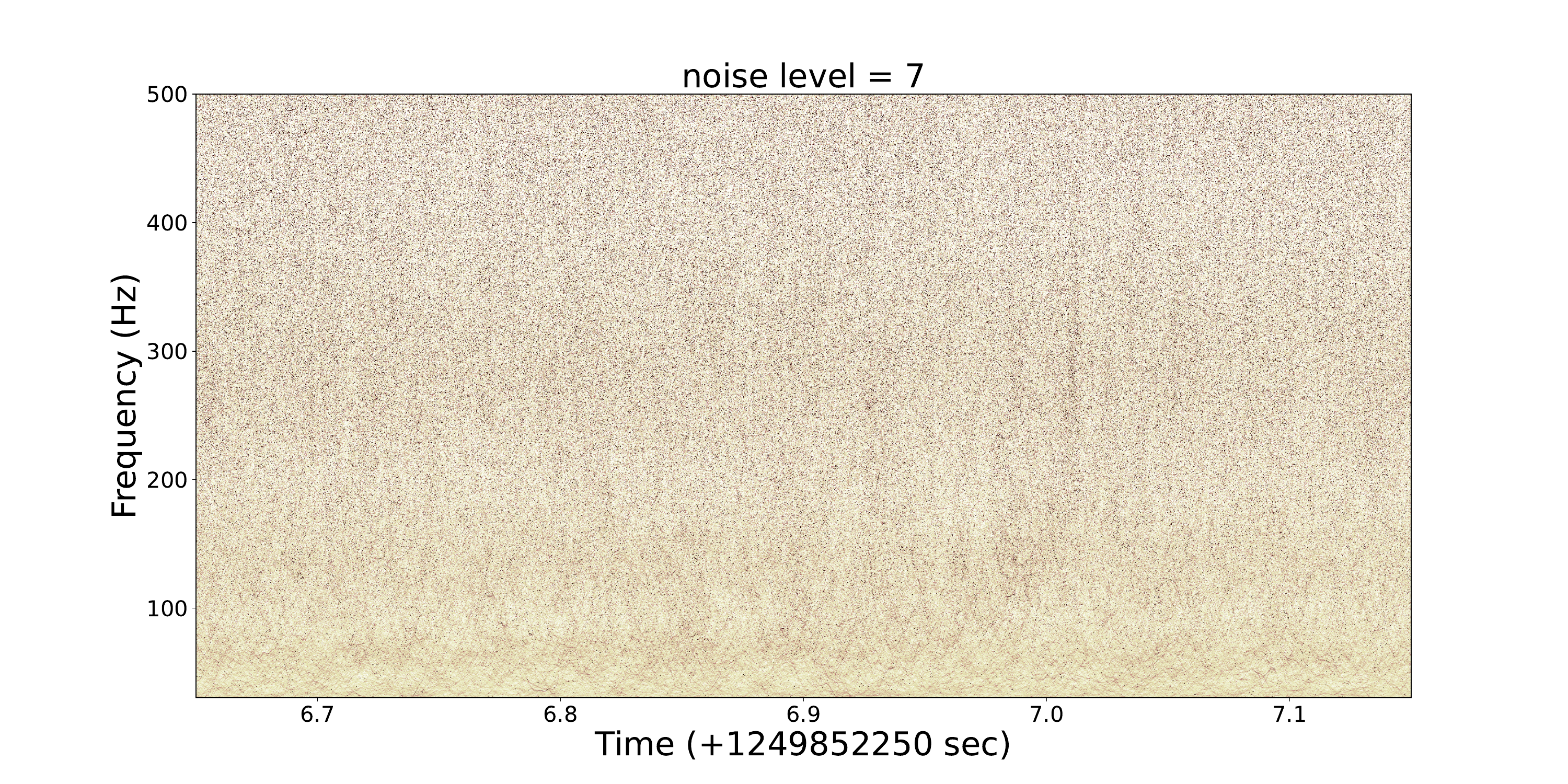}
\caption{{\footnotesize Stacked Hilbert spectra of GW190814 obtained by sHHT with different external noise levels. Here we accumulated the spectra 1000 times and showed the spectrograms generated by the Livingston data.}} 
\label{append1}
\end{figure*}

Since we have no information about the SNR of one GW whitened data in the beginning, the optimal noise level incorporated to perform the sHHT can only be decided from a range (i.e., 0.01--10) with a repeated investigation on the visual performance of the results. 
We created a random time series by multiplying the external noise level and the dataset's standard deviation (amplitude). 
This time series was then imposed on the original data to perform sHHT.
Here we present all the stacked Hilbert spectra of the Livingston data generated with different external noise levels (Fig.~\ref{append1}) as references.
Please note that the contour level of each plot in Fig.~\ref{append1} is flexibly adjusted to highlight the real signal. 
Results indicate that the mode-mixing phenomenon remains significant, and the Hilbert spectrum is heavily influenced by intrinsic/internal noise when the incorporated noise level is too low.
On the contrary, the real signal on the Hilbert spectra can be smeared by the strong external noise incorporated to perform the sHHT.

\begin{figure*}[htb]
\centering
\includegraphics[width=8.5cm,height=4.7cm]{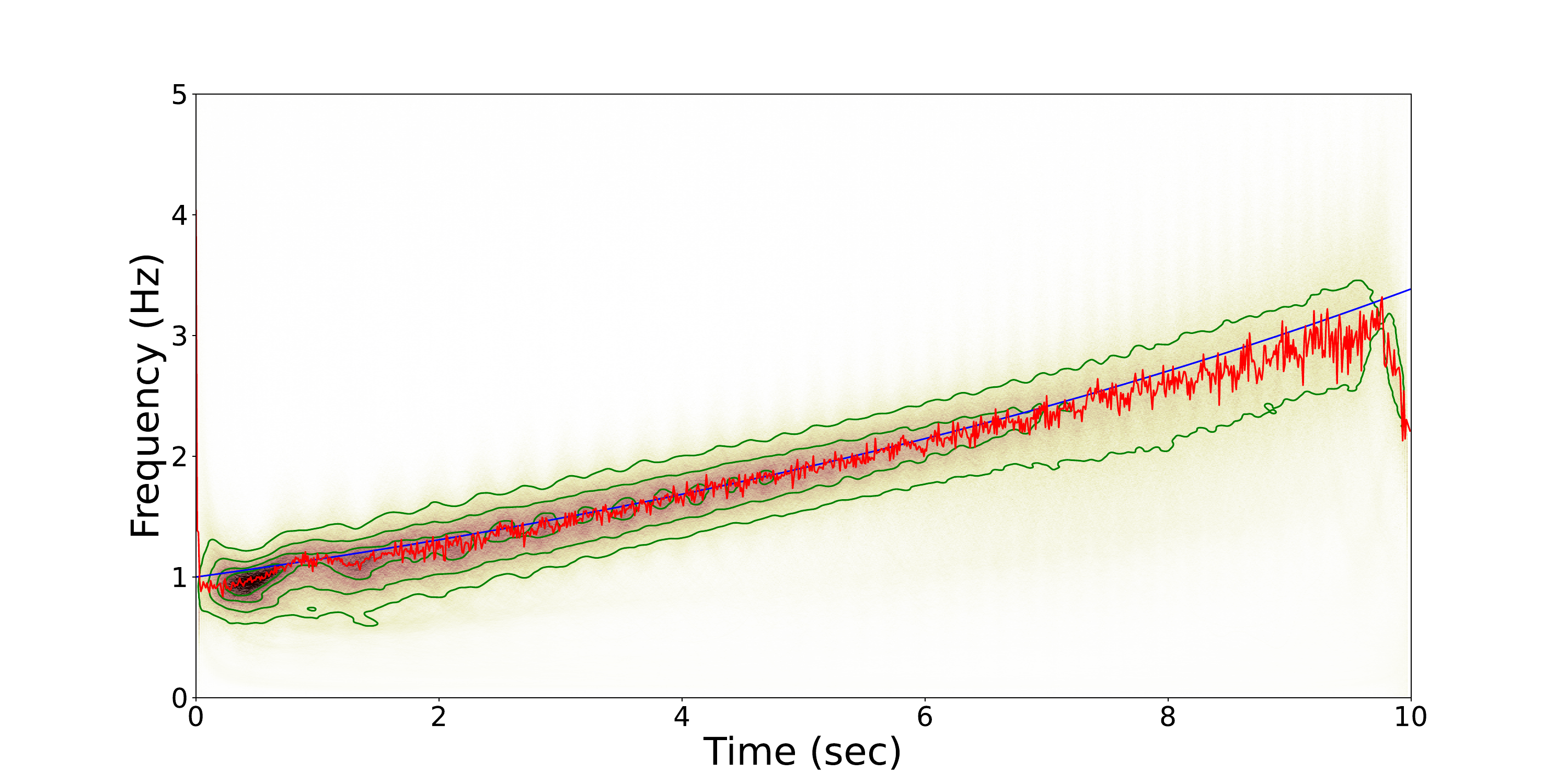}
\includegraphics[width=8.5cm,height=4.7cm]{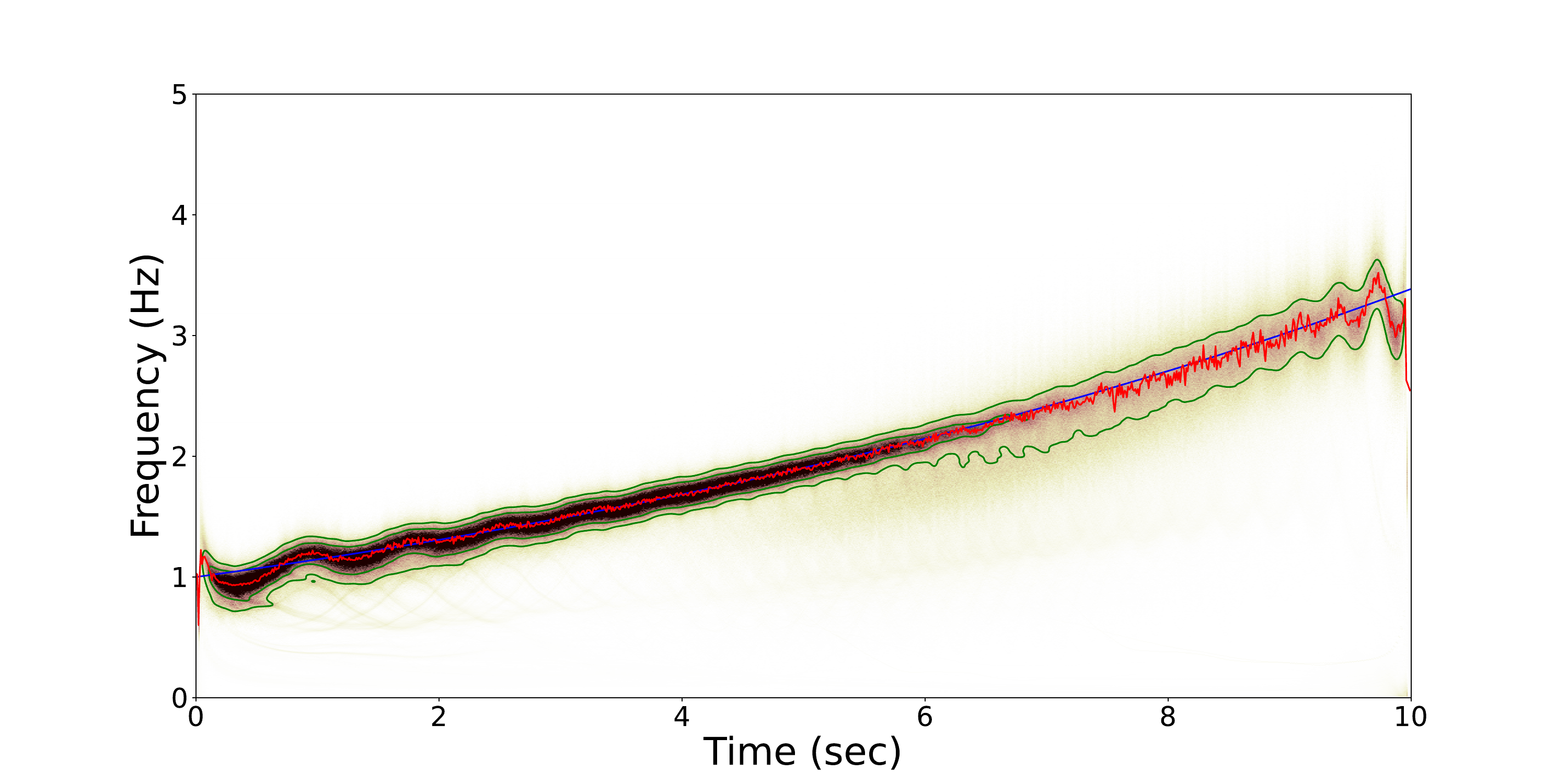}
\caption{{\footnotesize Pure exponential chirp signal the same as Fig.~\ref{linearchirp_disp1} resolved by the sHHT method. The stacked Hilbert spectrum was obtained with the incorporation of the external noise of 0.5 in the left panel and of 0.05 in the right panel. The blue curve shows the actual signal frequency, and the red curves in both panels represent the IF determined by the sHHT, accumulated 10000 times. The green contour exhibits the different Hilbert energy levels on the stacked Hilbert spectra.}} 
\label{append2}
\end{figure*}

\section{Different input noise levels imposed on the pure signal}\label{app:PS_investigation}

In order to confirm whether the oscillation modulations shown on the stacked Hilbert spectra are caused by the internal noise of the data, we directly apply the sHHT to investigate the pure exponential chirp signal described in \S~\ref{sec:dispersion}.C.
Fig.~\ref{append2} presents the results yielded with different input noise levels to conduct the sHHT.
Compared with the stacked Hilbert spectra shown in Figs.~\ref{logchirp_disp1} and~\ref{logchirp_disp2}, the contour to denote the signal behavior is almost uniformly distributed along the actual blue signals. 
For the IFs determined in two plots of Fig.~\ref{append2}, we can still see minor oscillations because we incorporate the external noise to perform the sHHT, and it also reflects on the wavy artifacts of the contours.
We notice that the oscillation of IF in the left panel is visually more significant than that shown in the right panel of Fig~\ref{append2} because the external noise level applied to perform the sHHT is one order of magnitude stronger. 
We therefore conclude that the oscillation features shown on the stacked Hilbert spectra are led by the intrinsic noises embedded inside the data, which cause significant intra- or inter-wave modulations to determine the IF.
The external noise incorporated to perform the sHHT can also lead to a similar minor effect, and to pick up the proper noise level is important to investigate a stationary signal with sHHT.


\begin{thebibliography}{58}%
\makeatletter
\providecommand \@ifxundefined [1]{%
 \@ifx{#1\undefined}
}%
\providecommand \@ifnum [1]{%
 \ifnum #1\expandafter \@firstoftwo
 \else \expandafter \@secondoftwo
 \fi
}%
\providecommand \@ifx [1]{%
 \ifx #1\expandafter \@firstoftwo
 \else \expandafter \@secondoftwo
 \fi
}%
\providecommand \natexlab [1]{#1}%
\providecommand \enquote  [1]{``#1''}%
\providecommand \bibnamefont  [1]{#1}%
\providecommand \bibfnamefont [1]{#1}%
\providecommand \citenamefont [1]{#1}%
\providecommand \href@noop [0]{\@secondoftwo}%
\providecommand \href [0]{\begingroup \@sanitize@url \@href}%
\providecommand \@href[1]{\@@startlink{#1}\@@href}%
\providecommand \@@href[1]{\endgroup#1\@@endlink}%
\providecommand \@sanitize@url [0]{\catcode `\\12\catcode `\$12\catcode
  `\&12\catcode `\#12\catcode `\^12\catcode `\_12\catcode `\%12\relax}%
\providecommand \@@startlink[1]{}%
\providecommand \@@endlink[0]{}%
\providecommand \url  [0]{\begingroup\@sanitize@url \@url }%
\providecommand \@url [1]{\endgroup\@href {#1}{\urlprefix }}%
\providecommand \urlprefix  [0]{URL }%
\providecommand \Eprint [0]{\href }%
\providecommand \doibase [0]{https://doi.org/}%
\providecommand \selectlanguage [0]{\@gobble}%
\providecommand \bibinfo  [0]{\@secondoftwo}%
\providecommand \bibfield  [0]{\@secondoftwo}%
\providecommand \translation [1]{[#1]}%
\providecommand \BibitemOpen [0]{}%
\providecommand \bibitemStop [0]{}%
\providecommand \bibitemNoStop [0]{.\EOS\space}%
\providecommand \EOS [0]{\spacefactor3000\relax}%
\providecommand \BibitemShut  [1]{\csname bibitem#1\endcsname}%
\let\auto@bib@innerbib\@empty
\bibitem [{\citenamefont {{Chatterji}}\ \emph {et~al.}(2004)\citenamefont
  {{Chatterji}}, \citenamefont {{Blackburn}}, \citenamefont {{Martin}},\ and\
  \citenamefont {{Katsavounidis}}}]{Chatterji2004}%
  \BibitemOpen
  \bibfield  {author} {\bibinfo {author} {\bibfnamefont {S.}~\bibnamefont
  {{Chatterji}}}, \bibinfo {author} {\bibfnamefont {L.}~\bibnamefont
  {{Blackburn}}}, \bibinfo {author} {\bibfnamefont {G.}~\bibnamefont
  {{Martin}}},\ and\ \bibinfo {author} {\bibfnamefont {E.}~\bibnamefont
  {{Katsavounidis}}},\ }\bibfield  {title} {\bibinfo {title} {{Multiresolution
  techniques for the detection of gravitational-wave bursts}},\ }\href
  {https://doi.org/10.1088/0264-9381/21/20/024} {\bibfield  {journal} {\bibinfo
   {journal} {Classical and Quantum Gravity}\ }\textbf {\bibinfo {volume}
  {21}},\ \bibinfo {pages} {S1809} (\bibinfo {year} {2004})},\ \Eprint
  {https://arxiv.org/abs/gr-qc/0412119} {arXiv:gr-qc/0412119 [gr-qc]}
  \BibitemShut {NoStop}%
\bibitem [{\citenamefont {{Lin}}\ \emph {et~al.}(2015)\citenamefont {{Lin}},
  \citenamefont {{Hu}}, \citenamefont {{Kong}}, \citenamefont {{Yen}},
  \citenamefont {{Takata}},\ and\ \citenamefont {{Chou}}}]{Lin2015}%
  \BibitemOpen
  \bibfield  {author} {\bibinfo {author} {\bibfnamefont {L.~C.-C.}\
  \bibnamefont {{Lin}}}, \bibinfo {author} {\bibfnamefont {C.-P.}\ \bibnamefont
  {{Hu}}}, \bibinfo {author} {\bibfnamefont {A.~K.~H.}\ \bibnamefont {{Kong}}},
  \bibinfo {author} {\bibfnamefont {D.~C.-C.}\ \bibnamefont {{Yen}}}, \bibinfo
  {author} {\bibfnamefont {J.}~\bibnamefont {{Takata}}},\ and\ \bibinfo
  {author} {\bibfnamefont {Y.}~\bibnamefont {{Chou}}},\ }\bibfield  {title}
  {\bibinfo {title} {{Long-term X-ray variability of ultraluminous X-ray
  sources}},\ }\href {https://doi.org/10.1093/mnras/stv2025} {\bibfield
  {journal} {\bibinfo  {journal} {MNRAS}\ }\textbf {\bibinfo {volume} {454}},\
  \bibinfo {pages} {1644} (\bibinfo {year} {2015})},\ \Eprint
  {https://arxiv.org/abs/1509.00547} {arXiv:1509.00547 [astro-ph.HE]}
  \BibitemShut {NoStop}%
\bibitem [{\citenamefont {{Yanagisawa}}\ \emph {et~al.}(2019)\citenamefont
  {{Yanagisawa}}, \citenamefont {{Jia}}, \citenamefont {{Hirobayashi}},
  \citenamefont {{Uchikata}}, \citenamefont {{Narikawa}}, \citenamefont
  {{Ueno}}, \citenamefont {{Takahashi}},\ and\ \citenamefont
  {{Tagoshi}}}]{Yanagisawa2019}%
  \BibitemOpen
  \bibfield  {author} {\bibinfo {author} {\bibfnamefont {K.}~\bibnamefont
  {{Yanagisawa}}}, \bibinfo {author} {\bibfnamefont {D.}~\bibnamefont {{Jia}}},
  \bibinfo {author} {\bibfnamefont {S.}~\bibnamefont {{Hirobayashi}}}, \bibinfo
  {author} {\bibfnamefont {N.}~\bibnamefont {{Uchikata}}}, \bibinfo {author}
  {\bibfnamefont {T.}~\bibnamefont {{Narikawa}}}, \bibinfo {author}
  {\bibfnamefont {K.}~\bibnamefont {{Ueno}}}, \bibinfo {author} {\bibfnamefont
  {H.}~\bibnamefont {{Takahashi}}},\ and\ \bibinfo {author} {\bibfnamefont
  {H.}~\bibnamefont {{Tagoshi}}},\ }\bibfield  {title} {\bibinfo {title} {{A
  time-frequency analysis of gravitational wave signals with non-harmonic
  analysis}},\ }\href {https://doi.org/10.1093/ptep/ptz043} {\bibfield
  {journal} {\bibinfo  {journal} {Progress of Theoretical and Experimental
  Physics}\ }\textbf {\bibinfo {volume} {2019}},\ \bibinfo {eid} {063F01}
  (\bibinfo {year} {2019})}\BibitemShut {NoStop}%
\bibitem [{\citenamefont {{Hu}}\ \emph {et~al.}(2023)\citenamefont {{Hu}},
  \citenamefont {{Dage}}, \citenamefont {{Clarkson}}, \citenamefont
  {{Brumback}}, \citenamefont {{Charles}}, \citenamefont {{Haggard}},
  \citenamefont {{Hickox}}, \citenamefont {{Mihara}}, \citenamefont
  {{Bahramian}}, \citenamefont {{Karam}}, \citenamefont {{Athukoralalage}},
  \citenamefont {{Altamirano}}, \citenamefont {{Neilsen}},\ and\ \citenamefont
  {{Kennea}}}]{HuDK2023}%
  \BibitemOpen
  \bibfield  {author} {\bibinfo {author} {\bibfnamefont {C.-P.}\ \bibnamefont
  {{Hu}}}, \bibinfo {author} {\bibfnamefont {K.~C.}\ \bibnamefont {{Dage}}},
  \bibinfo {author} {\bibfnamefont {W.~I.}\ \bibnamefont {{Clarkson}}},
  \bibinfo {author} {\bibfnamefont {M.}~\bibnamefont {{Brumback}}}, \bibinfo
  {author} {\bibfnamefont {P.~A.}\ \bibnamefont {{Charles}}}, \bibinfo {author}
  {\bibfnamefont {D.}~\bibnamefont {{Haggard}}}, \bibinfo {author}
  {\bibfnamefont {R.~C.}\ \bibnamefont {{Hickox}}}, \bibinfo {author}
  {\bibfnamefont {T.}~\bibnamefont {{Mihara}}}, \bibinfo {author}
  {\bibfnamefont {A.}~\bibnamefont {{Bahramian}}}, \bibinfo {author}
  {\bibfnamefont {R.}~\bibnamefont {{Karam}}}, \bibinfo {author} {\bibfnamefont
  {W.}~\bibnamefont {{Athukoralalage}}}, \bibinfo {author} {\bibfnamefont
  {D.}~\bibnamefont {{Altamirano}}}, \bibinfo {author} {\bibfnamefont
  {J.}~\bibnamefont {{Neilsen}}},\ and\ \bibinfo {author} {\bibfnamefont
  {J.}~\bibnamefont {{Kennea}}},\ }\bibfield  {title} {\bibinfo {title}
  {{Monitoring observations of SMC X-1's excursions (MOOSE) - II. A new
  excursion accompanies spin-up acceleration}},\ }\href
  {https://doi.org/10.1093/mnras/stad384} {\bibfield  {journal} {\bibinfo
  {journal} {MNRAS}\ }\textbf {\bibinfo {volume} {520}},\ \bibinfo {pages}
  {3436} (\bibinfo {year} {2023})},\ \Eprint {https://arxiv.org/abs/2302.00260}
  {arXiv:2302.00260 [astro-ph.HE]} \BibitemShut {NoStop}%
\bibitem [{\citenamefont {{Brown}}(1991)}]{Brown91}%
  \BibitemOpen
  \bibfield  {author} {\bibinfo {author} {\bibfnamefont {J.~C.}\ \bibnamefont
  {{Brown}}},\ }\bibfield  {title} {\bibinfo {title} {{Calculation of a
  constant Q spectral transform}},\ }\href {https://doi.org/10.1121/1.400476}
  {\bibfield  {journal} {\bibinfo  {journal} {Acoustical Society of America
  Journal}\ }\textbf {\bibinfo {volume} {89}},\ \bibinfo {pages} {425}
  (\bibinfo {year} {1991})}\BibitemShut {NoStop}%
\bibitem [{\citenamefont {Robinet}\ \emph {et~al.}(2020)\citenamefont
  {Robinet}, \citenamefont {Arnaud}, \citenamefont {Leroy}, \citenamefont
  {Lundgren}, \citenamefont {Macleod},\ and\ \citenamefont
  {McIver}}]{Robinet2020}%
  \BibitemOpen
  \bibfield  {author} {\bibinfo {author} {\bibfnamefont {F.}~\bibnamefont
  {Robinet}}, \bibinfo {author} {\bibfnamefont {N.}~\bibnamefont {Arnaud}},
  \bibinfo {author} {\bibfnamefont {N.}~\bibnamefont {Leroy}}, \bibinfo
  {author} {\bibfnamefont {A.}~\bibnamefont {Lundgren}}, \bibinfo {author}
  {\bibfnamefont {D.}~\bibnamefont {Macleod}},\ and\ \bibinfo {author}
  {\bibfnamefont {J.}~\bibnamefont {McIver}},\ }\bibfield  {title} {\bibinfo
  {title} {Omicron: A tool to characterize transient noise in
  gravitational-wave detectors},\ }\href
  {https://doi.org/https://doi.org/10.1016/j.softx.2020.100620} {\bibfield
  {journal} {\bibinfo  {journal} {SoftwareX}\ }\textbf {\bibinfo {volume}
  {12}},\ \bibinfo {pages} {100620} (\bibinfo {year} {2020})}\BibitemShut
  {NoStop}%
\bibitem [{\citenamefont {{Brady}}\ \emph {et~al.}(2022)\citenamefont
  {{Brady}}, \citenamefont {{Losurdo}},\ and\ \citenamefont
  {{Shinkai}}}]{BLS2022}%
  \BibitemOpen
  \bibfield  {author} {\bibinfo {author} {\bibfnamefont {P.}~\bibnamefont
  {{Brady}}}, \bibinfo {author} {\bibfnamefont {G.}~\bibnamefont {{Losurdo}}},\
  and\ \bibinfo {author} {\bibfnamefont {H.}~\bibnamefont {{Shinkai}}},\
  }\bibfield  {title} {\bibinfo {title} {{LIGO, VIRGO, and KAGRA as the
  International Gravitational Wave Network}},\ }in\ \href
  {https://doi.org/10.1007/978-981-15-4702-7_51-1} {\emph {\bibinfo {booktitle}
  {Handbook of Gravitational Wave Astronomy}}},\ \bibinfo {editor} {edited by\
  \bibinfo {editor} {\bibfnamefont {C.}~\bibnamefont {{Bambi}}}, \bibinfo
  {editor} {\bibfnamefont {S.}~\bibnamefont {{Katsanevas}}},\ and\ \bibinfo
  {editor} {\bibfnamefont {K.~D.}\ \bibnamefont {{Kokkotas}}}}\ (\bibinfo
  {year} {2022})\ p.~\bibinfo {pages} {51}\BibitemShut {NoStop}%
\bibitem [{\citenamefont {{Pines}}\ and\ \citenamefont
  {{Salvino}}(2002)}]{PS2002}%
  \BibitemOpen
  \bibfield  {author} {\bibinfo {author} {\bibfnamefont {D.~J.}\ \bibnamefont
  {{Pines}}}\ and\ \bibinfo {author} {\bibfnamefont {L.~W.}\ \bibnamefont
  {{Salvino}}},\ }\bibfield  {title} {\bibinfo {title} {{Health monitoring of
  one-dimensional structures using empirical mode decomposition and the
  Hilbert-Huang transform}},\ }in\ \href {https://doi.org/10.1117/12.474653}
  {\emph {\bibinfo {booktitle} {Smart Structures and Materials 2002: Smart
  Structures and Integrated Systems}}},\ \bibinfo {series} {Society of
  Photo-Optical Instrumentation Engineers (SPIE) Conference Series}, Vol.\
  \bibinfo {volume} {4701},\ \bibinfo {editor} {edited by\ \bibinfo {editor}
  {\bibfnamefont {L.~P.}\ \bibnamefont {{Davis}}}}\ (\bibinfo {year} {2002})\
  pp.\ \bibinfo {pages} {127--143}\BibitemShut {NoStop}%
\bibitem [{\citenamefont {Jan}\ and\ \citenamefont {Pachori}(2008)}]{JP2009}%
  \BibitemOpen
  \bibfield  {author} {\bibinfo {author} {\bibfnamefont {J.}~\bibnamefont
  {Jan}}\ and\ \bibinfo {author} {\bibfnamefont {R.~B.}\ \bibnamefont
  {Pachori}},\ }\bibfield  {title} {\bibinfo {title} {Discrimination between
  ictal and seizure-free eeg signals using empirical mode decomposition},\
  }\href {https://doi.org/10.1155/2008/293056} {\bibfield  {journal} {\bibinfo
  {journal} {Research Letters in Signal Processing}\ }\textbf {\bibinfo
  {volume} {2008}},\ \bibinfo {pages} {293056} (\bibinfo {year}
  {2008})}\BibitemShut {NoStop}%
\bibitem [{\citenamefont {Huang}\ \emph {et~al.}(2003)\citenamefont {Huang},
  \citenamefont {Wu}, \citenamefont {Qu}, \citenamefont {Long},\ and\
  \citenamefont {Shen}}]{Huang2003}%
  \BibitemOpen
  \bibfield  {author} {\bibinfo {author} {\bibfnamefont {N.~E.}\ \bibnamefont
  {Huang}}, \bibinfo {author} {\bibfnamefont {M.-L.}\ \bibnamefont {Wu}},
  \bibinfo {author} {\bibfnamefont {W.}~\bibnamefont {Qu}}, \bibinfo {author}
  {\bibfnamefont {S.~R.}\ \bibnamefont {Long}},\ and\ \bibinfo {author}
  {\bibfnamefont {S.~S.~P.}\ \bibnamefont {Shen}},\ }\bibfield  {title}
  {\bibinfo {title} {Applications of hilbert–huang transform to
  non-stationary financial time series analysis},\ }\href
  {https://doi.org/https://doi.org/10.1002/asmb.501} {\bibfield  {journal}
  {\bibinfo  {journal} {Applied Stochastic Models in Business and Industry}\
  }\textbf {\bibinfo {volume} {19}},\ \bibinfo {pages} {245} (\bibinfo {year}
  {2003})},\ \Eprint
  {https://arxiv.org/abs/https://onlinelibrary.wiley.com/doi/pdf/10.1002/asmb.501}
  {https://onlinelibrary.wiley.com/doi/pdf/10.1002/asmb.501} \BibitemShut
  {NoStop}%
\bibitem [{\citenamefont {Zhang}\ \emph {et~al.}(2003)\citenamefont {Zhang},
  \citenamefont {Ma}, \citenamefont {Safak},\ and\ \citenamefont
  {Hartzell}}]{ZSH2003}%
  \BibitemOpen
  \bibfield  {author} {\bibinfo {author} {\bibfnamefont {R.~R.}\ \bibnamefont
  {Zhang}}, \bibinfo {author} {\bibfnamefont {S.}~\bibnamefont {Ma}}, \bibinfo
  {author} {\bibfnamefont {E.}~\bibnamefont {Safak}},\ and\ \bibinfo {author}
  {\bibfnamefont {S.}~\bibnamefont {Hartzell}},\ }\bibfield  {title} {\bibinfo
  {title} {Hilbert-huang transform analysis of dynamic and earthquake motion
  recordings},\ }\href
  {https://doi.org/10.1061/(ASCE)0733-9399(2003)129:8(861)} {\bibfield
  {journal} {\bibinfo  {journal} {Journal of Engineering Mechanics}\ }\textbf
  {\bibinfo {volume} {129}},\ \bibinfo {pages} {861} (\bibinfo {year}
  {2003})}\BibitemShut {NoStop}%
\bibitem [{\citenamefont {Li}\ \emph {et~al.}(2016)\citenamefont {Li},
  \citenamefont {Li}, \citenamefont {Wang}, \citenamefont {Feng}, \citenamefont
  {Kong}, \citenamefont {Chen}, \citenamefont {Li},\ and\ \citenamefont
  {Li}}]{Li2016}%
  \BibitemOpen
  \bibfield  {author} {\bibinfo {author} {\bibfnamefont {X.}~\bibnamefont
  {Li}}, \bibinfo {author} {\bibfnamefont {Z.}~\bibnamefont {Li}}, \bibinfo
  {author} {\bibfnamefont {E.}~\bibnamefont {Wang}}, \bibinfo {author}
  {\bibfnamefont {J.}~\bibnamefont {Feng}}, \bibinfo {author} {\bibfnamefont
  {X.}~\bibnamefont {Kong}}, \bibinfo {author} {\bibfnamefont {L.}~\bibnamefont
  {Chen}}, \bibinfo {author} {\bibfnamefont {B.}~\bibnamefont {Li}},\ and\
  \bibinfo {author} {\bibfnamefont {N.}~\bibnamefont {Li}},\ }\bibfield
  {title} {\bibinfo {title} {Analysis of natural mineral earthquake and blast
  based on hilbert–huang transform (hht)},\ }\href
  {https://doi.org/https://doi.org/10.1016/j.jappgeo.2016.03.024} {\bibfield
  {journal} {\bibinfo  {journal} {Journal of Applied Geophysics}\ }\textbf
  {\bibinfo {volume} {128}},\ \bibinfo {pages} {79} (\bibinfo {year}
  {2016})}\BibitemShut {NoStop}%
\bibitem [{\citenamefont {Schlurmann}(2001)}]{Schlurmann2001}%
  \BibitemOpen
  \bibfield  {author} {\bibinfo {author} {\bibfnamefont {T.}~\bibnamefont
  {Schlurmann}},\ }\bibfield  {title} {\bibinfo {title} {{Spectral Analysis of
  Nonlinear Water Waves Based on the Hilbert-Huang Transformation }},\ }\href
  {https://doi.org/10.1115/1.1423911} {\bibfield  {journal} {\bibinfo
  {journal} {Journal of Offshore Mechanics and Arctic Engineering}\ }\textbf
  {\bibinfo {volume} {124}},\ \bibinfo {pages} {22} (\bibinfo {year} {2001})},\
  \Eprint
  {https://arxiv.org/abs/https://asmedigitalcollection.asme.org/offshoremechanics/article-pdf/124/1/22/5719251/22\_1.pdf}
  {https://asmedigitalcollection.asme.org/offshoremechanics/article-pdf/124/1/22/5719251/22\_1.pdf}
  \BibitemShut {NoStop}%
\bibitem [{\citenamefont {Veltcheva}(2002)}]{Veltcheva2002}%
  \BibitemOpen
  \bibfield  {author} {\bibinfo {author} {\bibfnamefont {A.~D.}\ \bibnamefont
  {Veltcheva}},\ }\bibfield  {title} {\bibinfo {title} {Wave and group
  transformation by a hilbert spectrum},\ }\href
  {https://doi.org/10.1142/S057856340200055X} {\bibfield  {journal} {\bibinfo
  {journal} {Coastal Engineering Journal}\ }\textbf {\bibinfo {volume} {44}},\
  \bibinfo {pages} {283} (\bibinfo {year} {2002})},\ \Eprint
  {https://arxiv.org/abs/https://doi.org/10.1142/S057856340200055X}
  {https://doi.org/10.1142/S057856340200055X} \BibitemShut {NoStop}%
\bibitem [{\citenamefont {{Hu}}\ \emph {et~al.}(2014)\citenamefont {{Hu}},
  \citenamefont {{Chou}}, \citenamefont {{Yang}},\ and\ \citenamefont
  {{Su}}}]{Hu2014}%
  \BibitemOpen
  \bibfield  {author} {\bibinfo {author} {\bibfnamefont {C.-P.}\ \bibnamefont
  {{Hu}}}, \bibinfo {author} {\bibfnamefont {Y.}~\bibnamefont {{Chou}}},
  \bibinfo {author} {\bibfnamefont {T.-C.}\ \bibnamefont {{Yang}}},\ and\
  \bibinfo {author} {\bibfnamefont {Y.-H.}\ \bibnamefont {{Su}}},\ }\bibfield
  {title} {\bibinfo {title} {{Tracking the Evolution of Quasi-periodic
  Oscillation in RE J1034+396 Using the Hilbert-Huang Transform}},\ }\href
  {https://doi.org/10.1088/0004-637X/788/1/31} {\bibfield  {journal} {\bibinfo
  {journal} {Astrophys. J.}\ }\textbf {\bibinfo {volume} {788}},\ \bibinfo
  {eid} {31} (\bibinfo {year} {2014})},\ \Eprint
  {https://arxiv.org/abs/1404.5108} {arXiv:1404.5108 [astro-ph.HE]}
  \BibitemShut {NoStop}%
\bibitem [{\citenamefont {{Lin}}\ \emph {et~al.}(2020)\citenamefont {{Lin}},
  \citenamefont {{Hu}}, \citenamefont {{Li}}, \citenamefont {{Takata}},
  \citenamefont {{Yen}}, \citenamefont {{Kwak}}, \citenamefont {{Kim}},\ and\
  \citenamefont {{Kong}}}]{Lin2020}%
  \BibitemOpen
  \bibfield  {author} {\bibinfo {author} {\bibfnamefont {L.~C.-C.}\
  \bibnamefont {{Lin}}}, \bibinfo {author} {\bibfnamefont {C.-P.}\ \bibnamefont
  {{Hu}}}, \bibinfo {author} {\bibfnamefont {K.-L.}\ \bibnamefont {{Li}}},
  \bibinfo {author} {\bibfnamefont {J.}~\bibnamefont {{Takata}}}, \bibinfo
  {author} {\bibfnamefont {D.~C.-C.}\ \bibnamefont {{Yen}}}, \bibinfo {author}
  {\bibfnamefont {K.}~\bibnamefont {{Kwak}}}, \bibinfo {author} {\bibfnamefont
  {Y.-M.}\ \bibnamefont {{Kim}}},\ and\ \bibinfo {author} {\bibfnamefont
  {A.~K.~H.}\ \bibnamefont {{Kong}}},\ }\bibfield  {title} {\bibinfo {title}
  {{Investigation of X-ray timing and spectral properties of ESO 243-49 HLX-1
  with long-term Swift monitoring}},\ }\href
  {https://doi.org/10.1093/mnras/stz3372} {\bibfield  {journal} {\bibinfo
  {journal} {MNRAS}\ }\textbf {\bibinfo {volume} {491}},\ \bibinfo {pages}
  {5682} (\bibinfo {year} {2020})},\ \Eprint {https://arxiv.org/abs/1911.13107}
  {arXiv:1911.13107 [astro-ph.HE]} \BibitemShut {NoStop}%
\bibitem [{\citenamefont {{Camp}}\ \emph {et~al.}(2007)\citenamefont {{Camp}},
  \citenamefont {{Cannizzo}},\ and\ \citenamefont {{Numata}}}]{Camp2007}%
  \BibitemOpen
  \bibfield  {author} {\bibinfo {author} {\bibfnamefont {J.~B.}\ \bibnamefont
  {{Camp}}}, \bibinfo {author} {\bibfnamefont {J.~K.}\ \bibnamefont
  {{Cannizzo}}},\ and\ \bibinfo {author} {\bibfnamefont {K.}~\bibnamefont
  {{Numata}}},\ }\bibfield  {title} {\bibinfo {title} {{Application of the
  Hilbert-Huang transform to the search for gravitational waves}},\ }\href
  {https://doi.org/10.1103/PhysRevD.75.061101} {\bibfield  {journal} {\bibinfo
  {journal} {\prd}\ }\textbf {\bibinfo {volume} {75}},\ \bibinfo {eid} {061101}
  (\bibinfo {year} {2007})},\ \Eprint {https://arxiv.org/abs/gr-qc/0701148}
  {arXiv:gr-qc/0701148 [gr-qc]} \BibitemShut {NoStop}%
\bibitem [{\citenamefont {{Kaneyama}}\ \emph {et~al.}(2016)\citenamefont
  {{Kaneyama}}, \citenamefont {{Oohara}}, \citenamefont {{Takahashi}},
  \citenamefont {{Sekiguchi}}, \citenamefont {{Tagoshi}},\ and\ \citenamefont
  {{Shibata}}}]{Kaneyama2016}%
  \BibitemOpen
  \bibfield  {author} {\bibinfo {author} {\bibfnamefont {M.}~\bibnamefont
  {{Kaneyama}}}, \bibinfo {author} {\bibfnamefont {K.-i.}\ \bibnamefont
  {{Oohara}}}, \bibinfo {author} {\bibfnamefont {H.}~\bibnamefont
  {{Takahashi}}}, \bibinfo {author} {\bibfnamefont {Y.}~\bibnamefont
  {{Sekiguchi}}}, \bibinfo {author} {\bibfnamefont {H.}~\bibnamefont
  {{Tagoshi}}},\ and\ \bibinfo {author} {\bibfnamefont {M.}~\bibnamefont
  {{Shibata}}},\ }\bibfield  {title} {\bibinfo {title} {{Analysis of
  gravitational waves from binary neutron star merger by Hilbert-Huang
  transform}},\ }\href {https://doi.org/10.1103/PhysRevD.93.123010} {\bibfield
  {journal} {\bibinfo  {journal} {\prd}\ }\textbf {\bibinfo {volume} {93}},\
  \bibinfo {eid} {123010} (\bibinfo {year} {2016})}\BibitemShut {NoStop}%
\bibitem [{\citenamefont {Sakai}\ \emph {et~al.}(2017)\citenamefont {Sakai},
  \citenamefont {Oohara}, \citenamefont {Nakano}, \citenamefont {Kaneyama},\
  and\ \citenamefont {Takahashi}}]{Sakai2017}%
  \BibitemOpen
  \bibfield  {author} {\bibinfo {author} {\bibfnamefont {K.}~\bibnamefont
  {Sakai}}, \bibinfo {author} {\bibfnamefont {K.-i.}\ \bibnamefont {Oohara}},
  \bibinfo {author} {\bibfnamefont {H.}~\bibnamefont {Nakano}}, \bibinfo
  {author} {\bibfnamefont {M.}~\bibnamefont {Kaneyama}},\ and\ \bibinfo
  {author} {\bibfnamefont {H.}~\bibnamefont {Takahashi}},\ }\bibfield  {title}
  {\bibinfo {title} {Estimation of starting times of quasinormal modes in
  ringdown gravitational waves with the hilbert-huang transform},\ }\href
  {https://doi.org/10.1103/PhysRevD.96.044047} {\bibfield  {journal} {\bibinfo
  {journal} {Phys. Rev. D}\ }\textbf {\bibinfo {volume} {96}},\ \bibinfo
  {pages} {044047} (\bibinfo {year} {2017})}\BibitemShut {NoStop}%
\bibitem [{\citenamefont {{Akhshi}}\ \emph {et~al.}(2021)\citenamefont
  {{Akhshi}}, \citenamefont {{Alimohammadi}}, \citenamefont {{Baghram}},
  \citenamefont {{Rahvar}}, \citenamefont {{Tabar}},\ and\ \citenamefont
  {{Arfaei}}}]{Akhshi2021}%
  \BibitemOpen
  \bibfield  {author} {\bibinfo {author} {\bibfnamefont {A.}~\bibnamefont
  {{Akhshi}}}, \bibinfo {author} {\bibfnamefont {H.}~\bibnamefont
  {{Alimohammadi}}}, \bibinfo {author} {\bibfnamefont {S.}~\bibnamefont
  {{Baghram}}}, \bibinfo {author} {\bibfnamefont {S.}~\bibnamefont {{Rahvar}}},
  \bibinfo {author} {\bibfnamefont {M.~R.~R.}\ \bibnamefont {{Tabar}}},\ and\
  \bibinfo {author} {\bibfnamefont {H.}~\bibnamefont {{Arfaei}}},\ }\bibfield
  {title} {\bibinfo {title} {{A template-free approach for waveform extraction
  of gravitational wave events}},\ }\href
  {https://doi.org/10.1038/s41598-021-98821-z} {\bibfield  {journal} {\bibinfo
  {journal} {Scientific Reports}\ }\textbf {\bibinfo {volume} {11}},\ \bibinfo
  {eid} {20507} (\bibinfo {year} {2021})},\ \Eprint
  {https://arxiv.org/abs/2005.11352} {arXiv:2005.11352 [astro-ph.IM]}
  \BibitemShut {NoStop}%
\bibitem [{\citenamefont {{Takeda}}\ \emph {et~al.}(2021)\citenamefont
  {{Takeda}}, \citenamefont {{Hiranuma}}, \citenamefont {{Kanda}},
  \citenamefont {{Kotake}}, \citenamefont {{Kuroda}}, \citenamefont
  {{Negishi}}, \citenamefont {{Oohara}}, \citenamefont {{Sakai}}, \citenamefont
  {{Sakai}}, \citenamefont {{Sawada}}, \citenamefont {{Takahashi}},
  \citenamefont {{Tsuchida}}, \citenamefont {{Watanabe}},\ and\ \citenamefont
  {{Yokozawa}}}]{Takeda2021}%
  \BibitemOpen
  \bibfield  {author} {\bibinfo {author} {\bibfnamefont {M.}~\bibnamefont
  {{Takeda}}}, \bibinfo {author} {\bibfnamefont {Y.}~\bibnamefont
  {{Hiranuma}}}, \bibinfo {author} {\bibfnamefont {N.}~\bibnamefont {{Kanda}}},
  \bibinfo {author} {\bibfnamefont {K.}~\bibnamefont {{Kotake}}}, \bibinfo
  {author} {\bibfnamefont {T.}~\bibnamefont {{Kuroda}}}, \bibinfo {author}
  {\bibfnamefont {R.}~\bibnamefont {{Negishi}}}, \bibinfo {author}
  {\bibfnamefont {K.}~\bibnamefont {{Oohara}}}, \bibinfo {author}
  {\bibfnamefont {K.}~\bibnamefont {{Sakai}}}, \bibinfo {author} {\bibfnamefont
  {Y.}~\bibnamefont {{Sakai}}}, \bibinfo {author} {\bibfnamefont
  {T.}~\bibnamefont {{Sawada}}}, \bibinfo {author} {\bibfnamefont
  {H.}~\bibnamefont {{Takahashi}}}, \bibinfo {author} {\bibfnamefont
  {S.}~\bibnamefont {{Tsuchida}}}, \bibinfo {author} {\bibfnamefont
  {Y.}~\bibnamefont {{Watanabe}}},\ and\ \bibinfo {author} {\bibfnamefont
  {T.}~\bibnamefont {{Yokozawa}}},\ }\bibfield  {title} {\bibinfo {title}
  {{Application of the Hilbert-Huang transform for analyzing
  standing-accretion-shock-instability induced gravitational waves in a
  core-collapse supernova}},\ }\href
  {https://doi.org/10.1103/PhysRevD.104.084063} {\bibfield  {journal} {\bibinfo
   {journal} {\prd}\ }\textbf {\bibinfo {volume} {104}},\ \bibinfo {eid}
  {084063} (\bibinfo {year} {2021})},\ \Eprint
  {https://arxiv.org/abs/2107.05213} {arXiv:2107.05213 [astro-ph.HE]}
  \BibitemShut {NoStop}%
\bibitem [{\citenamefont {{Hu}}\ \emph {et~al.}(2022)\citenamefont {{Hu}},
  \citenamefont {{Lin}}, \citenamefont {{Pan}}, \citenamefont {{Li}},
  \citenamefont {{Yen}}, \citenamefont {{Kong}},\ and\ \citenamefont
  {{Hui}}}]{Hu2022}%
  \BibitemOpen
  \bibfield  {author} {\bibinfo {author} {\bibfnamefont {C.-P.}\ \bibnamefont
  {{Hu}}}, \bibinfo {author} {\bibfnamefont {L.~C.-C.}\ \bibnamefont {{Lin}}},
  \bibinfo {author} {\bibfnamefont {K.-C.}\ \bibnamefont {{Pan}}}, \bibinfo
  {author} {\bibfnamefont {K.-L.}\ \bibnamefont {{Li}}}, \bibinfo {author}
  {\bibfnamefont {C.-C.}\ \bibnamefont {{Yen}}}, \bibinfo {author}
  {\bibfnamefont {A.~K.~H.}\ \bibnamefont {{Kong}}},\ and\ \bibinfo {author}
  {\bibfnamefont {C.~Y.}\ \bibnamefont {{Hui}}},\ }\bibfield  {title} {\bibinfo
  {title} {{A Comprehensive Analysis of the Gravitational Wave Events with the
  Stacked Hilbert-Huang Transform: From Compact Binary Coalescence to
  Supernova}},\ }\href {https://doi.org/10.3847/1538-4357/ac8165} {\bibfield
  {journal} {\bibinfo  {journal} {Astrophys. J.}\ }\textbf {\bibinfo {volume}
  {935}},\ \bibinfo {eid} {127} (\bibinfo {year} {2022})},\ \Eprint
  {https://arxiv.org/abs/2207.06714} {arXiv:2207.06714 [astro-ph.HE]}
  \BibitemShut {NoStop}%
\bibitem [{\citenamefont {{Son}}\ \emph {et~al.}(2018)\citenamefont {{Son}},
  \citenamefont {{Kim}}, \citenamefont {{Kim}}, \citenamefont {{McIver}},
  \citenamefont {{Oh}},\ and\ \citenamefont {{Oh}}}]{Son2018}%
  \BibitemOpen
  \bibfield  {author} {\bibinfo {author} {\bibfnamefont {E.~J.}\ \bibnamefont
  {{Son}}}, \bibinfo {author} {\bibfnamefont {W.}~\bibnamefont {{Kim}}},
  \bibinfo {author} {\bibfnamefont {Y.-M.}\ \bibnamefont {{Kim}}}, \bibinfo
  {author} {\bibfnamefont {J.}~\bibnamefont {{McIver}}}, \bibinfo {author}
  {\bibfnamefont {J.~J.}\ \bibnamefont {{Oh}}},\ and\ \bibinfo {author}
  {\bibfnamefont {S.~H.}\ \bibnamefont {{Oh}}},\ }\bibfield  {title} {\bibinfo
  {title} {{Generating Event Triggers Based on Hilbert-Huang Transform and Its
  Application to Gravitational-Wave Data}},\ }\href@noop {} {\bibfield
  {journal} {\bibinfo  {journal} {arXiv e-prints}\ ,\ \bibinfo {eid}
  {arXiv:1810.07555}} (\bibinfo {year} {2018})},\ \Eprint
  {https://arxiv.org/abs/1810.07555} {arXiv:1810.07555 [astro-ph.IM]}
  \BibitemShut {NoStop}%
\bibitem [{\citenamefont {Wu}\ and\ \citenamefont {Huang}(2009)}]{WH2009}%
  \BibitemOpen
  \bibfield  {author} {\bibinfo {author} {\bibfnamefont {Z.}~\bibnamefont
  {Wu}}\ and\ \bibinfo {author} {\bibfnamefont {N.~E.}\ \bibnamefont {Huang}},\
  }\bibfield  {title} {\bibinfo {title} {Ensemble empirical mode decomposition:
  A noise-assisted data analysis method},\ }\href
  {https://doi.org/10.1142/S1793536909000047} {\bibfield  {journal} {\bibinfo
  {journal} {AADA}\ }\textbf {\bibinfo {volume} {1}},\ \bibinfo {pages} {1}
  (\bibinfo {year} {2009})}\BibitemShut {NoStop}%
\bibitem [{\citenamefont {{Huang}}\ \emph {et~al.}(1998)\citenamefont
  {{Huang}}, \citenamefont {{Shen}}, \citenamefont {{Long}}, \citenamefont
  {{Wu}}, \citenamefont {{Shih}}, \citenamefont {{Zheng}}, \citenamefont
  {{Yen}}, \citenamefont {{Tung}},\ and\ \citenamefont {{Liu}}}]{Huang98}%
  \BibitemOpen
  \bibfield  {author} {\bibinfo {author} {\bibfnamefont {N.~E.}\ \bibnamefont
  {{Huang}}}, \bibinfo {author} {\bibfnamefont {Z.}~\bibnamefont {{Shen}}},
  \bibinfo {author} {\bibfnamefont {S.~R.}\ \bibnamefont {{Long}}}, \bibinfo
  {author} {\bibfnamefont {M.~C.}\ \bibnamefont {{Wu}}}, \bibinfo {author}
  {\bibfnamefont {H.~H.}\ \bibnamefont {{Shih}}}, \bibinfo {author}
  {\bibfnamefont {Q.}~\bibnamefont {{Zheng}}}, \bibinfo {author} {\bibfnamefont
  {N.-C.}\ \bibnamefont {{Yen}}}, \bibinfo {author} {\bibfnamefont {C.~C.}\
  \bibnamefont {{Tung}}},\ and\ \bibinfo {author} {\bibfnamefont {H.~H.}\
  \bibnamefont {{Liu}}},\ }\bibfield  {title} {\bibinfo {title} {{The empirical
  mode decomposition and the Hilbert spectrum for nonlinear and non-stationary
  time series analysis}},\ }\href {https://doi.org/10.1098/rspa.1998.0193}
  {\bibfield  {journal} {\bibinfo  {journal} {Royal Society of London
  Proceedings Series A}\ }\textbf {\bibinfo {volume} {454}},\ \bibinfo {pages}
  {903} (\bibinfo {year} {1998})}\BibitemShut {NoStop}%
\bibitem [{\citenamefont {{Huang}}\ and\ \citenamefont
  {{Wu}}(2008)}]{Huang2008}%
  \BibitemOpen
  \bibfield  {author} {\bibinfo {author} {\bibfnamefont {N.~E.}\ \bibnamefont
  {{Huang}}}\ and\ \bibinfo {author} {\bibfnamefont {Z.}~\bibnamefont {{Wu}}},\
  }\bibfield  {title} {\bibinfo {title} {{A review on Hilbert-Huang transform:
  Method and its applications to geophysical studies}},\ }\href
  {https://doi.org/10.1029/2007RG000228} {\bibfield  {journal} {\bibinfo
  {journal} {Reviews of Geophysics}\ }\textbf {\bibinfo {volume} {46}},\
  \bibinfo {eid} {RG2006} (\bibinfo {year} {2008})}\BibitemShut {NoStop}%
\bibitem [{\citenamefont {Huang}\ and\ \citenamefont {Shen}(2014)}]{Huang2014}%
  \BibitemOpen
  \bibfield  {author} {\bibinfo {author} {\bibfnamefont {N.}~\bibnamefont
  {Huang}}\ and\ \bibinfo {author} {\bibfnamefont {S.}~\bibnamefont {Shen}},\
  }\href {https://books.google.com.tw/books?id=aJ26CgAAQBAJ} {\emph {\bibinfo
  {title} {Hilbert-Huang Transform and Its Applications}}},\ EBSCO ebook
  academic collection\ (\bibinfo  {publisher} {World Scientific},\ \bibinfo
  {year} {2014})\BibitemShut {NoStop}%
\bibitem [{\citenamefont {{Mohapatra}}\ \emph {et~al.}(2012)\citenamefont
  {{Mohapatra}}, \citenamefont {{Nemtzow}}, \citenamefont
  {{Chassande-Mottin}},\ and\ \citenamefont {{Cadonati}}}]{Mohapatra2012}%
  \BibitemOpen
  \bibfield  {author} {\bibinfo {author} {\bibfnamefont {S.}~\bibnamefont
  {{Mohapatra}}}, \bibinfo {author} {\bibfnamefont {Z.}~\bibnamefont
  {{Nemtzow}}}, \bibinfo {author} {\bibfnamefont {{\'E}.}~\bibnamefont
  {{Chassande-Mottin}}},\ and\ \bibinfo {author} {\bibfnamefont
  {L.}~\bibnamefont {{Cadonati}}},\ }\bibfield  {title} {\bibinfo {title}
  {{Performance of a Chirplet-based analysis for gravitational-waves from
  binary black-hole mergers}},\ }in\ \href
  {https://doi.org/10.1088/1742-6596/363/1/012031} {\emph {\bibinfo {booktitle}
  {Journal of Physics Conference Series}}},\ \bibinfo {series} {Journal of
  Physics Conference Series}, Vol.\ \bibinfo {volume} {363}\ (\bibinfo {year}
  {2012})\ p.\ \bibinfo {pages} {012031},\ \Eprint
  {https://arxiv.org/abs/1111.3621} {arXiv:1111.3621 [gr-qc]} \BibitemShut
  {NoStop}%
\bibitem [{\citenamefont {Bernardino}\ and\ \citenamefont
  {Santos-Victor}(2005)}]{bernardino2005}%
  \BibitemOpen
  \bibfield  {author} {\bibinfo {author} {\bibfnamefont {A.}~\bibnamefont
  {Bernardino}}\ and\ \bibinfo {author} {\bibfnamefont {J.}~\bibnamefont
  {Santos-Victor}},\ }\bibfield  {title} {\bibinfo {title} {A real-time gabor
  primal sketch for visual attention},\ }in\ \href@noop {} {\emph {\bibinfo
  {booktitle} {Iberian Conference on Pattern Recognition and Image Analysis}}}\
  (\bibinfo {organization} {Springer},\ \bibinfo {year} {2005})\ pp.\ \bibinfo
  {pages} {335--342}\BibitemShut {NoStop}%
\bibitem [{Note1()}]{Note1}%
  \BibitemOpen
  \bibinfo {note} {Https://pycbc.org/}\BibitemShut {NoStop}%
\bibitem [{\citenamefont {{Biwer}}\ \emph {et~al.}(2019)\citenamefont
  {{Biwer}}, \citenamefont {{Capano}}, \citenamefont {{De}}, \citenamefont
  {{Cabero}}, \citenamefont {{Brown}}, \citenamefont {{Nitz}},\ and\
  \citenamefont {{Raymond}}}]{Biwer2019}%
  \BibitemOpen
  \bibfield  {author} {\bibinfo {author} {\bibfnamefont {C.~M.}\ \bibnamefont
  {{Biwer}}}, \bibinfo {author} {\bibfnamefont {C.~D.}\ \bibnamefont
  {{Capano}}}, \bibinfo {author} {\bibfnamefont {S.}~\bibnamefont {{De}}},
  \bibinfo {author} {\bibfnamefont {M.}~\bibnamefont {{Cabero}}}, \bibinfo
  {author} {\bibfnamefont {D.~A.}\ \bibnamefont {{Brown}}}, \bibinfo {author}
  {\bibfnamefont {A.~H.}\ \bibnamefont {{Nitz}}},\ and\ \bibinfo {author}
  {\bibfnamefont {V.}~\bibnamefont {{Raymond}}},\ }\bibfield  {title} {\bibinfo
  {title} {{PyCBC Inference: A Python-based Parameter Estimation Toolkit for
  Compact Binary Coalescence Signal}},\ }\href
  {https://doi.org/10.1088/1538-3873/aaef0b} {\bibfield  {journal} {\bibinfo
  {journal} {\pasp}\ }\textbf {\bibinfo {volume} {131}},\ \bibinfo {pages}
  {024503} (\bibinfo {year} {2019})},\ \Eprint
  {https://arxiv.org/abs/1807.10312} {arXiv:1807.10312 [astro-ph.IM]}
  \BibitemShut {NoStop}%
\bibitem [{\citenamefont {{Hu}}\ \emph {et~al.}(2011)\citenamefont {{Hu}},
  \citenamefont {{Chou}}, \citenamefont {{Wu}}, \citenamefont {{Yang}},\ and\
  \citenamefont {{Su}}}]{Hu2011}%
  \BibitemOpen
  \bibfield  {author} {\bibinfo {author} {\bibfnamefont {C.-P.}\ \bibnamefont
  {{Hu}}}, \bibinfo {author} {\bibfnamefont {Y.}~\bibnamefont {{Chou}}},
  \bibinfo {author} {\bibfnamefont {M.-C.}\ \bibnamefont {{Wu}}}, \bibinfo
  {author} {\bibfnamefont {T.-C.}\ \bibnamefont {{Yang}}},\ and\ \bibinfo
  {author} {\bibfnamefont {Y.-H.}\ \bibnamefont {{Su}}},\ }\bibfield  {title}
  {\bibinfo {title} {{Time-frequency Analysis of the Superorbital Modulation of
  the X-Ray Binary SMC X-1 Using the Hilbert-Huang Transform}},\ }\href
  {https://doi.org/10.1088/0004-637X/740/2/67} {\bibfield  {journal} {\bibinfo
  {journal} {Astrophys. J.}\ }\textbf {\bibinfo {volume} {740}},\ \bibinfo
  {eid} {67} (\bibinfo {year} {2011})},\ \Eprint
  {https://arxiv.org/abs/1107.5143} {arXiv:1107.5143 [astro-ph.HE]}
  \BibitemShut {NoStop}%
\bibitem [{\citenamefont {Cannon}\ \emph {et~al.}(2012)\citenamefont {Cannon},
  \citenamefont {Cariou}, \citenamefont {Chapman}, \citenamefont
  {Crispin-Ortuzar}, \citenamefont {Fotopoulos}, \citenamefont {Frei},
  \citenamefont {Hanna}, \citenamefont {Kara}, \citenamefont {Keppel},
  \citenamefont {Liao}, \citenamefont {Privitera}, \citenamefont {Searle},
  \citenamefont {Singer},\ and\ \citenamefont {Weinstein}}]{2012Cannon}%
  \BibitemOpen
  \bibfield  {author} {\bibinfo {author} {\bibfnamefont {K.}~\bibnamefont
  {Cannon}}, \bibinfo {author} {\bibfnamefont {R.}~\bibnamefont {Cariou}},
  \bibinfo {author} {\bibfnamefont {A.}~\bibnamefont {Chapman}}, \bibinfo
  {author} {\bibfnamefont {M.}~\bibnamefont {Crispin-Ortuzar}}, \bibinfo
  {author} {\bibfnamefont {N.}~\bibnamefont {Fotopoulos}}, \bibinfo {author}
  {\bibfnamefont {M.}~\bibnamefont {Frei}}, \bibinfo {author} {\bibfnamefont
  {C.}~\bibnamefont {Hanna}}, \bibinfo {author} {\bibfnamefont
  {E.}~\bibnamefont {Kara}}, \bibinfo {author} {\bibfnamefont {D.}~\bibnamefont
  {Keppel}}, \bibinfo {author} {\bibfnamefont {L.}~\bibnamefont {Liao}},
  \bibinfo {author} {\bibfnamefont {S.}~\bibnamefont {Privitera}}, \bibinfo
  {author} {\bibfnamefont {A.}~\bibnamefont {Searle}}, \bibinfo {author}
  {\bibfnamefont {L.}~\bibnamefont {Singer}},\ and\ \bibinfo {author}
  {\bibfnamefont {A.}~\bibnamefont {Weinstein}},\ }\bibfield  {title} {\bibinfo
  {title} {Toward early-warning detection of gravitational waves from compact
  binary coalescence},\ }\href {https://doi.org/10.1088/0004-637X/748/2/136}
  {\bibfield  {journal} {\bibinfo  {journal} {The Astrophysical Journal}\
  }\textbf {\bibinfo {volume} {748}},\ \bibinfo {pages} {136} (\bibinfo {year}
  {2012})}\BibitemShut {NoStop}%
\bibitem [{\citenamefont {{Krimm}}\ \emph {et~al.}(2013)\citenamefont
  {{Krimm}}, \citenamefont {{Holland}}, \citenamefont {{Corbet}}, \citenamefont
  {{Pearlman}}, \citenamefont {{Romano}}, \citenamefont {{Kennea}},
  \citenamefont {{Bloom}}, \citenamefont {{Barthelmy}}, \citenamefont
  {{Baumgartner}}, \citenamefont {{Cummings}}, \citenamefont {{Gehrels}},
  \citenamefont {{Lien}}, \citenamefont {{Markwardt}}, \citenamefont
  {{Palmer}}, \citenamefont {{Sakamoto}}, \citenamefont {{Stamatikos}},\ and\
  \citenamefont {{Ukwatta}}}]{KrimmHC2013}%
  \BibitemOpen
  \bibfield  {author} {\bibinfo {author} {\bibfnamefont {H.~A.}\ \bibnamefont
  {{Krimm}}}, \bibinfo {author} {\bibfnamefont {S.~T.}\ \bibnamefont
  {{Holland}}}, \bibinfo {author} {\bibfnamefont {R.~H.~D.}\ \bibnamefont
  {{Corbet}}}, \bibinfo {author} {\bibfnamefont {A.~B.}\ \bibnamefont
  {{Pearlman}}}, \bibinfo {author} {\bibfnamefont {P.}~\bibnamefont
  {{Romano}}}, \bibinfo {author} {\bibfnamefont {J.~A.}\ \bibnamefont
  {{Kennea}}}, \bibinfo {author} {\bibfnamefont {J.~S.}\ \bibnamefont
  {{Bloom}}}, \bibinfo {author} {\bibfnamefont {S.~D.}\ \bibnamefont
  {{Barthelmy}}}, \bibinfo {author} {\bibfnamefont {W.~H.}\ \bibnamefont
  {{Baumgartner}}}, \bibinfo {author} {\bibfnamefont {J.~R.}\ \bibnamefont
  {{Cummings}}}, \bibinfo {author} {\bibfnamefont {N.}~\bibnamefont
  {{Gehrels}}}, \bibinfo {author} {\bibfnamefont {A.~Y.}\ \bibnamefont
  {{Lien}}}, \bibinfo {author} {\bibfnamefont {C.~B.}\ \bibnamefont
  {{Markwardt}}}, \bibinfo {author} {\bibfnamefont {D.~M.}\ \bibnamefont
  {{Palmer}}}, \bibinfo {author} {\bibfnamefont {T.}~\bibnamefont
  {{Sakamoto}}}, \bibinfo {author} {\bibfnamefont {M.}~\bibnamefont
  {{Stamatikos}}},\ and\ \bibinfo {author} {\bibfnamefont {T.~N.}\ \bibnamefont
  {{Ukwatta}}},\ }\bibfield  {title} {\bibinfo {title} {{The Swift/BAT Hard
  X-Ray Transient Monitor}},\ }\href
  {https://doi.org/10.1088/0067-0049/209/1/14} {\bibfield  {journal} {\bibinfo
  {journal} {Astrophys. J. Suppl.}\ }\textbf {\bibinfo {volume} {209}},\
  \bibinfo {eid} {14} (\bibinfo {year} {2013})},\ \Eprint
  {https://arxiv.org/abs/1309.0755} {arXiv:1309.0755 [astro-ph.HE]}
  \BibitemShut {NoStop}%
\bibitem [{\citenamefont {{Abbott}}\ \emph
  {et~al.}(2023{\natexlab{a}})\citenamefont {{Abbott}}, \citenamefont {{Abe}},
  \citenamefont {{Acernese}}, \citenamefont {{et al.}}, \citenamefont {{LIGO
  Scientific Collaboration}},\ and\ \citenamefont {{Virgo
  Collaboration}}}]{O3data}%
  \BibitemOpen
  \bibfield  {author} {\bibinfo {author} {\bibfnamefont {R.}~\bibnamefont
  {{Abbott}}}, \bibinfo {author} {\bibfnamefont {H.}~\bibnamefont {{Abe}}},
  \bibinfo {author} {\bibfnamefont {F.}~\bibnamefont {{Acernese}}}, \bibinfo
  {author} {\bibnamefont {{et al.}}}, \bibinfo {author} {\bibnamefont {{LIGO
  Scientific Collaboration}}},\ and\ \bibinfo {author} {\bibnamefont {{Virgo
  Collaboration}}},\ }\bibfield  {title} {\bibinfo {title} {{Open Data from the
  Third Observing Run of LIGO, Virgo, KAGRA, and GEO}},\ }\href
  {https://doi.org/10.3847/1538-4365/acdc9f} {\bibfield  {journal} {\bibinfo
  {journal} {Astrophys. J. Suppl.}\ }\textbf {\bibinfo {volume} {267}},\
  \bibinfo {eid} {29} (\bibinfo {year} {2023}{\natexlab{a}})},\ \Eprint
  {https://arxiv.org/abs/2302.03676} {arXiv:2302.03676 [gr-qc]} \BibitemShut
  {NoStop}%
\bibitem [{Note2()}]{Note2}%
  \BibitemOpen
  \bibinfo {note} {Https://gwosc.org/eventapi/html/}\BibitemShut {NoStop}%
\bibitem [{\citenamefont {{Abbott}}\ \emph {et~al.}(2016)\citenamefont
  {{Abbott}}, \citenamefont {{Abbott}}, \citenamefont {{Abbott}}, \citenamefont
  {{Abernathy}}, \citenamefont {{Acernese}}, \citenamefont {{et al.}},
  \citenamefont {{LIGO Scientific Collaboration}},\ and\ \citenamefont {{Virgo
  Collaboration}}}]{2016LV}%
  \BibitemOpen
  \bibfield  {author} {\bibinfo {author} {\bibfnamefont {B.~P.}\ \bibnamefont
  {{Abbott}}}, \bibinfo {author} {\bibfnamefont {R.}~\bibnamefont {{Abbott}}},
  \bibinfo {author} {\bibfnamefont {T.~D.}\ \bibnamefont {{Abbott}}}, \bibinfo
  {author} {\bibfnamefont {M.~R.}\ \bibnamefont {{Abernathy}}}, \bibinfo
  {author} {\bibfnamefont {F.}~\bibnamefont {{Acernese}}}, \bibinfo {author}
  {\bibnamefont {{et al.}}}, \bibinfo {author} {\bibnamefont {{LIGO Scientific
  Collaboration}}},\ and\ \bibinfo {author} {\bibnamefont {{Virgo
  Collaboration}}},\ }\bibfield  {title} {\bibinfo {title} {{Observation of
  Gravitational Waves from a Binary Black Hole Merger}},\ }\href
  {https://doi.org/10.1103/PhysRevLett.116.061102} {\bibfield  {journal}
  {\bibinfo  {journal} {\prl}\ }\textbf {\bibinfo {volume} {116}},\ \bibinfo
  {eid} {061102} (\bibinfo {year} {2016})},\ \Eprint
  {https://arxiv.org/abs/1602.03837} {arXiv:1602.03837 [gr-qc]} \BibitemShut
  {NoStop}%
\bibitem [{\citenamefont {{Abbott}}\ \emph
  {et~al.}(2017{\natexlab{a}})\citenamefont {{Abbott}}, \citenamefont
  {{Abbott}}, \citenamefont {{Abbott}}, \citenamefont {{Acernese}},
  \citenamefont {{Ackley}}, \citenamefont {{et al.}}, \citenamefont {{LIGO
  Scientific Collaboration}},\ and\ \citenamefont {{Virgo
  Collaboration}}}]{2017LV}%
  \BibitemOpen
  \bibfield  {author} {\bibinfo {author} {\bibfnamefont {B.~P.}\ \bibnamefont
  {{Abbott}}}, \bibinfo {author} {\bibfnamefont {R.}~\bibnamefont {{Abbott}}},
  \bibinfo {author} {\bibfnamefont {T.~D.}\ \bibnamefont {{Abbott}}}, \bibinfo
  {author} {\bibfnamefont {F.}~\bibnamefont {{Acernese}}}, \bibinfo {author}
  {\bibfnamefont {K.}~\bibnamefont {{Ackley}}}, \bibinfo {author} {\bibnamefont
  {{et al.}}}, \bibinfo {author} {\bibnamefont {{LIGO Scientific
  Collaboration}}},\ and\ \bibinfo {author} {\bibnamefont {{Virgo
  Collaboration}}},\ }\href {https://doi.org/10.1103/PhysRevLett.119.161101}
  {\bibfield  {journal} {\bibinfo  {journal} {\prl}\ }\textbf {\bibinfo
  {volume} {119}},\ \bibinfo {eid} {161101} (\bibinfo {year}
  {2017}{\natexlab{a}})},\ \Eprint {https://arxiv.org/abs/1710.05832}
  {arXiv:1710.05832 [gr-qc]} \BibitemShut {NoStop}%
\bibitem [{Note3()}]{Note3}%
  \BibitemOpen
  \bibinfo {note} {Https://gwpy.github.io/docs/stable/}\BibitemShut {NoStop}%
\bibitem [{\citenamefont {{Macleod}}\ \emph {et~al.}(2021)\citenamefont
  {{Macleod}}, \citenamefont {{Areeda}}, \citenamefont {{Coughlin}},
  \citenamefont {{Massinger}},\ and\ \citenamefont {{Urban}}}]{gwpy}%
  \BibitemOpen
  \bibfield  {author} {\bibinfo {author} {\bibfnamefont {D.~M.}\ \bibnamefont
  {{Macleod}}}, \bibinfo {author} {\bibfnamefont {J.~S.}\ \bibnamefont
  {{Areeda}}}, \bibinfo {author} {\bibfnamefont {S.~B.}\ \bibnamefont
  {{Coughlin}}}, \bibinfo {author} {\bibfnamefont {T.~J.}\ \bibnamefont
  {{Massinger}}},\ and\ \bibinfo {author} {\bibfnamefont {A.~L.}\ \bibnamefont
  {{Urban}}},\ }\bibfield  {title} {\bibinfo {title} {{GWpy: A Python package
  for gravitational-wave astrophysics}},\ }\href
  {https://doi.org/10.1016/j.softx.2021.100657} {\bibfield  {journal} {\bibinfo
   {journal} {SoftwareX}\ }\textbf {\bibinfo {volume} {13}},\ \bibinfo {pages}
  {100657} (\bibinfo {year} {2021})}\BibitemShut {NoStop}%
\bibitem [{\citenamefont {{Abbott}}\ \emph {et~al.}(2020)\citenamefont
  {{Abbott}}, \citenamefont {{Abbott}}, \citenamefont {{Abraham}},
  \citenamefont {{Acernese}}, \citenamefont {{Ackley}}, \citenamefont {{et
  al.}}, \citenamefont {{LIGO Scientific Collaboration}},\ and\ \citenamefont
  {{Virgo Collaboration}}}]{GW190814}%
  \BibitemOpen
  \bibfield  {author} {\bibinfo {author} {\bibfnamefont {R.}~\bibnamefont
  {{Abbott}}}, \bibinfo {author} {\bibfnamefont {T.~D.}\ \bibnamefont
  {{Abbott}}}, \bibinfo {author} {\bibfnamefont {S.}~\bibnamefont {{Abraham}}},
  \bibinfo {author} {\bibfnamefont {F.}~\bibnamefont {{Acernese}}}, \bibinfo
  {author} {\bibfnamefont {K.}~\bibnamefont {{Ackley}}}, \bibinfo {author}
  {\bibnamefont {{et al.}}}, \bibinfo {author} {\bibnamefont {{LIGO Scientific
  Collaboration}}},\ and\ \bibinfo {author} {\bibnamefont {{Virgo
  Collaboration}}},\ }\bibfield  {title} {\bibinfo {title} {{GW190814:
  Gravitational Waves from the Coalescence of a 23 Solar Mass Black Hole with a
  2.6 Solar Mass Compact Object}},\ }\href
  {https://doi.org/10.3847/2041-8213/ab960f} {\bibfield  {journal} {\bibinfo
  {journal} {Astrophys. J. Lett.}\ }\textbf {\bibinfo {volume} {896}},\
  \bibinfo {eid} {L44} (\bibinfo {year} {2020})},\ \Eprint
  {https://arxiv.org/abs/2006.12611} {arXiv:2006.12611 [astro-ph.HE]}
  \BibitemShut {NoStop}%
\bibitem [{\citenamefont {{Pratten}}\ \emph {et~al.}(2021)\citenamefont
  {{Pratten}}, \citenamefont {{Garc{\'\i}a-Quir{\'o}s}}, \citenamefont
  {{Colleoni}}, \citenamefont {{Ramos-Buades}}, \citenamefont {{Estell{\'e}s}},
  \citenamefont {{Mateu-Lucena}}, \citenamefont {{Jaume}}, \citenamefont
  {{Haney}}, \citenamefont {{Keitel}}, \citenamefont {{Thompson}},\ and\
  \citenamefont {{Husa}}}]{Pratten2021}%
  \BibitemOpen
  \bibfield  {author} {\bibinfo {author} {\bibfnamefont {G.}~\bibnamefont
  {{Pratten}}}, \bibinfo {author} {\bibfnamefont {C.}~\bibnamefont
  {{Garc{\'\i}a-Quir{\'o}s}}}, \bibinfo {author} {\bibfnamefont
  {M.}~\bibnamefont {{Colleoni}}}, \bibinfo {author} {\bibfnamefont
  {A.}~\bibnamefont {{Ramos-Buades}}}, \bibinfo {author} {\bibfnamefont
  {H.}~\bibnamefont {{Estell{\'e}s}}}, \bibinfo {author} {\bibfnamefont
  {M.}~\bibnamefont {{Mateu-Lucena}}}, \bibinfo {author} {\bibfnamefont
  {R.}~\bibnamefont {{Jaume}}}, \bibinfo {author} {\bibfnamefont
  {M.}~\bibnamefont {{Haney}}}, \bibinfo {author} {\bibfnamefont
  {D.}~\bibnamefont {{Keitel}}}, \bibinfo {author} {\bibfnamefont {J.~E.}\
  \bibnamefont {{Thompson}}},\ and\ \bibinfo {author} {\bibfnamefont
  {S.}~\bibnamefont {{Husa}}},\ }\bibfield  {title} {\bibinfo {title}
  {{Computationally efficient models for the dominant and subdominant harmonic
  modes of precessing binary black holes}},\ }\href
  {https://doi.org/10.1103/PhysRevD.103.104056} {\bibfield  {journal} {\bibinfo
   {journal} {\prd}\ }\textbf {\bibinfo {volume} {103}},\ \bibinfo {eid}
  {104056} (\bibinfo {year} {2021})},\ \Eprint
  {https://arxiv.org/abs/2004.06503} {arXiv:2004.06503 [gr-qc]} \BibitemShut
  {NoStop}%
\bibitem [{\citenamefont {{Abbott}}\ \emph
  {et~al.}(2023{\natexlab{b}})\citenamefont {{Abbott}}, \citenamefont
  {{Abbott}}, \citenamefont {{Acernese}}, \citenamefont {{Ackley}},
  \citenamefont {{Adams}}, \citenamefont {{et al.}}, \citenamefont {{LIGO
  Scientific Collaboration}}, \citenamefont {{VIRGO Collaboration}},\ and\
  \citenamefont {{Kagra Collaboration}}}]{GWTC-3}%
  \BibitemOpen
  \bibfield  {author} {\bibinfo {author} {\bibfnamefont {R.}~\bibnamefont
  {{Abbott}}}, \bibinfo {author} {\bibfnamefont {T.~D.}\ \bibnamefont
  {{Abbott}}}, \bibinfo {author} {\bibfnamefont {F.}~\bibnamefont
  {{Acernese}}}, \bibinfo {author} {\bibfnamefont {K.}~\bibnamefont
  {{Ackley}}}, \bibinfo {author} {\bibfnamefont {C.}~\bibnamefont {{Adams}}},
  \bibinfo {author} {\bibnamefont {{et al.}}}, \bibinfo {author} {\bibnamefont
  {{LIGO Scientific Collaboration}}}, \bibinfo {author} {\bibnamefont {{VIRGO
  Collaboration}}},\ and\ \bibinfo {author} {\bibnamefont {{Kagra
  Collaboration}}},\ }\bibfield  {title} {\bibinfo {title} {{GWTC-3: Compact
  Binary Coalescences Observed by LIGO and Virgo during the Second Part of the
  Third Observing Run}},\ }\href {https://doi.org/10.1103/PhysRevX.13.041039}
  {\bibfield  {journal} {\bibinfo  {journal} {Physical Review X}\ }\textbf
  {\bibinfo {volume} {13}},\ \bibinfo {eid} {041039} (\bibinfo {year}
  {2023}{\natexlab{b}})},\ \Eprint {https://arxiv.org/abs/2111.03606}
  {arXiv:2111.03606 [gr-qc]} \BibitemShut {NoStop}%
\bibitem [{\citenamefont {{Ossokine}}\ \emph {et~al.}(2020)\citenamefont
  {{Ossokine}}, \citenamefont {{Buonanno}}, \citenamefont {{Marsat}},
  \citenamefont {{Cotesta}}, \citenamefont {{Babak}}, \citenamefont
  {{Dietrich}}, \citenamefont {{Haas}}, \citenamefont {{Hinder}}, \citenamefont
  {{Pfeiffer}}, \citenamefont {{P{\"u}rrer}}, \citenamefont {{Woodford}},
  \citenamefont {{Boyle}}, \citenamefont {{Kidder}}, \citenamefont {{Scheel}},\
  and\ \citenamefont {{Szil{\'a}gyi}}}]{Ossokine2020}%
  \BibitemOpen
  \bibfield  {author} {\bibinfo {author} {\bibfnamefont {S.}~\bibnamefont
  {{Ossokine}}}, \bibinfo {author} {\bibfnamefont {A.}~\bibnamefont
  {{Buonanno}}}, \bibinfo {author} {\bibfnamefont {S.}~\bibnamefont
  {{Marsat}}}, \bibinfo {author} {\bibfnamefont {R.}~\bibnamefont {{Cotesta}}},
  \bibinfo {author} {\bibfnamefont {S.}~\bibnamefont {{Babak}}}, \bibinfo
  {author} {\bibfnamefont {T.}~\bibnamefont {{Dietrich}}}, \bibinfo {author}
  {\bibfnamefont {R.}~\bibnamefont {{Haas}}}, \bibinfo {author} {\bibfnamefont
  {I.}~\bibnamefont {{Hinder}}}, \bibinfo {author} {\bibfnamefont {H.~P.}\
  \bibnamefont {{Pfeiffer}}}, \bibinfo {author} {\bibfnamefont
  {M.}~\bibnamefont {{P{\"u}rrer}}}, \bibinfo {author} {\bibfnamefont {C.~J.}\
  \bibnamefont {{Woodford}}}, \bibinfo {author} {\bibfnamefont
  {M.}~\bibnamefont {{Boyle}}}, \bibinfo {author} {\bibfnamefont {L.~E.}\
  \bibnamefont {{Kidder}}}, \bibinfo {author} {\bibfnamefont {M.~A.}\
  \bibnamefont {{Scheel}}},\ and\ \bibinfo {author} {\bibfnamefont
  {B.}~\bibnamefont {{Szil{\'a}gyi}}},\ }\bibfield  {title} {\bibinfo {title}
  {{Multipolar effective-one-body waveforms for precessing binary black holes:
  Construction and validation}},\ }\href
  {https://doi.org/10.1103/PhysRevD.102.044055} {\bibfield  {journal} {\bibinfo
   {journal} {\prd}\ }\textbf {\bibinfo {volume} {102}},\ \bibinfo {eid}
  {044055} (\bibinfo {year} {2020})},\ \Eprint
  {https://arxiv.org/abs/2004.09442} {arXiv:2004.09442 [gr-qc]} \BibitemShut
  {NoStop}%
\bibitem [{\citenamefont {{Messick}}\ \emph {et~al.}(2017)\citenamefont
  {{Messick}}, \citenamefont {{Blackburn}}, \citenamefont {{Brady}},
  \citenamefont {{Brockill}}, \citenamefont {{Cannon}}, \citenamefont
  {{Cariou}}, \citenamefont {{Caudill}}, \citenamefont {{Chamberlin}},
  \citenamefont {{Creighton}}, \citenamefont {{Everett}}, \citenamefont
  {{Hanna}}, \citenamefont {{Keppel}}, \citenamefont {{Lang}}, \citenamefont
  {{Li}}, \citenamefont {{Meacher}}, \citenamefont {{Nielsen}}, \citenamefont
  {{Pankow}}, \citenamefont {{Privitera}}, \citenamefont {{Qi}}, \citenamefont
  {{Sachdev}}, \citenamefont {{Sadeghian}}, \citenamefont {{Singer}},
  \citenamefont {{Thomas}}, \citenamefont {{Wade}}, \citenamefont {{Wade}},
  \citenamefont {{Weinstein}},\ and\ \citenamefont {{Wiesner}}}]{Messick2017}%
  \BibitemOpen
  \bibfield  {author} {\bibinfo {author} {\bibfnamefont {C.}~\bibnamefont
  {{Messick}}}, \bibinfo {author} {\bibfnamefont {K.}~\bibnamefont
  {{Blackburn}}}, \bibinfo {author} {\bibfnamefont {P.}~\bibnamefont
  {{Brady}}}, \bibinfo {author} {\bibfnamefont {P.}~\bibnamefont {{Brockill}}},
  \bibinfo {author} {\bibfnamefont {K.}~\bibnamefont {{Cannon}}}, \bibinfo
  {author} {\bibfnamefont {R.}~\bibnamefont {{Cariou}}}, \bibinfo {author}
  {\bibfnamefont {S.}~\bibnamefont {{Caudill}}}, \bibinfo {author}
  {\bibfnamefont {S.~J.}\ \bibnamefont {{Chamberlin}}}, \bibinfo {author}
  {\bibfnamefont {J.~D.~E.}\ \bibnamefont {{Creighton}}}, \bibinfo {author}
  {\bibfnamefont {R.}~\bibnamefont {{Everett}}}, \bibinfo {author}
  {\bibfnamefont {C.}~\bibnamefont {{Hanna}}}, \bibinfo {author} {\bibfnamefont
  {D.}~\bibnamefont {{Keppel}}}, \bibinfo {author} {\bibfnamefont {R.~N.}\
  \bibnamefont {{Lang}}}, \bibinfo {author} {\bibfnamefont {T.~G.~F.}\
  \bibnamefont {{Li}}}, \bibinfo {author} {\bibfnamefont {D.}~\bibnamefont
  {{Meacher}}}, \bibinfo {author} {\bibfnamefont {A.}~\bibnamefont
  {{Nielsen}}}, \bibinfo {author} {\bibfnamefont {C.}~\bibnamefont {{Pankow}}},
  \bibinfo {author} {\bibfnamefont {S.}~\bibnamefont {{Privitera}}}, \bibinfo
  {author} {\bibfnamefont {H.}~\bibnamefont {{Qi}}}, \bibinfo {author}
  {\bibfnamefont {S.}~\bibnamefont {{Sachdev}}}, \bibinfo {author}
  {\bibfnamefont {L.}~\bibnamefont {{Sadeghian}}}, \bibinfo {author}
  {\bibfnamefont {L.}~\bibnamefont {{Singer}}}, \bibinfo {author}
  {\bibfnamefont {E.~G.}\ \bibnamefont {{Thomas}}}, \bibinfo {author}
  {\bibfnamefont {L.}~\bibnamefont {{Wade}}}, \bibinfo {author} {\bibfnamefont
  {M.}~\bibnamefont {{Wade}}}, \bibinfo {author} {\bibfnamefont
  {A.}~\bibnamefont {{Weinstein}}},\ and\ \bibinfo {author} {\bibfnamefont
  {K.}~\bibnamefont {{Wiesner}}},\ }\bibfield  {title} {\bibinfo {title}
  {{Analysis framework for the prompt discovery of compact binary mergers in
  gravitational-wave data}},\ }\href
  {https://doi.org/10.1103/PhysRevD.95.042001} {\bibfield  {journal} {\bibinfo
  {journal} {\prd}\ }\textbf {\bibinfo {volume} {95}},\ \bibinfo {eid} {042001}
  (\bibinfo {year} {2017})},\ \Eprint {https://arxiv.org/abs/1604.04324}
  {arXiv:1604.04324 [astro-ph.IM]} \BibitemShut {NoStop}%
\bibitem [{\citenamefont {{Bailyn}}\ \emph {et~al.}(1998)\citenamefont
  {{Bailyn}}, \citenamefont {{Jain}}, \citenamefont {{Coppi}},\ and\
  \citenamefont {{Orosz}}}]{Bailyn98}%
  \BibitemOpen
  \bibfield  {author} {\bibinfo {author} {\bibfnamefont {C.~D.}\ \bibnamefont
  {{Bailyn}}}, \bibinfo {author} {\bibfnamefont {R.~K.}\ \bibnamefont
  {{Jain}}}, \bibinfo {author} {\bibfnamefont {P.}~\bibnamefont {{Coppi}}},\
  and\ \bibinfo {author} {\bibfnamefont {J.~A.}\ \bibnamefont {{Orosz}}},\
  }\bibfield  {title} {\bibinfo {title} {{The Mass Distribution of Stellar
  Black Holes}},\ }\href {https://doi.org/10.1086/305614} {\bibfield  {journal}
  {\bibinfo  {journal} {Astrophys. J.}\ }\textbf {\bibinfo {volume} {499}},\
  \bibinfo {pages} {367} (\bibinfo {year} {1998})},\ \Eprint
  {https://arxiv.org/abs/astro-ph/9708032} {arXiv:astro-ph/9708032 [astro-ph]}
  \BibitemShut {NoStop}%
\bibitem [{\citenamefont {{{\"O}zel}}\ \emph {et~al.}(2012)\citenamefont
  {{{\"O}zel}}, \citenamefont {{Psaltis}}, \citenamefont {{Narayan}},\ and\
  \citenamefont {{Santos Villarreal}}}]{Ozel2012}%
  \BibitemOpen
  \bibfield  {author} {\bibinfo {author} {\bibfnamefont {F.}~\bibnamefont
  {{{\"O}zel}}}, \bibinfo {author} {\bibfnamefont {D.}~\bibnamefont
  {{Psaltis}}}, \bibinfo {author} {\bibfnamefont {R.}~\bibnamefont
  {{Narayan}}},\ and\ \bibinfo {author} {\bibfnamefont {A.}~\bibnamefont
  {{Santos Villarreal}}},\ }\bibfield  {title} {\bibinfo {title} {{On the Mass
  Distribution and Birth Masses of Neutron Stars}},\ }\href
  {https://doi.org/10.1088/0004-637X/757/1/55} {\bibfield  {journal} {\bibinfo
  {journal} {Astrophys. J.}\ }\textbf {\bibinfo {volume} {757}},\ \bibinfo
  {eid} {55} (\bibinfo {year} {2012})},\ \Eprint
  {https://arxiv.org/abs/1201.1006} {arXiv:1201.1006 [astro-ph.HE]}
  \BibitemShut {NoStop}%
\bibitem [{\citenamefont {{Abbott}}\ \emph {et~al.}(2024)\citenamefont
  {{Abbott}}, \citenamefont {{Abbott}}, \citenamefont {{Acernese}},
  \citenamefont {{Ackley}}, \citenamefont {{Adams}}, \citenamefont {{et al.}},
  \citenamefont {{LIGO Scientific Collaboration}},\ and\ \citenamefont {{the
  Virgo Collaboration}}}]{GWTC-2.1}%
  \BibitemOpen
  \bibfield  {author} {\bibinfo {author} {\bibfnamefont {R.}~\bibnamefont
  {{Abbott}}}, \bibinfo {author} {\bibfnamefont {T.~D.}\ \bibnamefont
  {{Abbott}}}, \bibinfo {author} {\bibfnamefont {F.}~\bibnamefont
  {{Acernese}}}, \bibinfo {author} {\bibfnamefont {K.}~\bibnamefont
  {{Ackley}}}, \bibinfo {author} {\bibfnamefont {C.}~\bibnamefont {{Adams}}},
  \bibinfo {author} {\bibnamefont {{et al.}}}, \bibinfo {author} {\bibnamefont
  {{LIGO Scientific Collaboration}}},\ and\ \bibinfo {author} {\bibnamefont
  {{the Virgo Collaboration}}},\ }\bibfield  {title} {\bibinfo {title}
  {{GWTC-2.1: Deep extended catalog of compact binary coalescences observed by
  LIGO and Virgo during the first half of the third observing run}},\ }\href
  {https://doi.org/10.1103/PhysRevD.109.022001} {\bibfield  {journal} {\bibinfo
   {journal} {\prd}\ }\textbf {\bibinfo {volume} {109}},\ \bibinfo {eid}
  {022001} (\bibinfo {year} {2024})}\BibitemShut {NoStop}%
\bibitem [{\citenamefont {{Abbott}}\ \emph
  {et~al.}(2017{\natexlab{b}})\citenamefont {{Abbott}}, \citenamefont
  {{Abbott}}, \citenamefont {{Abbott}}, \citenamefont {{Acernese}},
  \citenamefont {{Ackley}}, \citenamefont {{et al.}}, \citenamefont {{LIGO
  Scientific Collaboration}},\ and\ \citenamefont {{Virgo
  Collaboration}}}]{GW170817}%
  \BibitemOpen
  \bibfield  {author} {\bibinfo {author} {\bibfnamefont {B.~P.}\ \bibnamefont
  {{Abbott}}}, \bibinfo {author} {\bibfnamefont {R.}~\bibnamefont {{Abbott}}},
  \bibinfo {author} {\bibfnamefont {T.~D.}\ \bibnamefont {{Abbott}}}, \bibinfo
  {author} {\bibfnamefont {F.}~\bibnamefont {{Acernese}}}, \bibinfo {author}
  {\bibfnamefont {K.}~\bibnamefont {{Ackley}}}, \bibinfo {author} {\bibnamefont
  {{et al.}}}, \bibinfo {author} {\bibnamefont {{LIGO Scientific
  Collaboration}}},\ and\ \bibinfo {author} {\bibnamefont {{Virgo
  Collaboration}}},\ }\bibfield  {title} {\bibinfo {title} {{GW170817:
  Observation of Gravitational Waves from a Binary Neutron Star Inspiral}},\
  }\href {https://doi.org/10.1103/PhysRevLett.119.161101} {\bibfield  {journal}
  {\bibinfo  {journal} {\prl}\ }\textbf {\bibinfo {volume} {119}},\ \bibinfo
  {eid} {161101} (\bibinfo {year} {2017}{\natexlab{b}})},\ \Eprint
  {https://arxiv.org/abs/1710.05832} {arXiv:1710.05832 [gr-qc]} \BibitemShut
  {NoStop}%
\bibitem [{Note4()}]{Note4}%
  \BibitemOpen
  \bibinfo {note} {Https://emd.readthedocs.io/en/stable/index.html}\BibitemShut
  {NoStop}%
\bibitem [{\citenamefont {Quinn}\ \emph {et~al.}(2021)\citenamefont {Quinn},
  \citenamefont {Lopes-dos Santos}, \citenamefont {Dupret}, \citenamefont
  {Nobre},\ and\ \citenamefont {Woolrich}}]{Quinn2021}%
  \BibitemOpen
  \bibfield  {author} {\bibinfo {author} {\bibfnamefont {A.~J.}\ \bibnamefont
  {Quinn}}, \bibinfo {author} {\bibfnamefont {V.}~\bibnamefont {Lopes-dos
  Santos}}, \bibinfo {author} {\bibfnamefont {D.}~\bibnamefont {Dupret}},
  \bibinfo {author} {\bibfnamefont {A.~C.}\ \bibnamefont {Nobre}},\ and\
  \bibinfo {author} {\bibfnamefont {M.~W.}\ \bibnamefont {Woolrich}},\
  }\bibfield  {title} {\bibinfo {title} {Emd: Empirical mode decomposition and
  hilbert-huang spectral analyses in python},\ }\href
  {https://doi.org/10.21105/joss.02977} {\bibfield  {journal} {\bibinfo
  {journal} {Journal of Open Source Software}\ }\textbf {\bibinfo {volume}
  {6}},\ \bibinfo {pages} {2977} (\bibinfo {year} {2021})}\BibitemShut
  {NoStop}%
\bibitem [{\citenamefont {{Klimenko}}\ \emph {et~al.}(2008)\citenamefont
  {{Klimenko}}, \citenamefont {{Yakushin}}, \citenamefont {{Mercer}},\ and\
  \citenamefont {{Mitselmakher}}}]{2008Klimenko}%
  \BibitemOpen
  \bibfield  {author} {\bibinfo {author} {\bibfnamefont {S.}~\bibnamefont
  {{Klimenko}}}, \bibinfo {author} {\bibfnamefont {I.}~\bibnamefont
  {{Yakushin}}}, \bibinfo {author} {\bibfnamefont {A.}~\bibnamefont
  {{Mercer}}},\ and\ \bibinfo {author} {\bibfnamefont {G.}~\bibnamefont
  {{Mitselmakher}}},\ }\bibfield  {title} {\bibinfo {title} {{A coherent method
  for detection of gravitational wave bursts}},\ }\href
  {https://doi.org/10.1088/0264-9381/25/11/114029} {\bibfield  {journal}
  {\bibinfo  {journal} {Classical and Quantum Gravity}\ }\textbf {\bibinfo
  {volume} {25}},\ \bibinfo {eid} {114029} (\bibinfo {year} {2008})},\ \Eprint
  {https://arxiv.org/abs/0802.3232} {arXiv:0802.3232 [gr-qc]} \BibitemShut
  {NoStop}%
\bibitem [{\citenamefont {{Klimenko}}\ \emph {et~al.}(2016)\citenamefont
  {{Klimenko}}, \citenamefont {{Vedovato}}, \citenamefont {{Drago}},
  \citenamefont {{Salemi}}, \citenamefont {{Tiwari}}, \citenamefont {{Prodi}},
  \citenamefont {{Lazzaro}}, \citenamefont {{Ackley}}, \citenamefont
  {{Tiwari}}, \citenamefont {{Da Silva}},\ and\ \citenamefont
  {{Mitselmakher}}}]{2016Klimenko}%
  \BibitemOpen
  \bibfield  {author} {\bibinfo {author} {\bibfnamefont {S.}~\bibnamefont
  {{Klimenko}}}, \bibinfo {author} {\bibfnamefont {G.}~\bibnamefont
  {{Vedovato}}}, \bibinfo {author} {\bibfnamefont {M.}~\bibnamefont {{Drago}}},
  \bibinfo {author} {\bibfnamefont {F.}~\bibnamefont {{Salemi}}}, \bibinfo
  {author} {\bibfnamefont {V.}~\bibnamefont {{Tiwari}}}, \bibinfo {author}
  {\bibfnamefont {G.~A.}\ \bibnamefont {{Prodi}}}, \bibinfo {author}
  {\bibfnamefont {C.}~\bibnamefont {{Lazzaro}}}, \bibinfo {author}
  {\bibfnamefont {K.}~\bibnamefont {{Ackley}}}, \bibinfo {author}
  {\bibfnamefont {S.}~\bibnamefont {{Tiwari}}}, \bibinfo {author}
  {\bibfnamefont {C.~F.}\ \bibnamefont {{Da Silva}}},\ and\ \bibinfo {author}
  {\bibfnamefont {G.}~\bibnamefont {{Mitselmakher}}},\ }\bibfield  {title}
  {\bibinfo {title} {{Method for detection and reconstruction of gravitational
  wave transients with networks of advanced detectors}},\ }\href
  {https://doi.org/10.1103/PhysRevD.93.042004} {\bibfield  {journal} {\bibinfo
  {journal} {\prd}\ }\textbf {\bibinfo {volume} {93}},\ \bibinfo {eid} {042004}
  (\bibinfo {year} {2016})},\ \Eprint {https://arxiv.org/abs/1511.05999}
  {arXiv:1511.05999 [gr-qc]} \BibitemShut {NoStop}%
\bibitem [{\citenamefont {{Vedovato}}\ \emph {et~al.}(2022)\citenamefont
  {{Vedovato}}, \citenamefont {{Milotti}}, \citenamefont {{Prodi}},
  \citenamefont {{Bini}}, \citenamefont {{Drago}}, \citenamefont {{Gayathri}},
  \citenamefont {{Halim}}, \citenamefont {{Lazzaro}}, \citenamefont {{Lopez}},
  \citenamefont {{Miani}}, \citenamefont {{O'Brien}}, \citenamefont {{Salemi}},
  \citenamefont {{Szczepanczyk}}, \citenamefont {{Tiwari}}, \citenamefont
  {{Virtuoso}},\ and\ \citenamefont {{Klimenko}}}]{2022high-order}%
  \BibitemOpen
  \bibfield  {author} {\bibinfo {author} {\bibfnamefont {G.}~\bibnamefont
  {{Vedovato}}}, \bibinfo {author} {\bibfnamefont {E.}~\bibnamefont
  {{Milotti}}}, \bibinfo {author} {\bibfnamefont {G.~A.}\ \bibnamefont
  {{Prodi}}}, \bibinfo {author} {\bibfnamefont {S.}~\bibnamefont {{Bini}}},
  \bibinfo {author} {\bibfnamefont {M.}~\bibnamefont {{Drago}}}, \bibinfo
  {author} {\bibfnamefont {V.}~\bibnamefont {{Gayathri}}}, \bibinfo {author}
  {\bibfnamefont {O.}~\bibnamefont {{Halim}}}, \bibinfo {author} {\bibfnamefont
  {C.}~\bibnamefont {{Lazzaro}}}, \bibinfo {author} {\bibfnamefont
  {D.}~\bibnamefont {{Lopez}}}, \bibinfo {author} {\bibfnamefont
  {A.}~\bibnamefont {{Miani}}}, \bibinfo {author} {\bibfnamefont
  {B.}~\bibnamefont {{O'Brien}}}, \bibinfo {author} {\bibfnamefont
  {F.}~\bibnamefont {{Salemi}}}, \bibinfo {author} {\bibfnamefont
  {M.}~\bibnamefont {{Szczepanczyk}}}, \bibinfo {author} {\bibfnamefont
  {S.}~\bibnamefont {{Tiwari}}}, \bibinfo {author} {\bibfnamefont
  {A.}~\bibnamefont {{Virtuoso}}},\ and\ \bibinfo {author} {\bibfnamefont
  {S.}~\bibnamefont {{Klimenko}}},\ }\bibfield  {title} {\bibinfo {title}
  {{Minimally-modeled search of higher multipole gravitational-wave radiation
  in compact binary coalescences}},\ }\href
  {https://doi.org/10.1088/1361-6382/ac45da} {\bibfield  {journal} {\bibinfo
  {journal} {Classical and Quantum Gravity}\ }\textbf {\bibinfo {volume}
  {39}},\ \bibinfo {eid} {045001} (\bibinfo {year} {2022})},\ \Eprint
  {https://arxiv.org/abs/2108.13384} {arXiv:2108.13384 [gr-qc]} \BibitemShut
  {NoStop}%
\bibitem [{\citenamefont {{London}}\ \emph {et~al.}(2018)\citenamefont
  {{London}}, \citenamefont {{Khan}}, \citenamefont {{Fauchon-Jones}},
  \citenamefont {{Garc{\'\i}a}}, \citenamefont {{Hannam}}, \citenamefont
  {{Husa}}, \citenamefont {{Jim{\'e}nez-Forteza}}, \citenamefont
  {{Kalaghatgi}}, \citenamefont {{Ohme}},\ and\ \citenamefont
  {{Pannarale}}}]{2018London}%
  \BibitemOpen
  \bibfield  {author} {\bibinfo {author} {\bibfnamefont {L.}~\bibnamefont
  {{London}}}, \bibinfo {author} {\bibfnamefont {S.}~\bibnamefont {{Khan}}},
  \bibinfo {author} {\bibfnamefont {E.}~\bibnamefont {{Fauchon-Jones}}},
  \bibinfo {author} {\bibfnamefont {C.}~\bibnamefont {{Garc{\'\i}a}}}, \bibinfo
  {author} {\bibfnamefont {M.}~\bibnamefont {{Hannam}}}, \bibinfo {author}
  {\bibfnamefont {S.}~\bibnamefont {{Husa}}}, \bibinfo {author} {\bibfnamefont
  {X.}~\bibnamefont {{Jim{\'e}nez-Forteza}}}, \bibinfo {author} {\bibfnamefont
  {C.}~\bibnamefont {{Kalaghatgi}}}, \bibinfo {author} {\bibfnamefont
  {F.}~\bibnamefont {{Ohme}}},\ and\ \bibinfo {author} {\bibfnamefont
  {F.}~\bibnamefont {{Pannarale}}},\ }\bibfield  {title} {\bibinfo {title}
  {{First Higher-Multipole Model of Gravitational Waves from Spinning and
  Coalescing Black-Hole Binaries}},\ }\href
  {https://doi.org/10.1103/PhysRevLett.120.161102} {\bibfield  {journal}
  {\bibinfo  {journal} {\prl}\ }\textbf {\bibinfo {volume} {120}},\ \bibinfo
  {eid} {161102} (\bibinfo {year} {2018})},\ \Eprint
  {https://arxiv.org/abs/1708.00404} {arXiv:1708.00404 [gr-qc]} \BibitemShut
  {NoStop}%
\bibitem [{\citenamefont {{Roy}}\ \emph {et~al.}(2021)\citenamefont {{Roy}},
  \citenamefont {{Sengupta}},\ and\ \citenamefont {{Arun}}}]{2021Roy}%
  \BibitemOpen
  \bibfield  {author} {\bibinfo {author} {\bibfnamefont {S.}~\bibnamefont
  {{Roy}}}, \bibinfo {author} {\bibfnamefont {A.~S.}\ \bibnamefont
  {{Sengupta}}},\ and\ \bibinfo {author} {\bibfnamefont {K.~G.}\ \bibnamefont
  {{Arun}}},\ }\bibfield  {title} {\bibinfo {title} {{Unveiling the spectrum of
  inspiralling binary black holes}},\ }\href
  {https://doi.org/10.1103/PhysRevD.103.064012} {\bibfield  {journal} {\bibinfo
   {journal} {\prd}\ }\textbf {\bibinfo {volume} {103}},\ \bibinfo {eid}
  {064012} (\bibinfo {year} {2021})}\BibitemShut {NoStop}%
\bibitem [{\citenamefont {{Necula}}\ \emph {et~al.}(2012)\citenamefont
  {{Necula}}, \citenamefont {{Klimenko}},\ and\ \citenamefont
  {{Mitselmakher}}}]{2012NKM}%
  \BibitemOpen
  \bibfield  {author} {\bibinfo {author} {\bibfnamefont {V.}~\bibnamefont
  {{Necula}}}, \bibinfo {author} {\bibfnamefont {S.}~\bibnamefont
  {{Klimenko}}},\ and\ \bibinfo {author} {\bibfnamefont {G.}~\bibnamefont
  {{Mitselmakher}}},\ }\bibfield  {title} {\bibinfo {title} {{Transient
  analysis with fast Wilson-Daubechies time-frequency transform}},\ }in\ \href
  {https://doi.org/10.1088/1742-6596/363/1/012032} {\emph {\bibinfo {booktitle}
  {Journal of Physics Conference Series}}},\ \bibinfo {series} {Journal of
  Physics Conference Series}, Vol.\ \bibinfo {volume} {363}\ (\bibinfo
  {publisher} {IOP},\ \bibinfo {year} {2012})\ p.\ \bibinfo {pages}
  {012032}\BibitemShut {NoStop}%
\bibitem [{Note5()}]{Note5}%
  \BibitemOpen
  \bibinfo {note} {Https://github.com/linlupin/sHHT}\BibitemShut {NoStop}%
\end{thebibliography}

%

\end{document}